\documentclass[lettersize,journal]{IEEEtran}

\usepackage{amsmath,amsfonts,bm}

\def\eqref#1{equation~\ref{#1}}

\def\1{\bm{1}}

\def\rve{{\mathbf{e}}}

\def\rvu{{\mathbf{i}}}

\def\rvs{{\mathbf{s}}}

\def\rvu{{\mathbf{u}}}
\def\rvv{{\mathbf{v}}}

\def\rmA{{\mathbf{A}}}

\def\rmC{{\mathbf{C}}}
\def\rmD{{\mathbf{D}}}

\def\rmI{{\mathbf{I}}}

\def\rmS{{\mathbf{S}}}

\def\rmU{{\mathbf{U}}}
\def\rmV{{\mathbf{V}}}
\def\rmW{{\mathbf{W}}}
\def\rmX{{\mathbf{X}}}
\def\rmY{{\mathbf{Y}}}
\def\rmZ{{\mathbf{Z}}}

\DeclareMathAlphabet{\mathsfit}{\encodingdefault}{\sfdefault}{m}{sl}
\SetMathAlphabet{\mathsfit}{bold}{\encodingdefault}{\sfdefault}{bx}{n}

\def\gI{{\mathcal{I}}}

\def\gL{{\mathcal{L}}}

\def\gN{{\mathcal{N}}}

\def\gP{{\mathcal{P}}}

\def\gZ{{\mathcal{Z}}}

\def\sB{{\mathbb{B}}}

\def\sI{{\mathbb{I}}}

\def\sR{{\mathbb{R}}}

\def\sU{{\mathbb{U}}}
\def\sV{{\mathbb{V}}}

\newcommand{\KL}{D_{\mathrm{KL}}}

\usepackage{hyperref}

\usepackage{graphicx}
\usepackage{graphics}
\usepackage{subcaption}

\usepackage{mathtools}
\usepackage{amssymb}
\usepackage{amsthm}
\usepackage{amsmath}

\usepackage{multirow}
\usepackage{booktabs}
\usepackage{algorithm}
\usepackage{algorithmic}
\usepackage{wrapfig}
\usepackage{caption}
\usepackage{float}
\usepackage[normalem]{ulem}
\useunder{\uline}{\ul}{}
\usepackage{bbding}
\usepackage{pifont}
\usepackage[most]{tcolorbox}

\theoremstyle{plain}

\newtheorem{proposition}{Proposition}

\theoremstyle{definition}
\newtheorem{definition}{Definition}
\newtheorem{assumption}{Assumption}

\theoremstyle{remark}
\newtheorem{remark}{Remark}
\usepackage{thm-restate}

\begin{document}

\title{Towards Robust Cross-Domain Recommendation with Joint Identifiability of User Preference}

\author{Jing Du*, Zesheng Ye*, Bin Guo, Zhiwen Yu, Jia Wu, Jian Yang, Michael Sheng, Lina Yao
\thanks{J. Du is with the School of Computing, Macquarie University, Sydney, NSW 2109, Australia (e-mail: jing.du@mq.edu.au).}
\thanks{Z. Ye is with the School of Computing and Information Systems, University of Melbourne, Melbourne, VIC 3052, Australia (e-mail: zesheng.ye@unimelb.edu.au).}
\thanks{B. Guo and Z. Yu are with the School of Computer Science, Northwestern Polytechnical University, Xi'an, Shaanxi, China (e-mail: guobin.keio, zhiweny@gmail.com).}
\thanks{J. Wu, J. Yang, and M. Sheng are with the School of Computing, Macquarie University, Sydney, NSW 2109, Australia (e-mail: jia.wu, jian.yang, michael.sheng@mq.edu.au).}
\thanks{L. Yao is with Data 61, CSIRO, Sydney, NSW 2122, Australia, and also with the School of Computer Science and Engineering, University of New South Wales, Sydney, NSW 2015, Australia. She is also affiliated with the School of Computing, Macquarie University, Sydney, NSW 2109, Australia (e-mail: lina.yao@data61.csiro.au).}
\thanks{* J. Du and Z. Ye share equal contributions as co-first authors.}
\thanks{Manuscript received November 5, 2024; revised December 5, 2024.}
}

\markboth{Journal of \LaTeX\ Class Files,~Vol.~14, No.~8, August~2021}%
{Shell \MakeLowercase{\textit{et al.}}: A Sample Article Using IEEEtran.cls for IEEE Journals}


\maketitle

\begin{abstract}
Recent cross-domain recommendation (CDR) studies assume that disentangled domain-shared and domain-specific user representations can mitigate domain gaps and facilitate effective knowledge transfer.
However, achieving perfect disentanglement is challenging in practice, because user behaviors in CDR are highly complex, and the true underlying user preferences cannot be fully captured through observed user-item interactions alone.
Given this impracticability, we instead propose to model {\it joint identifiability} that establishes unique correspondence of user representations across domains, ensuring consistent preference modeling even when user behaviors exhibit shifts in different domains.
To achieve this, we introduce a hierarchical user preference modeling framework that organizes user representations by the neural network encoder's depth, allowing separate treatment of shallow and deeper subspaces.
In the shallow subspace, our framework models the interest centroids for each user within each domain, probabilistically determining the users' interest belongings and selectively aligning these centroids across domains to ensure fine-grained consistency in domain-irrelevant features.
For deeper subspace representations, we enforce joint identifiability by decomposing it into a shared cross-domain stable component and domain-variant components, linked by a bijective transformation for unique correspondence.
Empirical studies on real-world CDR tasks with varying domain correlations demonstrate that our method consistently surpasses state-of-the-art, even with weakly correlated tasks, highlighting the importance of joint identifiability in achieving robust CDR.

\end{abstract}


\section{Introduction}
\IEEEPARstart{C}{ross}-Domain Recommendation~(CDR) explores strategies to leverage knowledge from a source domain to improve prediction accuracy in a target domain~\cite{ming2015domain}.
In CDR, the user representations, denoted by $\rmU_{X}$ and $\rmU_{Y}$, capture latent features driving user behaviors in the source and target domains based on the user-item interactions (\textit{a.k.a} user behaviors). 
These representations are modeled through a joint distribution $p(\rmU_X, \rmU_Y)$, which captures cross-domain relationships of user behaviors, enabling us to derive $p(\rmU_X)$ and $p(\rmU_Y)$ that reflect user behaviors in each domain.
\begin{figure}
    \centering
    \includegraphics[width=0.7\linewidth]{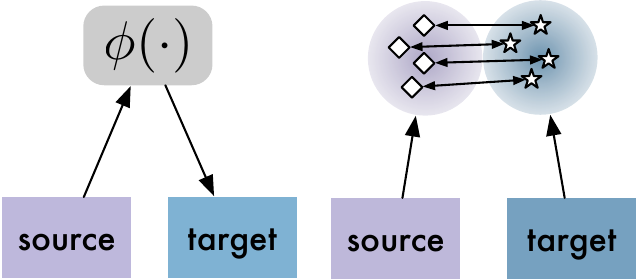}
    \caption{Bridge method (left) and Alignment method (right).}
    \label{fig:example}
\end{figure}
As Fig.~\ref{fig:example} shows, previous CDR methods primarily differ in their strategies for deriving $p(\rmU_{X})$ and $p(\rmU_Y)$ to achieve knowledge transfer.

\emph{Bridge} methods propose that user representations can be effectively transferred across source and target domains through a parameterized mapping $\phi(\cdot)$~\cite{man2017cross, zhu2021transfer}, aiming to align the marginal distributions of transformed user representations by enforcing $p(\rmU_{\rmY}) = p(\phi(\rmU_{\rmX}))$.
In this way, the latent features modeled within each domain are translated across domains. 
However, these methods treat user representations as a unified whole and rely on a blanket transformation across the entire representation. 
This indiscriminate strategy may risk negative transfer, especially when certain components exhibit lower transferability~\cite{zhao2024alignment}.
For example in the book-movie CDR context, while a user may share a common interest in the sci-fi genre in both domains, the user might prefer detailed and in-depth books but opt for lighter and more entertaining movies. 
Such discrepancies are often overlooked by {\it bridge} methods.

Alternatively, \emph{alignment} methods~\cite{liu2023chiasmal} argue that \emph{domain-invariant} information can be modeled through the behaviors from overlapping users in both domains.
Recent progress posits that user behaviors arise from latent variables $\rmZ$ and attempts to disentangle ``domain-shared'' features $\rmZ_{\rm sh}$ from ``domain-specific'' features $\rmZ_{\rm sp}^X$ and $\rmZ_{\rm sp}^Y$. 
By focusing on the shared component $\rmZ_{\rm sh}$, these methods seek to impose alignment on $p(\rmU_{\rmX} | \rmZ_{\rm sh}) = p(\rmU_{\rmY} | \rmZ_{\rm sh})$. 
This disentanglement proves useful in yielding more consistent representations of user behaviors between domains~\cite{cao2022disencdr, cao2022cross, du2023distributional}.
Nevertheless, this is predicated on--those ``domain-specific'' features are uncorrelated, such that $p(\rmZ_{\rm sp}^X, \rmZ_{\rm sp}^Y | \rmZ_{\rm sh})=p(\rmZ_{\rm sp}^X | \rmZ_{\rm sh})p(\rmZ_{\rm sp}^Y | \rmZ_{\rm sh})$--an assumption that may not hold in practice, because users tend to exhibit complex and multifaceted preferences that cannot be simply divided into domain-shared and domain-specific portions.
For instance, while writing style in books and cinematography in movies are viewed as domain-specific, a user who favors nonlinear narratives and complex character arcs in literature may similarly appreciate movies using advanced storytelling techniques, such as non-chronological timelines or layered plotlines.
This implies that the so-called ``domain-specific features'' are not strictly independent, but still reflect underlying user preferences that traverse domain boundaries. 
Thus, it is essential to \emph{capture the dependencies between these features}, even if they are modeled within respective domains.
Accordingly, instead of pursuing perfect disentanglement, we propose to establish \emph{joint identifiability for CDR}, namely
\begin{tcolorbox}[toprule=0.5pt, bottomrule=0.5pt, leftrule=0.5pt, rightrule=0.5pt, top=0.1pt, bottom=0.1pt, left=0.5pt, right=0.5pt]
    The joint distribution $p(\rmU_{\rmX}, \rmU_{\rmY})$ is {\it uniquely} determined by the latent representation $\rmZ$, such that marginal distributions $p(\rmU_{\rmX})$ and $p(\rmU_{\rmY})$, derived from $p(\rmU_{\rmX}, \rmU_{\rmY})$, {\it consistently} reflect the underlying user preferences across both domains.
\end{tcolorbox}
\begin{remark}
    By modeling $p(\rmU_{\rmX}, \rmU_{\rmY})$ in this way, {\it joint identifiability} guarantees 1) consistent interpretation of user preferences across domains: a specific $\rmZ$ always corresponds to the same underlying preference, regardless of whether the user is interacting with books or movies; and 2) consistent cross-domain behaviors, such that the same user preferences, captured by the unique $\rmZ$, always lead to consistent predictions of user behaviors, despite that shift occurs across the marginal densities $p(\rmU_\rmX)$ and $p(\rmU_\rmY)$ in different domains.
\end{remark}

Building on our prior study~\cite{du2024identifiability}, this paper presents \textbf{CIDER}, a user interest \textit{C}entroids-based framework for joint \textit{IDE}ntifiable cross-domain \textit{R}ecommendation, via a deep adaptation architecture~\cite{long2015learning}.
CIDER is designed around \emph{feature hierarchy}~(FH)~\cite{zeiler2014visualizing}, which suggests that shallow neural network layers capture general and domain-agnostic, while deeper layers adapt to more domain-relevant factors. 
Motivated by this, CIDER encodes the overall user representation in two stages. 
First, shallow layers model and align \emph{domain-irrelevant component} of both domains, capturing general user preferences and promoting cross-domain consistency. 
Then, deeper layers learn \emph{domain-relevant components} to understand behavioral variations across domains~\cite{zhang2020domain}, where we further apply the \emph{minimal change}~(MC) principle~\cite{wang2016efficient}, which favors minimal disruption to model's existing knowledge when adapting from one domain to another.
Thus, for deep-layer features, CIDER retains a shared facet, while modeling the variant facets to reflect shifts between domains\footnote{To illustrate, consider two Gaussians representing user preferences across different domains with the same mean--shared facet--consistency across domains, while the differing variances--variation facets--shifts between domains.}. 
By establishing a bijective mapping between them, CIDER ensures {\it joint identifiability} of user representations.

Concretely, CIDER implements the FH and MC principles through two key design choices.
For the domain-irrelevant component, CIDER includes a Centroids-based Probabilistic Alignment strategy (CPA) (detailed in Section~\ref{sec:shallow_cpa}) to enforce cross-domain alignment of fine-grained shallow-level features, in terms of the ``interest classes''--e.g., action or musical for ``movies'', adventure or prose for ``books''.
Rather than simply aligning all shallow-level features~\cite{du2024identifiability}, CPA computes the centroids of each interest class and measures the distance from user representations to these centroids to interpret user interests probabilistically.
When aligning user preferences across domains, CPA selectively emphasizes related interests, for instance, associating action in ``movies'' with adventure in ``books'', while reducing the correlation with unrelated interests like prose.
This targeted alignment yields a more consistent cross-domain representation of user interests.
Then, following the MC principle, CIDER further defines causal connections between shared and variant facets of domain-relevant components, to establish {\it joint identifiability}.
Specifically, CIDER focuses on identifying the unique cross-domain variation in user preferences, i.e., the variant facets, by learning a bijective transformation with a normalizing flow-based generative model~\cite{huang2018neural}, ensuring that variations in user preferences across different domains are consistently captured.

The contributions of our work are summarized as follows:
\begin{itemize}
    \item We present CIDER, a new framework striving to generate identifiable cross-domain user representations, addressing the impracticability of existing perfect disentanglement-based CDR strategy.
    \item We model user representation by decoupling domain-irrelevant feature alignment from the identifiability modeling of domain-relevant features. Using an interest centroid alignment strategy, we achieve more fine-grained domain-irrelevant interest alignment, while a normalizing flow-based approach ensures reversible transformations of domain-relevant features, structured by causal dependencies between shared and variant facets of these features.
    \item We conduct extensive experiments on six CDR tasks, showing that CIDER consistently outperforms state-of-the-art methods. Notably, CIDER exhibits pronounced benefits under weaker domain correlations, where more substantial domain shifts occur, highlighting joint identifiability as an essential factor for robust CDR performance.
\end{itemize}

\section{Related Work}
Depending on how knowledge is transferred between domains, existing CDR studies can be primarily categorized as either \emph{bridge} or \emph{alignment} methods.

\noindent \textbf{Bridge methods} focus on constructing transformation functions to facilitate information transfer across domains, often through bespoke neural architectures~\cite{jiang2015social,zhu2023unified}.
EMCDR~\cite{man2017cross} pioneers the embedding-and-mapping approach and applies a fully connected neural network to map representations from the source to the target domain.
Subsequent methods, including CDRVAE\cite{zhang2023vae}, HCDIR\cite{bi2020heterogeneous}, TMCDR\cite{zhu2021transfer}, PTUPCDR\cite{zhu2022personalized}, and DML\cite{pan2023dml} adopt similar procedure to capture cross-domain correlations.
For example, HCDIR models user relationships in a shared latent space, while TMCDR employs a meta-network for effective few-shot generalization. 
PTUPCDR further refines this meta-based network by introducing personalized transformation bridges.
DML complements these efforts with an orthogonal mapping that preserves user-item similarity, enforcing minimal distortion in CDR.

\noindent \textbf{Alignment method}. 
Existing works in this category primarily focus on sharing or aligning partial user representations in the latent space.
For example, CMF~\cite{singh2008relational} pioneered joint factorization of interaction matrices of different domains, sharing embeddings for common users across domains, while LFM~\cite{agarwal2011localized} assumes a global user-specific latent factor for user representations.
However, these methods assume complete representation sharing, potentially leading to negative transfer from non-transferable features.
To refine this, recent studies have introduced \emph{constraints} to isolate transferable components within user representations~\cite{zhang2023deep,zhao2024alignment}.
DisenCDR~\cite{cao2022disencdr} to distinguish domain-shared from unique information among users, while CDRIB~\cite{cao2022cross} further encourages the model to prioritize domain-shared information during optimization.
DPMCDR~\cite{du2023distributional} encodes user preferences into a domain-level preference distribution, aligning invariant preferences across domains using Jensen-Shannon divergence,
UniCDR\cite{cao2023towards} employs contrastive learning with domain masking to improve domain-invariant learning.
By focusing on domain-shared information, \emph{constraints}-based alignment methods report state-of-the-art performances in CDR tasks.

\noindent \textbf{CIDER differs from previous studies}.
While recent alignment methods have shown performance gains, they rely on the assumption that domain-shared and domain-specific information can be perfectly disentangled.
CIDER challenges this assumption, particularly within CDR, where it is difficult to truly distinguish domain-shared from ``domain-specific'' factors because we cannot observe the true data-generating structure of user behaviors.
Instead, CIDER acknowledges the dependencies between domain-relevant and seeks to establish identifiability between them.
In this way, CIDER achieves unique recovery of user preference across domains, delivering robust performance regardless of domain correlations.

\section{Preliminaries}
\noindent \textbf{CDR Setup}.
Define two recommendation domains $X$ and $Y$, each with distinct user, item, and interaction sets.
For domain $X$, let $\sU_X=\{ u_i \}^{|\mathbb{U}_{x}|}_{i=1}$ be the user set and $\sV_{X}=\{ v_j \}^{|\mathbb{V}_{x}|}_{j=1}$ be the item set.
Denote the user-item interaction set as $\sI_X = \{ \langle u_i, v_j \rangle | u_i \in \sU_X, v_j \in \sV_X \}$ with $\rmI_X \in \{0, 1\}^{|\mathbb{U}_{x}| \times |\mathbb{V}_{x}|}$, 
where $\langle u_i, v_j \rangle =1$ indicates user $i$ has interacted with item $j$, otherwise $ \langle u_i, v_j \rangle =0$.
Likewise, domain $Y$ includes its own $\sU_Y, \sV_Y$ and $\sI_Y$ defined similarly to those in domain $X$. 
To address the task of next-item prediction, CDR models~\cite{du2023distributional} map each user $u_i \in \sU$ and item $v_j \in \sV$ in each domain to a joint latent representation space, such that $\rvu_i \in \sR^d$ and $\rvv_j \in \sR^d$, collectively leading to $\rmU$ and $\rmV$ for $\forall u_i \in \sU$ and $\forall v_j \in \sV$.

By optimizing these representations jointly in both domains, CDR models learn to represent the \textit{underlying user preferences} in one domain to inform recommendations in the other.
The training involves maximizing the similarity (e.g., inner product $\rvu_i^{\top} \rvv_j$) for observed interactions in $\sI_X$ and $\sI_Y$\footnote{Hereafter, user behavior $\rmI$ and representation $\rmU$ may be used interchangeably, provided that $\rmI$ is essentially $\rmU \rmV^{\top}$ given computed item representations.} using a loss function.
For prediction, given a user $u_i$ in either domain, the model ranks candidate items based on their similarity to $\rvu_i$, recommending the item with the highest score as the next likely interaction.
Ultimately, learning accurate and \emph{transferable user representations} is key to effective CDR and remains a primary focus in state-of-the-art CDR studies.

\

\noindent \textbf{Disentanglement in CDR}.
To identify transferrable knowledge across domains, recent studies seek to disentangle domain-shared components from domain-specific components in user representations, based on the assumption detailed below.
\begin{assumption}\label{ass:disentanglement}
    Consider user behaviors $\rmI_X$ and $\rmI_Y$ in domains $X$ and $Y$, generated by \textit{true underlying user preference} $\rmZ^*$.
    A perfect disentanglement $(\rmZ_{\text{sh}}, \rmZ_{\text{sp}}^{X}, \rmZ_{\text{sp}}^{Y})$ characterizes the data generative process as
    \begin{equation}\label{eq:data_gen}
        \begin{aligned}
            p (\rmI_X, \rmI_Y) & = \int p(\rmI_X | \rmZ_{\text{sh}}, \rmZ_{\text{sp}}^{X}) p(\rmI_Y | \rmZ_{\text{sh}}, \rmZ_{\text{sp}}^{Y}) \\ 
                             & \qquad \quad p(\rmZ_{\text{sh}}, \rmZ_{\text{sp}}^{X}, \rmZ_{\text{sp}}^{Y}) {\rm d} \rmZ_{\text{sh}} {\rm d} \rmZ_{\text{sp}}^{X} {\rm d} \rmZ_{\text{sp}}^{Y},
        \end{aligned}
    \end{equation}
    with the following properties hold
    \begin{align}
        \rmZ_{\text{sp}}^{X} \perp\!\!\!\perp \rmZ_{\text{sp}}^{X} & \mid \rmZ_{\text{sh}}, \tag{P1} \label{ass:indep} \\
        I(\rmI_X; \rmI_Y \mid \rmZ_{\text{sh}}, \rmZ_{\text{sp}}^{X}; \rmZ_{\text{sp}}^{Y}) & = 0, \tag{P2} \label{ass:sufficiency} \\
        I(\sI_K; \rmZ_{\text{sp}}^{K^\prime} \mid \rmZ_{\text{sh}}; \rmZ_{\text{sp}}^{K}) & = 0 \; \mathrm{for} \; K, K^\prime \in \{ X, Y \}, \tag{P3} \label{ass:separation}
    \end{align}
    where $\rmZ_{\text{sh}}$ is the domain-shared and $\rmZ_{\text{sp}}^{X}$, $\rmZ_{\text{sp}}^{Y}$ are domain-specific components.
    Under perfect disentanglement, $\rmZ_{\text{sh}}$ sufficiently captures information needed to establish cross-domain relationships (\ref{ass:sufficiency}-\ref{ass:separation}), and $\rmZ_{\text{sp}}^{X}$ and $\rmZ_{\text{sp}}^{Y}$ being conditionally independent (\ref{ass:indep}).
    This forms an ideal framework to \emph{uniquely} model user preferences across domains:
    if a disentanglement truly captures the fundamental structure of user behaviors, it must also remain invariant under transformations that preserve the independence of $(\rmZ_{\text{sh}}, \rmZ_{\text{sp}}^{X}, \rmZ_{\text{sp}}^{Y})$ (i.e., \ref{ass:indep}-\ref{ass:separation}) and recovery of the original user behavior (Eq.~(\ref{eq:data_gen})).
    Achieving such conditions, however, is rarely feasible unless the data-generating process strictly follows certain specific generative structures~\cite{besserve2021theory}.
    In CDR, it is critical as we can only observe limited user-item interactions, not the exact true user preference, making perfect disentanglement of complex cross-domain user behaviors unachievable~\cite{cao2022disencdr}. 
    Addressing this gap requires us to explore alternative strategies.
\end{assumption}

\section{Joint Identifiability}\label{sec:joint_id}
Since perfect disentanglement is unachievable, in this paper, we redirect our focus to {\it joint identifiability} of representations.
\begin{definition}[Joint Identifiability]\label{def:joint_id}
    Let $\gZ$ be the space of latent representations.
    We say a representation $\rmZ$ is joint identifiable if for any two $\rmZ, \tilde{\rmZ} \in \gZ$, there exists a bijective transformation $f$ generating observationally equivalent representations up to invertible transformation: $\tilde{\rmZ} = f(\rmZ)$, such that
    \begin{equation*}
        p(\rmI_X, \rmI_Y | \rmZ) = p(\rmI_X, \rmI_Y | \tilde{\rmZ}) \cdot | \det(J_{\phi}) |^{-1}.
    \end{equation*}
\end{definition}
\noindent By Definition~\ref{def:joint_id}, we are seeking latent representations uniquely determined by observed cross-domain behaviors $\sI_X$ and $\sI_Y$. 
That is, by learning latent user preferences that yield consistent likelihoods--up to invertible transformations achieving one-to-one correspondence of parameterizations in different domains--{\it joint identifiability} ensures reliable knowledge transfer and enhances robustness to domain shift (of observed data).
We further show that it is {\it practically achievable} under mild conditions of the joint distribution of user behaviors $p(\rmI_X, \rmI_Y)$.

\begin{proposition}\label{prop:joint_id}
    Joint identifiability is achievable specifically if the following conditions hold. 
    $\mathrm{(C1)}$ The mapping $g: \gZ \to \gP(\gI_X, \gI_Y)$ defined by $p(\rmI_X, \rmI_Y | \rmZ) = g(\rmZ)$ is injective up to bijective transformation; and $\mathrm{(C2)}$ The latent representation captures a compact encoding of the cross-domain joint user behaviors, i.e., $\mathrm{dim}(\rmZ) \leq \mathrm{dim}(\mathrm{supp}(p(\rmI_X, \rmI_Y))$, where $\mathrm{supp}(\cdot)$ is the support dimensionality.
\end{proposition}
\begin{proof}
    We prove (C1) by contradiction.
    Suppose $g$ is not injective up to bijection, then there exist $\rmZ, \tilde{\rmZ} \in \gZ$ such that $\rmZ \neq f^{\circ}(\tilde{\rmZ})$ for any bijective $f^{\circ}$ but $g(\rmZ) = g(\tilde{\rmZ})$.
    This leads to $p(\rmI_X, \rmI_Y | \rmZ) = p(\rmI_X, \rmI_Y | \tilde{\rmZ})$.
    By Bayes' Rule, we also have $p(\rmZ | \rmI_X, \rmI_Y) p(\rmI_X, \rmI_Y)/p(\rmZ) = p(\tilde{\rmZ} | \rmI_X, \rmI_Y) p(\rmI_X, \rmI_Y)/p(\tilde{\rmZ})$, implying $p(\rmZ | \rmI_X, \rmI_Y) = c \cdot p(\tilde{\rmZ} | \rmI_X, \rmI_Y)$, where $c$ is a constant.
    This contradicts joint identifiability, as we cannot uniquely recover $\rmZ$ or $\tilde{\rmZ}$ from observations.
    For (C2), the support of data can be captured through optimizing representation learning models (e.g., neural networks)~\cite{ansuini2019intrinsic}.
\end{proof}
\begin{remark}
    In CDR, if Proposition~\ref{prop:joint_id} is satisfied, the user behavior prediction, i.e., $g(\rmZ)$, guarantees that the latent representation $\rmZ$ maintains a consistent, i.e., one-to-one relationship of the same user's behavior in different domains.
    This implies that the CDR model can still identify cross-domain stable factors unaffected by domain shifts of observed behaviors--whether in $p(\rmI_X)$ or $p(\rmI_Y)$--without requiring perfect disentanglement of domain-shared and specific preferences.
\end{remark}


\begin{figure}[t]
    \centering
    \includegraphics[width=\linewidth]{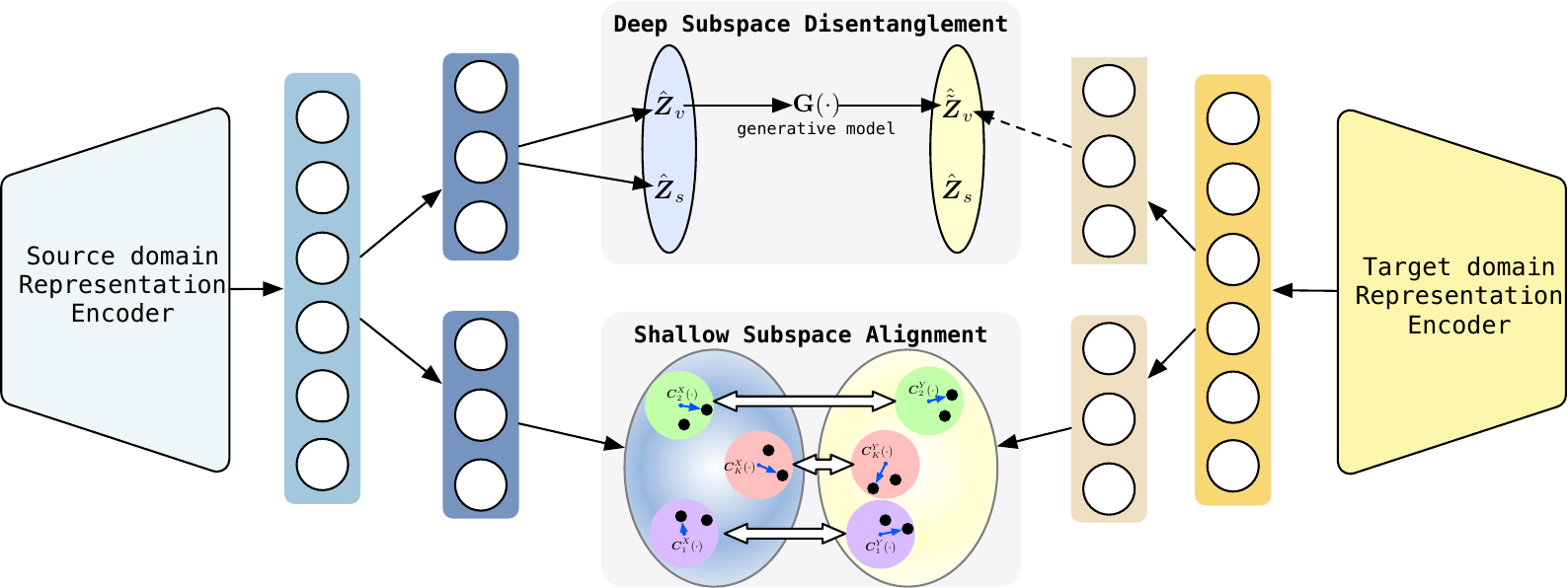}
    \caption{Overview of CIDER. It separates the user representations into shallow and deep subspaces corresponding to the depth of the neural encoder layers. The shallow subspace encodes domain-irrelevant features, enforced to be aligned between domains, while the deep subspace models domain-relevant factors, where a shared portion and variant portions are further identified by a causal data-generating structure.}
    \label{fig:overview}
\end{figure}

\section{Methodology}
\noindent \textbf{Overview}.
In this section, we introduce CIDER (conceptually depicted in Fig. (\ref{fig:overview})) that models joint identifiability for unique recovery of user behaviors in different domains with respect to the latent user representation.
Inspired by deep adaptation-like architecture~\cite{long2015learning, long2017deep}, CIDER’s structure consists of three key stages: \emph{encoding} (Section~\ref{sec:encode}), \emph{shallow-subspace alignment} (Section~\ref{sec:shallow_cpa}) and \emph{deep-subspace identification} (Section~\ref{sec:deep_id}).
First, CIDER views user-item interactions as bi-partite graphs and encodes user and item representations for each domain using Graph Neural Networks~\cite{cao2022disencdr, du2023distributional}.
Following the FH principle, user representations are then divided into shallow and deep subspaces.
In the shallow subspace, CIDER aligns domain-irrelevant representations across domains using a probabilistic strategy called user interest centroid probabilistic alignment (CPA), achieving \emph{shallow-subspace alignment}.
For the deep subspace (i.e., domain-relevant representation), CIDER further distinguishes a stable portion from variations within each domain according to the MC principle.
This separation is supported by a causal data generation graph that isolates the cross-domain shared latent factor from those within-domain variant factors, capturing shifts in user behaviors across domains.
By enabling unique recovery of the variant factors between domains, CIDER establishes \emph{joint identifiability} as demonstrated in Definition~\ref{def:joint_id}. 
It employs a generative Flow model~\cite{rezende2015variational} to parameterize a bijective transformation, ensuring a one-to-one mapping of variant factors across domains.

\begin{figure*}[h]
    \centering
    \begin{subfigure}[b]{.43\linewidth}
        \centering
        \includegraphics[width=\textwidth]{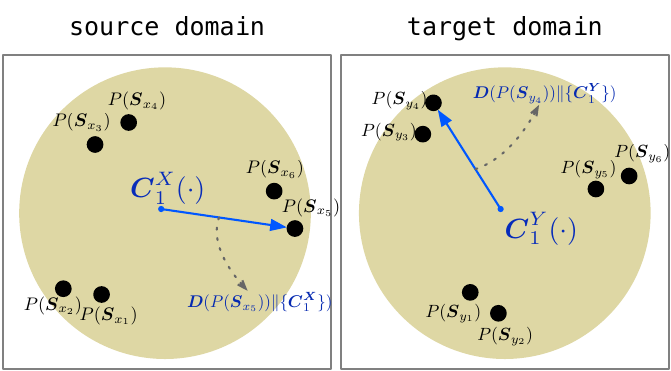}
        \caption{One centroid (K=1)}
        \label{fig:k=1}
    \end{subfigure}
    \begin{subfigure}[b]{.43\linewidth}
        \centering
        \includegraphics[width=\textwidth]{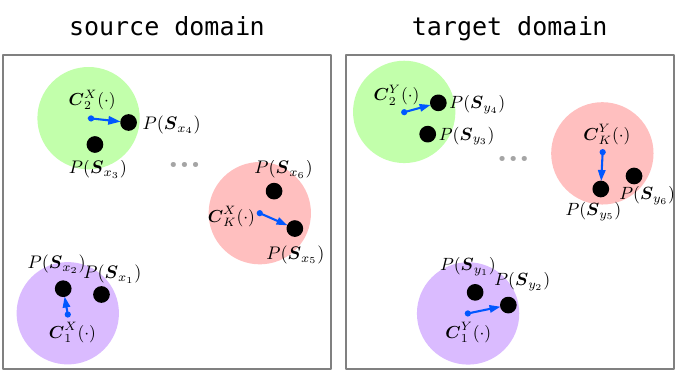}
        \caption{Multiple centroids (K \textgreater 1)}
        \label{fig:k=3}
    \end{subfigure}
    \caption{Concept of CPA. The discrepancy between domains is determined by the distance between their respective centroids. 
    We embed shallow user representations into a latent space $P(\cdot)$ characterized by multivariate Gaussian distributions. 
    Compressing $P(\cdot)$ corresponds to finding the centroids ($\boldsymbol{C}_k^{\boldsymbol{X}}(\cdot), \boldsymbol{C}_k^{\boldsymbol{Y}}(\cdot)$) using a statistical distance $\rmD_p$. 
    The discrepancy between the source and target domains is measured by the Kullback-Leibler divergence $\KL(\cdot)$ between the centroids. 
    For more complex datasets, multiple centroids are required to capture the full range of variability within the data~(\ref{fig:k=3}).}
    \label{fig: CPA}
\end{figure*}

\subsection{Representation Encoding}\label{sec:encode}

\noindent \textbf{Layered Encoding}.
We will temporarily omit domain identifier $X$ (or $Y$) and user/item indices $i$ (or $j$) for ease of notation.
Denote randomly initialized embedding for each user and item by $\rve_u$ and $\rve_v$,
CIDER stacks $K$-layer VBGEs~\cite{cao2022disencdr, du2023distributional} to encode user and item representations in both domains:
\begin{equation}
    \rmU^{k} \leftarrow \left\{ 
    \begin{aligned}
        & \mathrm{VBGE}_k(\overline{\rmA}, \rve_u),       & k = 1  \\
        & \mathrm{VBGE}_k(\overline{\rmA}, \rmU^{k-1}), & k \geq 2
    \end{aligned}
    \right. \;\; \forall u \in \sU,
\end{equation}
\begin{equation}
    \rmV^{k} \leftarrow \left\{ 
    \begin{aligned}
        & \mathrm{VBGE}_k(\overline{\rmA}^{\top}_{D}, \rve_v),       & k = 1 \\
        & \mathrm{VBGE}_k(\overline{\rmA}^{\top}_{D}, \rmV^{k-1}), & k \geq 2
    \end{aligned}
    \right.  \;\; \forall v \in \sV,
\end{equation}
where $\overline{\rmA}$ is the normalized matrix form of the interaction set $\sI$ of each domain, and $\overline{\rmA}$ is the transpose,
$\rmU$ and $\rmV$ collectively represent the users and items in each domain.
Since this is not the focus of this study, we refer interested readers to~\cite{cao2022disencdr, du2023distributional} for details of VBGE.

\

\noindent \textbf{Decoupling User Representations}.
The FH principle suggests that neural networks learn features hierarchically, with shallow layers capturing general features and deeper layers encoding more task-oriented information.
Applying this concept, we propose that shallow subspace of the user representations encodes shared and domain-irrelevant features, while the deep subspace captures more domain-relevant features.
Specifically, consider the $K$-layer VGBE, we hypothesize that domain-irrelevant features are encoded in the initial $k$ layers, with the subsequent $K-k$ layers focusing on domain-oriented features.
Accordingly, user representations $\rmU$ are formed by concatenating the shallow features $\rmS$ with deeper features $\rmD$, expressed as $\rmU_X = [ \rmS_X \| \rmD_X ]$ for domain $X$ and $\rmU_Y = [ \rmS_Y \| \rmD_Y ]$, with
\begin{equation}
    \begin{aligned}
        \rmS_X & = 
        \left[ \rmU_X^{1} \| \cdots \| \rmU_X^k \right]
        , \rmD_X = 
        \left[ \rmU_X^{k+1} \| \cdots \| \rmU_X^K \right] \\
        \rmS_Y & = 
        \left[ \rmU_Y^1 \| \cdots \| \rmU_Y^k \right]
        , \rmD_Y = 
        \left[ \rmU_Y^{k+1} \| \cdots \| \rmU_Y^K \right]
    \end{aligned}
\end{equation}
where $\|$ denotes vector concatenation.

\subsection{Shallow Subspace Alignment}
\label{sec:shallow_cpa}
Shallow subspaces aim to capture general user preferences, but variations in each domain's data distribution can shift these features due to unique statistical properties.
For example, the way certain themes like action or adventure co-occur in movie reviews may differ from book reviews, affecting the representation of these general interests. 
While a preference for action movies might align more with adventure in books than with poetry, the alignment requires finer-grained feature matching.
To achieve this, we propose user interest centroids-based probabilistic alignment (CPA) (see Fig.~\ref{fig: CPA}), which models interest classes within each domain by defining centroids based on shallow user representations.
Consider that $\rmS_X = [\rvs_X^{(i)}]_{i=1}^{N}$ and $\rmS_Y = [\rvs_Y^{(i)}]_{i=1}^{N}$ collect the representations of $N$ users in two domains.
Unlike previous approach~\cite{du2024identifiability} that imposes global alignment of all the $N$ users in $p(\rmS_X)$ and $p(\rmS_Y)$, CPA instead seeks ``cluster''-driven fine-grained alignment, which align users with those who share more consistent interests in another domain.
CPA achieves this by learning a number of interest centroids $\{ \rmC_t^{X}\}^{T}_{t=1}$ and $\{ \rmC_t^{Y}\}^{T}_{t=1}$, in each domain, as a structured compression of $p(\rmS_X)$ and $p(\rmS_Y)$.
For each user, CPA calculates the relative distances to these centroids.
The expected minimum distance indicates the probability of a user belonging to a particular interest class, serving as a matching score for alignment later.
We detail the steps of CPA below.

\

\noindent \textbf{Centroids Initialization}.
For domain $X$, we first define $T$ Gaussians, denoting a prior for each of $T$ centroids, such that $\rmC_t^{X} \triangleq \gN(\mu_t^X, \Sigma_t^X; \phi_t^X)$.
Each centroid is a statistical distribution of a particular interest.
Thus, for each user $u_X^{(i)}$, we quantify the tendency towards an interest by the Kullback-Leibler (KL) divergence (i.e., a distance measure) between the posterior $q(\rvs_{X}^{(i)} | \rve_u^{(i)})$ and each of $T$ centroids.
The expected distance across all centroids is expressed as
\begin{equation}
    \xi^{X}_{\rve^{(i)}} = \sum_{t=1}^{T} \pi_{X}^{(i)} (t) \cdot \KL \left( q(\rvs_{X}^{(i)} | \rve_u^{(i)}) \Vert \rmC_t^X \right),
\end{equation}
where $\rve_u^{(i)}$ is the initial user embedding input the encoder, the posterior $q(\rvs_{X}^{(i)} | \rve_u^{(i)})$ is also modeled as a Gaussian~\cite{apostolopoulourate}.
$\pi_{X}^{(i)}(t) \in [0, 1]$ is a soft assignment value calculated by
\begin{equation*}
    \pi_{X}^{(i)}(t) \gets \frac{ \pi_{X}(t) \cdot \exp \left( -\alpha \KL \left( q(\rvs_{X}^{(i)} | \rve_u^{(i)}) \Vert \rmC_t^X \right) \right) }{ \sum_{t^{\prime}=1}^{T} \pi_{X}(t^{\prime}) \cdot \exp \left( -\alpha \KL \left( q(\rvs_{X}^{(i)} | \rve_u^{(i)}) \Vert \rmC_{t^{\prime}}^X \right) \right) }
\end{equation*}
with $\pi_{X}(t) = \sum_{i=1}^{N} \pi_{X}^{(i)}(t)$, which is initially set to $1/T$, implying each centroid is equally likely to represent an input.
$\alpha$ is the temperature that controls the model’s confidence in assignment.
Collectively, we define the expected distance for all $N$ users as the within-domain matching score for domain $X$, denoted as $\mathcal{L}_{mX}$.
Following the same procedure, we obtain the within-domain matching score for domain $Y$ as well.
Formally,
\begin{equation}
        \mathcal{L}_{mX} = \sum_{i=1}^{N} \xi^{X}_{\rve^{(i)}}, \quad \mathcal{L}_{mY} = \sum_{i=1}^{N} \xi^{Y}_{\rve^{(i)}}
\end{equation}

\noindent \textbf{Iterative Update}.
As the training progresses, the encoder better describes the data (we will also have a better posterior), and the centroids and assignment values are iteratively updated accordingly.
At each iteration, we recalculate the assignment probabilities $\pi_{X}^{(i)}(t)$ and $\pi_{X}(t)$ over all users of domain $X$ (the same applies to domain $Y$).
Following, the parameters of Gaussian centroids in both domains $\phi^X = \{ \phi_t^{X} \}_{t=1}^{T}$ and $\phi^Y = \{ \phi_t^{Y} \}_{t=1}^{T}$ are updated with respect to the gradients of $\gL_{mX}$ and $\gL_{mY}$, respectively:
\begin{equation}\label{eq:iter_update}
    \phi^X \gets \phi^X - \eta \nabla(\mathcal{L}_{mX}), \; \phi^Y \gets \phi^Y - \eta \nabla(\mathcal{L}_{mY}),
\end{equation}
where $\eta$ is the learning rate.

\noindent \textbf{Centroids Alignment}.
Upon optimization, the centroids have encoded fine-grained interests in both domains.
To achieve domain-irrelevant interests alignment, we compute and minimize the centroid-wise distances between domains as
\begin{equation}\label{eq:centroid_alignment}
    \gL_{sa} = \sum_{t=1}^{T} \KL \left( \rmC_t^{X} \Vert \rmC_t^{Y} \right).
\end{equation}

\subsection{Deep Subspace Identification}
\label{sec:deep_id}
In line with the MC principle, CIDER further decomposes domain-relevant representations in the deep subspace, into a shared (i.e., cross-domain stable) component and variant (i.e., accounting for user behavioral shifts) components in both domain $X$ and $Y$. 
Although this resembles existing disentanglement-based alignment methods~\cite{cao2022cross, du2023distributional} at first glance, achieving strict separation in CDR is often challenging.
Therefore, we propose a causal data-generating graph to focus on joint identifiability~\cite{kong2022partial} of variant components, which provides a practical alternative to conventional disentanglement.

\begin{figure}
    \centering
    \includegraphics[width=0.9\linewidth]{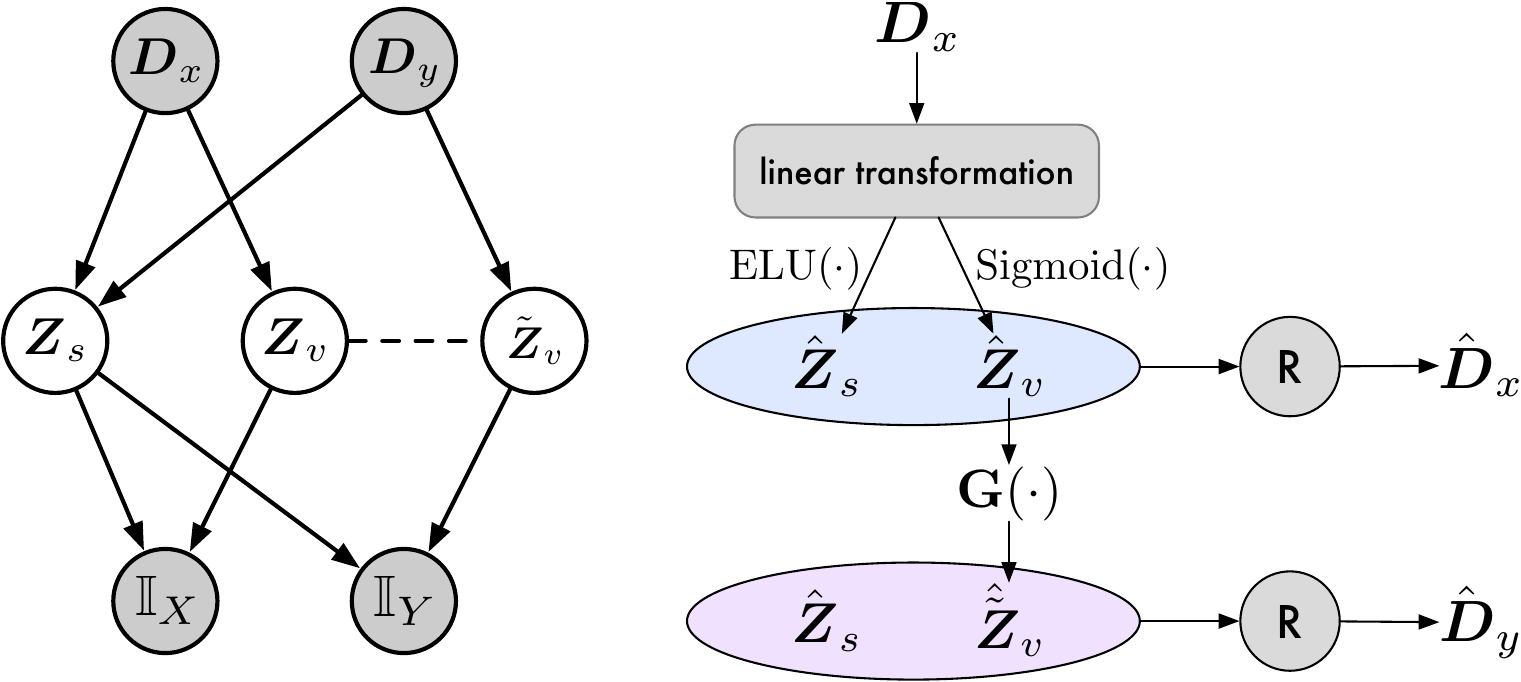}
    \caption{The data generation graph (left) and framework (right) for deep subspace disentanglement. The grey circles represent observed variables, while the dashed lines indicate correlations between variables. $R$ denotes the reparameterization trick.}
    \label{fig:deep}
\end{figure}

\

\noindent \textbf{Data Generation Graph}.
Assume, the observed user behavior data $\rmI_X$ in domain $X$ is generated from the latent mapping $g: \gZ_X \to \gI_X$, with a parallel process $g: \gZ_X \to \gI_Y$ for domain $Y$.
As Fig.~\ref{fig:deep} shows, we consider that user behaviors in each domain are jointly shaped by both: a stable part that keeps unchanged regardless of which domain, and a variant part that reflects interest variability tied to each specific domain.
Thus, $\rmZ_X$ is decomposed into $\rmZ_s$ and $\rmZ_v^{X}$, and $\rmZ_Y$ is decomposed into $\rmZ_s$ and $\rmZ_v^{Y}$ accordingly.
With the stable component $\rmZ_s$ shared between $\rmZ_X$ and $\rmZ_Y$, we need only to model the relationship between $\rmZ_v^{X}$ and $\rmZ_v^{Y}$, which serves as a surrogate for understanding the full dynamics of $\rmZ_X$ and $\rmZ_Y$.

\

\noindent \textbf{Latent Decomposition}.
Let $\rmZ_X \sim \mathcal{N}(\rmZ_s, \rmZ_v^{X})$ be the sample drawn from a Gaussian with mean $\rmZ_s$ and variance $\rmZ_v^{X}$, where $\rmZ_s = \mathrm{ELU}(\rmW_s^X \rmD_X)$ and $\rmZ_v^{X} = \mathrm{sigmoid}(\rmW_v^X \rmD_X)$ are non-linear transformations of the deep subspace features $\rmD_X$ with trainable weights $\rmW_s^X$ and $\rmW_v^X$.
This formulation also applies to $\rmZ_Y$, with trainable weight matrices $\rmW_s^Y$ and $\rmW_v^Y$.

\

\noindent \textbf{Joint Identifiability}.
Since $\rmZ_v^{X}$ and $\rmZ_v^{Y}$ represent behavior variances of the same users, we hope that $\rmZ_v^{X}$ can be uniquely mapped from $\rmZ_v^{Y}$, and vice versa.
We thus take advantage of the reversible transformation ability of Normalizing Flow~\cite{rezende2015variational} to capture a bijective relationship between $p(\rmZ_v^{X})$ and $p(\rmZ_v^{Y})$, which establishes joint identifiability between $\rmZ_v^{X}$ and $\rmZ_v^{Y}$.
Importantly, the joint identifiability between $\rmZ_v^{X}$ and $\rmZ_v^{Y}$ also generalizes to the entire $\rmZ_X$ and $\rmZ_Y$.
Specifically, given marginal distributions $p(\rmZ_v^X)$ and $p(\rmZ_v^Y)$, we derive the correlations between them through
\begin{equation}
\label{eq:flow}
\begin{aligned}
    p(\rmZ_v^Y) & = p(f^{-1}(\rmZ_v^X)) \cdot | \det (J_{f^{-1}}( \rmZ_v^X )) | \\
                & = p(\rmZ_v^X) | \det (J_f(\rmZ_v^Y) |^{-1}.
\end{aligned}
\end{equation}
Here,
$J_f(\cdot)=\frac{\partial f}{\partial \cdot}$ denotes the Jacobian matrix, 
and $\text{det}(\cdot)$ is the matrix determinant.
To overcome the limitation of restricted expressiveness caused by the bijective mapping requirement, we apply a composition of $L$ bijective layers, $\mathbf{F}=f_1 \circ \cdots \circ f_L$, enabling the model to flexibly represent complex, non-linear structures.
Eq.~(\ref{eq:flow}) now equivalently becomes:
\begin{equation}
    \begin{aligned}
        p(\rmZ_v^Y)
        = p(\rmZ_v^X) \cdot \left[\frac{\partial \rmZ_{v, 1}^{X}}{\partial \rmZ_{v, 0}^{X}}\right] \left[\frac{\partial \rmZ_{v, 2}^{X}}{\partial \rmZ_{v, 1}^{X} }\right] \cdots \left[\frac{\partial \rmZ_{v}^{X} }{\partial \rmZ_{v, {L-1}}^{X} }\right].
    \end{aligned}
\end{equation}
The specific Normalizing Flow model can be implemented as any of~\cite{papamakarios2017masked, huang2018neural, chen2018neural, rezende2020normalizing}.
Once the unique mapping relationship is established, we refine $\hat{\rmD}_X, \hat{\rmD}_Y$ using the reparameterization trick~\cite{kingma2013auto} with $\rmZ_s$, $\rmZ_{v}^X$ and $\rmZ_{v}^{Y}$,
\begin{equation}
    \begin{aligned}
        \hat{\rmD}_X = \rmZ_s + \rmZ_{v}^X \odot \mathbf{\epsilon}_X, \;
        \hat{\rmD}_Y = \rmZ_s + \rmZ_{v}^{Y} \odot \mathbf{\epsilon}_Y,
    \end{aligned}
\end{equation}
with $\mathbf{\epsilon}_X, \mathbf{\epsilon}_Y \sim N(\mathbf{0}, \mathbf{I})$, 
$\odot$ denotes element-wise multiplication.
Substituting $\hat{\rmD}_X$ and $\hat{\rmD}_Y$ back, we construct the user representations as $\hat{\rmU}_X = [\rmS_X \| \hat{\rmD}_X]$ and $\hat{\rmU}_Y = [\rmS_Y \| \hat{\rmD}_Y]$.

\subsection{Optimization}
Following the bi-directional CDR setting~\cite{cao2022cross, du2023distributional}, CIDER operates in a mini-batch manner. 
For each batch, we randomly sample groups of $B$ users $\sB_{X}$ and $\sB_Y$ from both domains with $N$ users per domain.
Each batch of sampled users, $\sB_{X}$ and $\sB_Y$, are passed through CIDER, yielding user representations $\rmU_X$ and $\rmU_Y$.
The optimization of CIDER is then driven by three objectives: shallow subspace alignment, deeper subspace identification, and recommendation prediction.

\noindent \textbf{Shallow subspace alignment}.
We integrate the within-domain matching scores~(Eq.~(\ref{eq:iter_update})) with cross-domain alignment~(Eq.~(\ref{eq:centroid_alignment})).
The loss is then defined as,
\begin{equation}
    \mathcal{L}_s = \mathcal{L}_{sa} + \mathcal{L}_{mX} + \mathcal{L}_{mY}.
\end{equation}

\noindent \textbf{Deeper subspace identification}.
The Normalizing Flow model $\mathbf{F}(\cdot)$ maximizes the predictive likelihood on domain $Y$ by applying a series of bijective transformations to $\rmZ_{v}^X$, resulting
\begin{equation}
    \mathcal{L}_d = \log p \left( \mathbf{F} \left( \rmZ_v^X \right) \right) - \sum_{l=0}^{L} \log \det \left( J_{f_{l}} \left( \rmZ_v^X \right) \right).
\end{equation}

\noindent \textbf{Recommendation Prediction}.
We use the generated $\hat{\rmD}_X$ and $\hat{\rmD}_Y$ for next-item prediction.
Additionally, the \textit{variational information bottleneck} (VIB)~\cite{alemi2016deep} is applied to maximize mutual information between $\rmZ_X$ and $\hat{\rmD}_X$ in domain $X$ (and $\rmZ_X$ and $\hat{\rmD}_X$ in domain $Y$), because in CDR contexts, VIB has demonstrated effectiveness in obtaining compact and informative representations by~\cite{cao2022disencdr, du2023distributional}.
In practice, an empirical lower bound of VIB can be approximated by minimizing binary cross-entropy between the reconstructed user representations and ground-truth interacted item representations, which should rank top in similarity among all item candidates, thereby enhancing prediction accuracy and relevance.
Formally,
\begin{equation}
    \begin{aligned}
        I(\rmZ_X; \hat{\rmD}_X) & \geq \log \sigma(\langle \rmV_X, \hat{\rmD}_X \rangle) + \log ( 1 - \sigma( \langle  \bar{\rmV}_X, \hat{\rmD}_X  \rangle ) ),
        \\
        I(\rmZ_Y; \hat{\rmD}_Y) & \geq \log \sigma(\langle \rmV_Y, \hat{\rmD}_Y \rangle) + \log ( 1 - \sigma( \langle \bar{\rmV}_Y, \hat{\rmD}_Y  \rangle ) ),
    \end{aligned}
\end{equation}
where $\sigma(\cdot)$ the sigmoid function. 
$\rmV_X$ means the representations of interacted items for each user in domain $X$, while $\bar{\rmV}_X$ denotes the non-interacted items.

The overall objective takes into account all loss terms as,
\begin{equation}
        \mathcal{L} = \mathcal{L}_s + \mathcal{L}_G + I((\boldsymbol{Z}_s, \boldsymbol{Z}_v); \hat{\boldsymbol{D}}_x) + I((\boldsymbol{Z}_s, \tilde{\boldsymbol{Z}}_v); \hat{\boldsymbol{D}}_y).
\end{equation}


\section{Experiments}
Here are the research questions we aim to answer:
\begin{itemize}
    \itemsep0em
    \item[RQ1)]
    Does CIDER outperform existing CDR methods?
    \item[RQ2)]
    How does the ratio of overlapping users affect CIDER?
    Does CIDER still demonstrate efficacy even with challenging non-overlap CDR predictions?
    \item[RQ3)]
    Do individual components of CIDER contribute positively?
    \item[RQ4)]
    How do hyperparameters impact model performance?
\end{itemize}

\subsection{Experimental setup}
\subsubsection{Datasets}
All empirical studies are conducted on the Amazon reviews datasets\footnote{https://cseweb.ucsd.edu/~jmcauley/datasets/amazon/links.html}.
We further group up 6 pairwise CDR settings based on Pearson Correlations of overlapped users to showcase the impact of domain correlations: Game-Video and Cloth-Sport are strong-correlation tasks, and Game-Cloth, Game-Sport, Video-cloth, and Video-Game are weak-correlation tasks.
All the baselines and our model are evaluated on 6 tasks to evaluate the model performance comprehensively.
The detailed information is shown in Table.\ref{tab: data}.

\subsubsection{Baselines}
We include 7 representative \textbf{Alignment} methods as baselines: \textbf{CMF}\cite{singh2008relational}, \textbf{LFM}\cite{agarwal2011localized}, \textbf{DisenCDR}~\cite{cao2022disencdr}, \textbf{UniCDR}~\cite{cao2023towards}, \textbf{CDRIB}~\cite{cao2022cross}, \textbf{DPMCDR}~\cite{du2023distributional}, and \textbf{HJID}~\cite{du2024identifiability}.
For \textbf{Bridge} methods, we employ 3 methods: \textbf{EMCDR}~\cite{man2017cross}, implemented with two variants: \textbf{EMCDR-MF} in Matrix Factorization and \textbf{EMCDR-NGCF} in Neural Graph Collaborative Filtering\cite{wang2019neural}, and \textbf{PTUPCDR}~\cite{zhu2022personalized}.

\subsubsection{Implementation details}
All overlapped users are divided into training(80\%), testing(10\%), and validation(10\%) sets. 
Non-overlapped users are exclusively used for training.
Following standard practices~\cite{krichene2020sampled}, we evaluate our model and baselines using a {\it leave-one-out} strategy. 
For each user, we test one positive item and 999 randomly selected negatives. 
3 evaluation metrics, Mean Reciprocal Rank (MRR), Hit Rate (HR@{10, 20, 30}), and Normalized Discounted Cumulative Gain (NDCG@{10, 20, 30}), are reported to measure predictive performance.
The implementations of \textbf{CMF}, \textbf{LFM}, \textbf{EMCDR}, \textbf{PTUPCDR}, \textbf{DisenCDR},  \textbf{UniCDR}, and \textbf{CDRIB} follow the open-sourced codebases, implemented with PyTorch.
We implement \textbf{DPMCDR}, \textbf{HJID} according to~\cite{du2023distributional, du2024identifiability}.
The following are the default settings.
We employ a 3-layer VBGE ($K$=3) model with 2 shallow layers ($k$=2) and one deep layer. 
The user representations are divided as $\rmS_X=[ \boldsymbol{U}_x^1 \| \boldsymbol{U}_x^2]$ for shallow subspace and $\rmD_X=[\boldsymbol{U}_x^3]$ for deep subspace ($\rmS_Y=[ \boldsymbol{U}_y^1 \| \boldsymbol{U}_y^2]$ and $\rmD_Y=[\boldsymbol{U}_y^3]$ for the target domain).
We tune embedding size $d \in$ \{8, 16, 32, 64, 128\} and user group size $N \in$ \{16, 32, 64, 128, 256, 512, 1024\} through grid search.
For shallow subspace alignment, the number of centroids is searched in \{5, 10, 15, 20\}.
We also investigate the influence of different shallow depths $k$=\{1, 2, 3, 4\}.
For deep subspace disentanglement, we use ELU activations for $\boldsymbol{Z}_s$ and Sigmoid activations for $\boldsymbol{Z}_v$ to ensure smooth saturation~\cite{grelsson2018improved} and sensitivity to different distributions~\cite{zhang2016dnn}, respectively.
In terms of Normalizing Flow, we employ 4 transformations, MAF\cite{papamakarios2017masked}, NAF\cite{huang2018neural}, NODE\cite{chen2018neural}, and NCSF\cite{rezende2020normalizing}, based on open-sourced codebases\footnote{https://github.com/francois-rozet/zuko} to assess the impact of transformation techniques.
We optimize the overall model with the Adam optimizer.

\begin{table}
    \caption{Dataset statistics.}
    \label{tab: data}
    \centering
    \resizebox{0.9\linewidth}{!}{
    \begin{tabular}{l|c|c|c|c|c|c|c}
    \hline
    \textbf{No.}                     & \textbf{Domain} & \textbf{user}   & \textbf{\#Overlap}               & \textbf{item}   & \textbf{interactions} & \textbf{Testing}                & \textbf{validation}             \\
    \hline
    \multirow{2}{*}{Task 1} & Game   & 25,025 & \multirow{2}{*}{2,179} & 12,319 & 157,721      & \multirow{2}{*}{443}   & \multirow{2}{*}{434}   \\
                            & Video  & 19,457 &                        & 8,751  & 158,984      &                        &                        \\
    \hline
    \multirow{2}{*}{Task 2} & Cloth  & 41,829 & \multirow{2}{*}{9,828} & 17,943 & 194,121      & \multirow{2}{*}{1,972} & \multirow{2}{*}{1,964} \\
                            & Sport  & 27,328 &                        & 12,655 & 170,426      &                        &                        \\
    \hline
    \multirow{2}{*}{Task 3} & Game   & 17,299 & \multirow{2}{*}{5,968} & 9,125  & 104,034      & \multirow{2}{*}{1,200} & \multirow{2}{*}{1,192} \\
                            & Cloth  & 57,929 &                        & 25,196 & 297,241      &                        &                        \\
    \hline
    \multirow{2}{*}{Task 4} & Game   & 19,465 & \multirow{2}{*}{3,442} & 9,834  & 116,914      & \multirow{2}{*}{690}   & \multirow{2}{*}{688}   \\
                            & Sport  & 38,996 &                        & 16,964 & 255,959      &                        &                        \\
    \hline
    \multirow{2}{*}{Task 5} & Video  & 19,623 & \multirow{2}{*}{2,001} & 8,989  & 167,615      & \multirow{2}{*}{401}   & \multirow{2}{*}{400}   \\
                            & Cloth  & 70,183 &                        & 31,282 & 388,224      &                        &                        \\
    \hline
    \multirow{2}{*}{Task 6} & Video  & 19,648 & \multirow{2}{*}{2,159} & 8,781  & 164,457      & \multirow{2}{*}{439}   & \multirow{2}{*}{430}   \\
                            & Sport  & 41,939 &                        & 18,208 & 278,036      &                        &                        \\
    \hline
    \end{tabular}
    }
\end{table}

\begin{table*}
    \caption{Overall performance(\%).
    Improved shows the improvement over the runner-up results.
    * indicates that the improvement is statistically significant when the CIDER and the best baseline (HJID) are compared with a paired t-test level of p$<0.05$.}
    \label{tab: overall}
    \centering
    \resizebox{\linewidth}{!}{
    \begin{tabular}{l|ccccccc|ccccccc}
    \hline
    \multicolumn{1}{c|}{\multirow{2}{*}{\textbf{Methods}}} & \multicolumn{7}{c|}{\textbf{Game}}                                                                                                                                        & \multicolumn{7}{c}{\textbf{Video}}                                                                                                                                        \\ \cline{2-15} 
    \multicolumn{1}{c|}{}                         & MRR                  & NDCG@10              & NDCG@20              & NDCG@30              & HR@10                & HR@20                 & HR@30                 & MRR                  & NDCG@10              & NDCG@20              & NDCG@30              & HR@10                & HR@20                 & HR@30                 \\ \hline
    CMF                                           & 0.91($\pm$0.08)          & 0.41($\pm$0.15)          & 0.61($\pm$0.10)          & 1.05($\pm$0.08)          & 1.04($\pm$0.31)          & 1.82($\pm$0.32)           & 3.91($\pm$0.55)           & 0.84($\pm$0.07)          & 0.45($\pm$0.08)          & 0.82($\pm$0.08)          & 0.98($\pm$0.08)          & 1.30($\pm$0.10)          & 2.86($\pm$0.14)           & 3.65($\pm$0.80)           \\
    LFM                                           & 1.68($\pm$0.13)              & 1.18($\pm$0.13)              & 1.54($\pm$0.10)               & 1.9($\pm$0.12)              & 2.44($\pm$0.16)              & 3.84($\pm$0.16)               & 5.6($\pm$0.37)                & 1.86($\pm$0.24)              & 1.36($\pm$0.22)              & 1.82($\pm$0.23)              & 2.14($\pm$0.23)              & 2.76($\pm$0.25)              & 4.56($\pm$0.27)               & 6.08($\pm$0.29) 
    \\ \hline
    EMCDR-MF                                      & 1.46($\pm$0.37)          & 0.89($\pm$0.35)          & 1.46($\pm$0.36)          & 1.80($\pm$0.41)          & 1.56($\pm$0.55)          & 3.91($\pm$0.55)           & 5.47($\pm$0.55)           & 0.98($\pm$0.55)          & 0.64($\pm$0.13)          & 0.97($\pm$0.07)          & 1.25($\pm$0.39)          & 1.82($\pm$0.38)          & 3.13($\pm$0.32)           & 4.43($\pm$0.39)           \\
    EMCDR-NGCF                                    & 1.51($\pm$0.08)          & 1.07($\pm$0.23)          & 1.65($\pm$0.10)          & 2.26($\pm$0.34)          & 2.34($\pm$1.10)          & 4.69($\pm$0.55)           & 7.55($\pm$1.69)           & 1.11($\pm$0.11)          & 0.70($\pm$0.27)          & 0.98($\pm$0.46)          & 1.14($\pm$0.36)          & 1.56($\pm$0.84)          & 2.86($\pm$0.66)           & 3.65($\pm$0.85)           \\
    PTUPCDR                                       & 1.76($\pm$0.45)          & 1.23($\pm$0.69)          & 2.00($\pm$0.55)          & 2.61($\pm$0.43)          & 2.34($\pm$1.46)          & 5.47($\pm$0.96)           & 8.33($\pm$0.32)           & 1.39($\pm$0.28)          & 0.89($\pm$0.19)          & 1.34($\pm$0.39)          & 1.72($\pm$0.27)          & 1.56($\pm$0.46)          & 3.39($\pm$0.59)           & 5.73($\pm$0.76)           \\ \hline
    DisenCDR                                      & 2.15($\pm$0.10)          & 1.52($\pm$0.21)          & 2.32($\pm$0.22)          & 3.31($\pm$0.37)          & 3.20($\pm$0.60)          & 6.41($\pm$0.81)           & 11.06($\pm$1.41)          & 2.26($\pm$0.31)          & 1.59($\pm$0.41)          & 2.81($\pm$0.52)          & 3.79($\pm$0.54)          & 3.57($\pm$0.89)          & 8.43($\pm$1.40)           & 13.02($\pm$1.45)          \\
    UniCDR                                        & 3.02($\pm$0.10)          & 2.93($\pm$0.21)          & 3.74($\pm$0.22)          & 4.21($\pm$0.34)          & 5.05($\pm$0.70)          & 8.63($\pm$0.47)           & 11.18($\pm$0.74)          & 2.51($\pm$0.38)          & 2.56($\pm$0.30)          & 3.23($\pm$0.26)          & 3.81($\pm$0.28)          & 5.14($\pm$0.44)          & 7.80($\pm$0.62)           & 10.51($\pm$0.72)          \\
    CDRIB                                         & 4.01($\pm$0.23)          & 4.52($\pm$0.27)          & 5.80($\pm$0.35)          & 6.65($\pm$0.28)          & 9.32($\pm$0.41)          & 14.65($\pm$0.74)          & 18.64($\pm$0.38)          & 3.78($\pm$0.46)          & 3.84($\pm$0.51)          & 5.22($\pm$0.51)          & 6.12($\pm$0.64)          & 7.60($\pm$0.74)          & 13.09($\pm$0.73)          & 17.33($\pm$1.36)          \\
    DPMCDR                                        & 4.45($\pm$0.36)          & 4.80($\pm$0.41)          & 6.27($\pm$0.28)          & 7.14($\pm$0.36)          & 9.51($\pm$0.67)          & 15.22($\pm$0.57)          & 19.31($\pm$0.72)          & 3.67($\pm$0.14)          & 3.83($\pm$0.14)          & 5.06($\pm$0.30)          & 5.96($\pm$0.30)          & 7.89($\pm$0.32)          & 12.80($\pm$0.93)          & 17.04($\pm$1.00)          \\
    HJID                                          & {\ul 4.63($\pm$0.10)}          & {\ul 4.97($\pm$0.06)}          & {\ul 6.57($\pm$0.10)}          & {\ul 7.43($\pm$0.11)}          & {\ul 9.61($\pm$0.21)}          & {\ul 16.07($\pm$0.20)}          & {\ul 20.11($\pm$0.17)}          & {\ul 4.39($\pm$0.21)}          & {\ul 4.67($\pm$0.19)}         & {\ul 6.00($\pm$0.12)}           & {\ul 6.95($\pm$0.22)}          & {\ul 9.16($\pm$0.15)}          & {\ul 14.48($\pm$0.22)}          & {\ul 18.95($\pm$0.41)}   \\  \hline
    \textbf{CIDER*}                                          & \textbf{4.91($\pm$0.16)}          & \textbf{5.16($\pm$0.21)}          & \textbf{6.94($\pm$0.16)}          & \textbf{7.76($\pm$0.14)}          & \textbf{9.66($\pm$0.17)}          & \textbf{16.69($\pm$0.25)}          & \textbf{20.54($\pm$0.22)}          & \textbf{4.42($\pm$0.07)}          & \textbf{4.82($\pm$0.11)}         & \textbf{6.12($\pm$0.20)}          & \textbf{7.01($\pm$0.23)}           & \textbf{9.29($\pm$0.20)}          & \textbf{14.94($\pm$0.21)}          & \textbf{19.24($\pm$0.27) }         \\
    Improved(\%)                                      & 5.84\%                & 3.75\%                & 5.45\%                & 4.37\%                & 0.60\%                & 3.72\%                 & 2.10\%                 & 0.84\%               & 3.14\%               & 2.08\%               & 0.92\%               & 1.42\%               & 3.09\%                &1.55\%                \\ \hline
    \end{tabular}
    }

    \centering
    \resizebox{\linewidth}{!}{
    \begin{tabular}{l|ccccccc|ccccccc}
    \hline
    \multicolumn{1}{c|}{\multirow{2}{*}{\textbf{Methods}}} & \multicolumn{7}{c|}{\textbf{Cloth}}                                                                                                                                       & \multicolumn{7}{c}{\textbf{Sport}}                                                                                                                                        \\ \cline{2-15} 
    \multicolumn{1}{c|}{}                         & MRR                  & NDCG@10              & NDCG@20              & NDCG@30              & HR@10                & HR@20                 & HR@30                 & MRR                  & NDCG@10              & NDCG@20              & NDCG@30              & HR@10                & HR@20                 & HR@30                 \\ \hline
    CMF                                           & 1.42($\pm$0.17)          & 1.06($\pm$0.15)          & 1.73($\pm$0.10)          & 2.00($\pm$0.21)          & 2.60($\pm$0.43)          & 4.95($\pm$0.64)           & 6.25($\pm$0.32)           & 1.29($\pm$0.12)          & 1.00($\pm$0.30)          & 1.59($\pm$0.23)          & 1.86($\pm$0.12)          & 2.60($\pm$0.96)          & 4.95($\pm$0.64)           & 6.25($\pm$0.32)           \\
    LFM                                           & 1.48($\pm$0.02)              & 1.02($\pm$0.06)              & 1.46($\pm$0.04)              & 1.84($\pm$0.03)              & 2.42($\pm$0.10)              & 4.16($\pm$0.13)               & 5.94($\pm$0.21)               & 1.56($\pm$0.22)              & 1.06($\pm$0.23)              & 1.52($\pm$0.24)              & 1.96($\pm$0.22)              & 2.44($\pm$0.31)              & 4.32($\pm$0.38)               & 6.40($\pm$0.33)
    \\ \hline
    EMCDR-MF                                      & 1.55($\pm$0.69)          & 1.27($\pm$0.11)          & 1.40($\pm$0.04)          & 2.08($\pm$0.11)          & 2.60($\pm$0.32)          & 4.17($\pm$0.32)           & 6.16($\pm$0.91)           & 2.04($\pm$0.46)          & 1.49($\pm$0.46)          & 1.94($\pm$0.38)          & 2.22($\pm$0.44)          & 2.08($\pm$0.32)          & 4.17($\pm$0.31)           & 5.21($\pm$0.32)           \\
    EMCDR-NGCF                                    & 1.98($\pm$0.69)          & 1.64($\pm$0.85)          & 1.98($\pm$0.70)          & 2.37($\pm$0.42)          & 2.34($\pm$0.96)          & 3.65($\pm$0.32)           & 5.47($\pm$0.96)           & 2.39($\pm$0.68)          & 1.92($\pm$0.73)          & 2.50($\pm$0.72)          & 2.90($\pm$0.86)          & 3.13($\pm$0.96)          & 5.47($\pm$0.96)           & 7.29($\pm$1.59)           \\
    PTUPCDR                                       & 2.09($\pm$0.30)          & 2.15($\pm$0.21)          & 2.47($\pm$0.35)          & 2.53($\pm$0.41)          & 3.47($\pm$0.32)          & 3.65($\pm$0.64)           & 6.77($\pm$0.96)           & 1.14($\pm$0.31)          & 0.55($\pm$0.29)          & 0.89($\pm$0.52)          & 1.89($\pm$0.65)          & 1.56($\pm$0.32)          & 2.86($\pm$1.28)           & 7.55($\pm$1.91)           \\ \hline
    DisenCDR                                      & 2.13($\pm$0.12)          & 1.42($\pm$0.03)          & 2.29($\pm$0.23)          & 3.11($\pm$0.52)          & 3.08($\pm$0.13)          & 6.58($\pm$0.78)           & 10.44($\pm$1.14)          & 2.58($\pm$0.18)          & 1.54($\pm$0.08)          & 2.51($\pm$0.12)          & 3.58($\pm$0.08)          & 3.22($\pm$0.07)          & 7.09($\pm$0.27)           & 12.14($\pm$0.65)          \\
    UniCDR                                        & 2.70($\pm$0.18)          & 2.70($\pm$0.23)          & 3.48($\pm$0.29)          & 4.03($\pm$0.31)          & 5.20($\pm$0.48)          & 8.31($\pm$0.74)           & 10.93($\pm$0.83)          & 1.70($\pm$0.15)          & 1.55($\pm$0.20)          & 2.11($\pm$0.23)          & 2.51($\pm$0.25)          & 3.09($\pm$0.46)          & 5.33($\pm$0.50)           & 7.20($\pm$0.57)           \\
    CDRIB                                        & 3.24($\pm$0.16)    & 3.40($\pm$0.16)    & 4.39($\pm$0.13)    & 5.14($\pm$0.17)    & 6.54($\pm$0.24)    & 10.80($\pm$0.17)    & 14.35($\pm$0.24)    & 4.17($\pm$0.13)    & 4.44($\pm$0.12)    & 5.75($\pm$0.17)    & 6.61($\pm$0.16)    & 8.68($\pm$0.25)    & 13.91($\pm$0.31)    & 17.95($\pm$0.33)    \\
    DPMCDR                                       & 2.86($\pm$0.02)          & 2.88($\pm$0.04)          & 3.97($\pm$0.05)          & 4.79($\pm$0.12)          & 6.00($\pm$0.23)          & 11.03($\pm$0.12)          & 14.19($\pm$0.27)          & 4.12($\pm$0.27)          & 4.27($\pm$0.27)          & 5.54($\pm$0.37)          & 6.46($\pm$0.29)          & 8.19($\pm$0.34)          & 13.27($\pm$0.48)          & 17.63($\pm$0.37)          \\ 
    HJID                                          & {\ul 3.43($\pm$0.24)}         & {\ul 3.50($\pm$0.30)}          & {\ul 4.53($\pm$0.31)} & {\ul 5.31($\pm$0.28)} & {\ul 6.79($\pm$0.47)} & {\ul 11.90($\pm$0.53)} & {\ul 14.54($\pm$0.39)} & {\ul 4.61($\pm$0.04)} & {\ul 4.85($\pm$0.08)} & {\ul 6.26($\pm$0.06)} & {\ul 7.15($\pm$0.07)} & {\ul 9.17($\pm$0.30)} & {\ul 14.77($\pm$0.16)} & {\ul 19.08($\pm$0.23)} \\ \hline
    \textbf{CIDER*}                                        & \textbf{3.61($\pm$0.17)}  & \textbf{3.68($\pm$0.20)} & \textbf{4.90($\pm$0.44)} & \textbf{5.63($\pm$0.16)} & \textbf{6.99($\pm$0.27)} & \textbf{12.11($\pm$0.23)} & \textbf{15.21($\pm$0.39)} & \textbf{4.88($\pm$0.06)} & \textbf{5.24($\pm$0.19)} & \textbf{6.43($\pm$0.36)} & \textbf{7.28($\pm$0.19)} & \textbf{9.74($\pm$0.27)} & \textbf{15.10($\pm$0.15)} & \textbf{19.56($\pm$0.16)} \\
    Improved(\%)                                    & 4.91\%       & 14.44\%      & 7.63\%       & 5.69\%       & 2.83\%       & 1.67\%        & 4.41\%        & 5.53\%       & 7.48\%       & 2.74\%       & 4.40\%       & 5.92\%       & 2.20\%        & 2.43\%   \\     \hline
    \end{tabular}
    }
    
    \centering
    \resizebox{\linewidth}{!}{
    \begin{tabular}{l|ccccccc|ccccccc}
    \hline
    \multicolumn{1}{c|}{\multirow{2}{*}{\textbf{Methods}}} & \multicolumn{7}{c|}{\textbf{Game}}                                                                                                                                        & \multicolumn{7}{c}{\textbf{Cloth}}                                                                                                                                        \\ \cline{2-15} 
    \multicolumn{1}{c|}{}                         & MRR                  & NDCG@10              & NDCG@20              & NDCG@30              & HR@10                & HR@20                 & HR@30                 & MRR                  & NDCG@10              & NDCG@20              & NDCG@30              & HR@10                & HR@20                 & HR@30                 \\ \hline
    CMF                                           & 1.02($\pm$0.07)          & 0.74($\pm$0.07)          & 1.05($\pm$0.09)          & 1.22($\pm$0.15)          & 2.08($\pm$0.10)          & 3.39($\pm$0.16)           & 4.17($\pm$0.47)           & 0.98($\pm$0.34)          & 0.49($\pm$0.02)          & 0.96($\pm$0.04)          & 1.28($\pm$0.09)          & 1.30($\pm$0.13)          & 3.13($\pm$0.19)           & 4.95($\pm$0.18)           \\
    LFM                                           & 1.60($\pm$0.04)              & 1.06($\pm$0.06)              & 1.52($\pm$0.07)              & 1.96($\pm$0.09)              & 2.28($\pm$0.17)              & 4.20($\pm$0.20)               & 6.28($\pm$0.40)               & 1.62($\pm$0.04)              & 1.10($\pm$0.17)              & 1.58($\pm$0.17)              & 2.00($\pm$0.15)              & 2.40($\pm$0.13)              & 4.34($\pm$0.14)               & 6.30($\pm$0.50) 
    \\ \hline
    EMCDR-MF                                      & 1.11($\pm$0.06)          & 0.69($\pm$0.06)          & 1.02($\pm$0.18)          & 1.41($\pm$0.19)          & 2.08($\pm$0.32)          & 2.86($\pm$0.48)           & 4.69($\pm$0.84)           & 1.02($\pm$0.13)          & 0.48($\pm$0.06)          & 1.01($\pm$0.24)          & 1.45($\pm$0.13)          & 1.30($\pm$0.06)          & 3.39($\pm$0.05)           & 6.51($\pm$0.32)           \\
    EMCDR-NGCF                                    & 1.22($\pm$0.08)          & 0.91($\pm$0.04)          & 1.35($\pm$0.12)          & 1.58($\pm$0.19)          & 2.86($\pm$0.64)          & 4.43($\pm$0.32)           & 5.73($\pm$0.64)           & 1.06($\pm$0.09)          & 0.46($\pm$0.25)          & 1.02($\pm$0.08)          & 1.52($\pm$0.09)          & 1.30($\pm$0.16)          & 4.17($\pm$0.32)           & 5.73($\pm$0.28)           \\
    PTUPCDR                                       & 1.57($\pm$0.04)          & 1.19($\pm$0.02)          & 1.45($\pm$0.07)          & 2.74($\pm$0.19)          & 3.39($\pm$0.64)          & 5.73($\pm$0.32)           & 7.29($\pm$0.96)           & 2.03($\pm$0.21)          & 1.55($\pm$0.23)          & 2.21($\pm$0.34)          & 3.15($\pm$0.13)          & 2.86($\pm$0.32)          & 5.47($\pm$0.32)           & 9.90($\pm$0.36)           \\ \hline
    DisenCDR                                      & 2.15($\pm$0.02)          & 1.47($\pm$0.16)          & 2.31($\pm$0.09)          & 3.04($\pm$0.35)          & 3.23($\pm$0.34)          & 6.61($\pm$0.37)           & 10.07($\pm$1.27)          & 3.20($\pm$0.54)          & 2.82($\pm$0.27)          & 3.44($\pm$0.17)          & 4.03($\pm$0.69)          & 4.95($\pm$0.84)          & 7.40($\pm$0.46)           & 10.22($\pm$0.70)          \\
    UniCDR                                        & 2.14($\pm$0.12)          & 2.10($\pm$0.19)          & 2.79($\pm$0.21)          & 3.18($\pm$0.25)          & 4.16($\pm$0.50)          & 6.92($\pm$0.61)           & 8.79($\pm$0.77)           & 2.28($\pm$0.11)          & 2.23($\pm$0.12)          & 2.95($\pm$0.17)          & 3.49($\pm$0.22)          & 4.39($\pm$0.25)          & 7.28($\pm$0.44)           & 9.80($\pm$0.71)           \\
    CDRIB                                        & 3.30($\pm$0.08)     & 3.40($\pm$0.07)          & 4.43($\pm$0.10)          & 5.15($\pm$0.08)           & 6.71($\pm$0.08)          & 10.83($\pm$0.18)          & 14.18($\pm$0.09)    & 
    3.58($\pm$0.07)          & 3.63($\pm$0.08)          & 4.70($\pm$0.09)          & 5.48($\pm$0.13)          & 6.86($\pm$0.12)          & 11.10($\pm$0.18)          & 14.79($\pm$0.42)          \\
    DPMCDR                                       &  3.89($\pm$0.19)           &  4.06($\pm$0.24)        &  5.26($\pm$0.28)         &  6.00($\pm$0.27)         &  7.76($\pm$0.44)         &  12.54($\pm$0.60)         &  15.94($\pm$0.59)          &  3.61($\pm$0.10)          &  3.72($\pm$0.13)         &  4.82($\pm$0.09)         &  5.56($\pm$0.10)         &  7.19($\pm$0.27)         &  11.59($\pm$0.08)         &  15.05($\pm$0.13)         \\
    HJID                                         & {\ul 4.10($\pm$0.06)} & {\ul 4.48($\pm$0.04)} & {\ul 5.55($\pm$0.05)} & {\ul 6.28($\pm$0.05)} & {\ul 8.76($\pm$0.07)} & {\ul 13.04($\pm$0.18)} & {\ul 16.45($\pm$0.15)} & {\ul 3.81($\pm$0.05)} & {\ul 4.00($\pm$0.09)} & {\ul 5.14($\pm$0.05)} & {\ul 5.83($\pm$0.08)} & {\ul 7.72($\pm$0.20)} & {\ul 12.27($\pm$0.12)} & {\ul 16.52($\pm$0.23)} \\ \hline
    \textbf{CIDER*}                                        & \textbf{4.24($\pm$0.08)} & \textbf{4.59($\pm$0.07)} & \textbf{5.76($\pm$0.15)} & \textbf{6.43($\pm$0.07)} & \textbf{8.98($\pm$0.09)} & \textbf{13.59($\pm$0.34)} & \textbf{16.74($\pm$0.25)} & \textbf{3.87($\pm$0.26)} & \textbf{4.19($\pm$0.11)} & \textbf{5.49($\pm$0.14)} & \textbf{5.98($\pm$0.19)} & \textbf{7.84($\pm$0.23)} & \textbf{13.14($\pm$0.15)} & \textbf{16.88($\pm$0.17)} \\
    Improved(\%)                                    & 3.26\%                   & 2.34\%                   & 3.72\%                   & 2.40\%                   & 2.42\%                   & 4.07\%                    & 1.75\%                    & 1.62\%                   & 4.53\%                   & 6.45\%                   & 2.50\%                   & 1.50\%                   & 6.67\%                    & 2.11\%           \\ \hline
    \end{tabular}
    }

    \centering
    \resizebox{\linewidth}{!}{
    \begin{tabular}{l|ccccccc|ccccccc}
    \hline
    \multicolumn{1}{c|}{\multirow{2}{*}{\textbf{Methods}}} & \multicolumn{7}{c|}{\textbf{Game}}                                                                                                                                        & \multicolumn{7}{c}{\textbf{Sport}}                                                                                                                                        \\ \cline{2-15} 
    \multicolumn{1}{c|}{}                         & MRR                  & NDCG@10              & NDCG@20              & NDCG@30              & HR@10                & HR@20                 & HR@30                 & MRR                  & NDCG@10              & NDCG@20              & NDCG@30              & HR@10                & HR@20                 & HR@30                 \\ \hline
    CMF                                           & 1.45($\pm$0.28)          & 1.60($\pm$0.13)          & 2.12($\pm$0.19)          & 2.56($\pm$0.23)          & 4.13($\pm$0.61)          & 5.88($\pm$0.72)           & 8.29($\pm$0.84)           & 1.52($\pm$0.21)          & 1.17($\pm$0.26)          & 1.43($\pm$0.42)          & 1.76($\pm$0.37)          & 2.49($\pm$0.14)          & 4.21($\pm$0.19)           & 5.43($\pm$0.85)           \\
    LFM                                           & 1.48($\pm$0.05)              & 1.02($\pm$0.04)              & 1.46($\pm$0.16)              & 1.84($\pm$0.12)              & 2.42($\pm$0.15)              & 4.16($\pm$0.25)               & 5.94($\pm$0.33)               & 1.56($\pm$0.11)              & 1.06($\pm$0.14)              & 1.52($\pm$0.24)              & 1.96($\pm$0.28)              & 2.44($\pm$0.24)              & 4.32($\pm$0.27)               & 6.40($\pm$0.54) 
    \\ \hline
    EMCDR-MF                                      & 1.58($\pm$0.35)          & 1.60($\pm$0.13)          & 2.12($\pm$0.19)          & 2.12($\pm$0.54)          & 3.34($\pm$0.15)          & 5.09($\pm$0.19)           & 5.99($\pm$0.23)           & 1.39($\pm$0.25)          & 1.44($\pm$0.21)          & 1.89($\pm$0.47)          & 2.50($\pm$0.52)          & 3.01($\pm$0.32)          & 5.83($\pm$0.28)           & 7.36($\pm$0.71)           \\
    EMCDR-NGCF                                    & 1.35($\pm$0.27)          & 0.97($\pm$0.33)          & 1.62($\pm$0.50)          & 2.12($\pm$0.54)          & 2.68($\pm$0.28)          & 7.95($\pm$0.66)           & 8.63($\pm$0.54)           & 1.86($\pm$0.06)          & 1.15($\pm$0.09)          & 1.66($\pm$0.29)          & 2.11($\pm$0.36)          & 2.86($\pm$0.14)          & 4.95($\pm$0.08)           & 7.36($\pm$0.27)           \\
    PTUPCDR                                       & 1.54($\pm$0.30)          & 1.35($\pm$0.14)          & 2.05($\pm$0.06)          & 2.27($\pm$0.08)          & 2.42($\pm$0.09)          & 5.02($\pm$0.55)           & 6.92($\pm$0.14)           & 1.56($\pm$0.23)          & 1.33($\pm$0.14)          & 2.45($\pm$0.20)          & 2.78($\pm$0.02)          & 2.90($\pm$0.18)          & 7.02($\pm$0.37)           & 8.94($\pm$0.50)           \\ \hline
    DisenCDR                                      & 2.18($\pm$0.08)          & 1.50($\pm$0.03)          & 2.80($\pm$0.38)          & 3.63($\pm$0.57)          & 3.32($\pm$0.19)          & 8.52($\pm$1.59)           & 12.42($\pm$1.43)          & 2.33($\pm$0.21)          & 1.72($\pm$0.22)          & 2.82($\pm$0.39)          & 3.69($\pm$0.48)          & 3.87($\pm$0.60)          & 8.28($\pm$1.28)           & 12.40($\pm$1.72)          \\
    UniCDR                                        & 2.65($\pm$0.11)          & 2.62($\pm$0.08)          & 3.35($\pm$0.06)          & 3.92($\pm$0.10)          & 4.95($\pm$0.03)          & 7.87($\pm$0.46)           & 10.56($\pm$0.51)          & 2.60($\pm$0.42)          & 2.59($\pm$0.50)          & 3.38($\pm$0.58)          & 3.83($\pm$0.60)          & 4.94($\pm$0.85)          & 8.07($\pm$1.16)           & 10.21($\pm$1.24)          \\
    CDRIB                                        & 3.64($\pm$0.10)    & 3.88($\pm$0.10)    & 4.97($\pm$0.11)    & 5.71($\pm$0.09)    & 7.71($\pm$0.08)    & 12.04($\pm$0.15)    & 15.52($\pm$0.15)    & 4.25($\pm$0.29)    & 4.50($\pm$0.29)    & 5.83($\pm$0.36)    & 6.64($\pm$0.44)    & 8.67($\pm$0.42)    & 13.95($\pm$0.72)    & 17.76($\pm$1.07)    \\
    DPMCDR                                       & 3.43($\pm$0.09)          & 3.61($\pm$0.12)          & 4.82($\pm$0.08)          & 5.58($\pm$0.08)          & 7.28($\pm$0.24)          & 12.07($\pm$0.24)          & 15.67($\pm$0.38)          & 4.01($\pm$0.22)          & 4.22($\pm$0.38)          & 5.56($\pm$0.25)          & 6.39($\pm$0.21)          & 8.36($\pm$0.43)          & 13.69($\pm$0.31)          & 17.63($\pm$0.16)          \\ 
    HJID                                          & {\ul 3.92($\pm$0.06)} & {\ul 4.17($\pm$0.09)} & {\ul 5.23($\pm$0.06)} & {\ul 5.97($\pm$0.09)} & {\ul 7.96($\pm$0.18)} & {\ul 12.22($\pm$0.07)} & {\ul 15.70($\pm$0.26)} & {\ul 4.60($\pm$0.10)} & {\ul 4.91($\pm$0.10)} & {\ul 6.32($\pm$0.08)} & {\ul 7.20($\pm$0.07)} & {\ul 9.45($\pm$0.21)} & {\ul 15.05($\pm$0.08)} & {\ul 19.22($\pm$0.19)} \\  \hline
    \textbf{CIDER*}                                        & \textbf{4.08($\pm$0.13)} & \textbf{4.27($\pm$0.14)} & \textbf{5.49($\pm$0.09)} & \textbf{6.07($\pm$0.15)} & \textbf{8.18($\pm$0.14)} & \textbf{12.56($\pm$0.18)} & \textbf{16.24($\pm$0.25)} & \textbf{4.67($\pm$0.07)} & \textbf{4.96($\pm$0.16)} & \textbf{6.39($\pm$0.05)} & \textbf{7.35($\pm$0.17)} & \textbf{9.63($\pm$0.33)} & \textbf{15.41($\pm$0.13)} & \textbf{19.72($\pm$0.34)} \\
    Improved(\%)                                    & 4.02\%                & 2.44\%                & 2.97\%                & 1.69\%                & 2.71\%                & 2.72\%                 & 3.36\%                 & 1.41\%                & 1.03\%                & 1.17\%                & 2.14\%                & 1.90\%                & 2.34\%                 & 2.55\%                  \\ \hline
    \end{tabular}
    }
    
    \centering
    \resizebox{\linewidth}{!}{
    \begin{tabular}{l|ccccccc|ccccccc}
    \hline
    \multicolumn{1}{c|}{\multirow{2}{*}{\textbf{Methods}}} & \multicolumn{7}{c|}{\textbf{Video}}                                                                                                                                       & \multicolumn{7}{c}{\textbf{Cloth}}                                                                                                                                        \\ \cline{2-15} 
    \multicolumn{1}{c|}{}                         & MRR                  & NDCG@10              & NDCG@20              & NDCG@30              & HR@10                & HR@20                 & HR@30                 & MRR                  & NDCG@10              & NDCG@20              & NDCG@30              & HR@10                & HR@20                 & HR@30                 \\ \hline
    CMF                                           & 1.01($\pm$0.04)          & 0.45($\pm$0.14)          & 0.70($\pm$0.09)          & 1.20($\pm$0.11)          & 1.04($\pm$0.32)          & 2.08($\pm$0.64)           & 4.43($\pm$0.84)           & 1.20($\pm$0.13)          & 0.73($\pm$0.07)          & 1.45($\pm$0.20)          & 1.89($\pm$0.14)          & 1.82($\pm$0.32)          & 4.69($\pm$0.55)           & 6.77($\pm$0.32)           \\
    LFM                                           & 1.55($\pm$0.08)              & 1.07($\pm$0.08)              & 1.53($\pm$0.17)              & 1.93($\pm$0.13)              & 2.54($\pm$0.21)              & 4.37($\pm$0.31)               & 6.23($\pm$0.30)               & 1.64($\pm$0.12)              & 1.11($\pm$0.13)              & 1.59($\pm$0.15)              & 2.06($\pm$0.25)              & 2.56($\pm$0.16)              & 4.536($\pm$0.27)              & 6.72($\pm$0.46) 
    \\ \hline
    EMCDR-MF                                      & 0.87($\pm$0.49)          & 0.49($\pm$0.14)          & 0.80($\pm$0.09)          & 1.19($\pm$0.49)          & 1.56($\pm$0.32)          & 2.86($\pm$0.55)           & 4.69($\pm$0.32)           & 1.01($\pm$0.09)          & 0.70($\pm$0.07)          & 1.03($\pm$0.21)          & 1.36($\pm$0.17)          & 2.08($\pm$0.32)          & 3.39($\pm$0.84)           & 4.95($\pm$0.64)           \\
    EMCDR-NGCF                                    & 2.11($\pm$0.29)          & 1.63($\pm$0.36)          & 1.88($\pm$0.43)          & 2.27($\pm$0.45)          & 2.34($\pm$0.55)          & 3.39($\pm$0.84)           & 6.25($\pm$0.91)           & 1.00($\pm$0.12)          & 0.51($\pm$0.03)          & 1.15($\pm$0.16)          & 1.53($\pm$0.24)          & 1.56($\pm$0.09)          & 4.17($\pm$0.32)           & 5.99($\pm$0.84)           \\
    PTUPCDR                                       & 1.86($\pm$0.15)          & 1.77($\pm$0.19)          & 2.01($\pm$0.21)          & 2.52($\pm$0.29)          & 3.91($\pm$0.55)          & 4.95($\pm$0.78)           & 7.29($\pm$1.09)           & 1.45($\pm$0.04)          & 1.37($\pm$0.06)          & 2.01($\pm$0.33)          & 2.23($\pm$0.27)          & 3.65($\pm$0.32)          & 6.25($\pm$0.59)           & 7.29($\pm$0.91)           \\ \hline
    DisenCDR                                      & 2.12($\pm$0.04)          & 1.50($\pm$0.10)          & 2.49($\pm$0.07)          & 3.30($\pm$0.07)          & 3.45($\pm$0.26)          & 7.40($\pm$0.15)           & 11.23($\pm$0.26)          & 2.16($\pm$0.01)          & 1.60($\pm$0.05)          & 2.44($\pm$0.04)          & 3.26($\pm$0.05)          & 3.53($\pm$0.12)          & 6.88($\pm$0.16)           & 10.78($\pm$0.31)          \\
    UniCDR                                        & 1.71($\pm$0.19)          & 1.63($\pm$0.21)          & 2.14($\pm$0.20)          & 2.48($\pm$0.21)          & 3.29($\pm$0.32)          & 5.32($\pm$0.26)           & 6.92($\pm$0.31)           & 2.93($\pm$0.16)          & 2.94($\pm$0.18)          & 3.89($\pm$0.25)          & 4.57($\pm$0.27)          & 5.84($\pm$0.39)          & 9.60($\pm$0.67)           & 12.82($\pm$0.77)          \\
    CDRIB                                        & 4.13($\pm$0.09)    & 4.30($\pm$0.07)    & 5.70($\pm$0.11)    & 6.65($\pm$0.13)    & 8.56($\pm$0.07)    & 14.18($\pm$0.20)    & 18.63($\pm$0..36)   & 3.42($\pm$0.11)    & 3.52($\pm$0.14)    & 4.54($\pm$0.10)    & 5.26($\pm$0.11)    & 6.84($\pm$0.24)    & 10.79($\pm$0.12)    & 14.21($\pm$0.06)    \\
    DPMCDR                                       & 3.94($\pm$0.13)          & 4.07($\pm$0.16)          & 5.44($\pm$0.25)          & 6.45($\pm$0.25)          & 8.27($\pm$0.34)          & 13.76($\pm$0.72)          & 18.48($\pm$0.69)          & 3.31($\pm$0.26)          & 3.32($\pm$0.30)          & 4.40($\pm$0.26)          & 5.16($\pm$0.21)          & 6.41($\pm$0.39)          & 10.59($\pm$0.28)          & 14.17($\pm$0.16)          \\ 
    HJID                                          & {\ul 4.56($\pm$0.08)} & {\ul 4.82($\pm$0.08)} & {\ul 6.19($\pm$0.08)} & {\ul 7.15($\pm$0.07)} & {\ul 9.46($\pm$0.06)} & {\ul 14.92($\pm$0.10)} & {\ul 19.42($\pm$0.14)} & {\ul 3.69($\pm$0.22)} & {\ul 3.82($\pm$0.22)} & {\ul 4.82($\pm$0.24)} & {\ul 5.55($\pm$0.21)} & {\ul 7.22($\pm$0.17)} & {\ul 11.22($\pm$0.29)} & {\ul 14.66($\pm$0.20)} \\  \hline

    \textbf{CIDER*}                                         & \textbf{4.88($\pm$0.05)} & \textbf{5.18($\pm$0.07)} & \textbf{6.68($\pm$0.17)} & \textbf{7.71($\pm$0.23)} & \textbf{10.00($\pm$0.14)} & \textbf{15.95($\pm$0.29)} & \textbf{20.79($\pm$0.39)} & \textbf{3.96($\pm$0.16)} & \textbf{4.04($\pm$0.35)} & \textbf{5.10($\pm$0.07)} & \textbf{5.80($\pm$0.24)} & \textbf{7.43($\pm$0.17)} & \textbf{11.66($\pm$0.06)} & \textbf{15.11($\pm$0.24)}     \\
    Improved(\%)                                    & 6.66\%                & 7.07\%                & 7.30\%                & 7.25\%                & 5.40\%                 & 6.43\%                 & 6.61\%                 & 6.93\%                & 5.53\%                & 5.50\%                & 4.38\%                & 2.92\%                & 2.86\%                 & 3.01\%                \\ \hline
    \end{tabular}
    }
    
    \centering
    \resizebox{\linewidth}{!}{
    \begin{tabular}{l|ccccccc|ccccccc}
    \hline
    \multicolumn{1}{c|}{\multirow{2}{*}{\textbf{Methods}}} & \multicolumn{7}{c|}{\textbf{Video}}                                                                                                                                        & \multicolumn{7}{c}{\textbf{Sport}}                                                                                                                                        \\ \cline{2-15} 
    \multicolumn{1}{c|}{}                         & MRR                  & NDCG@10              & NDCG@20              & NDCG@30              & HR@10                 & HR@20                 & HR@30                 & MRR                  & NDCG@10              & NDCG@20              & NDCG@30              & HR@10                & HR@20                 & HR@30                 \\ \hline
    CMF                                           & 1.32($\pm$0.41)          & 0.85($\pm$0.22)          & 1.12($\pm$0.17)          & 1.45($\pm$0.15)          & 1.56($\pm$0.28)           & 2.60($\pm$0.24)           & 4.17($\pm$0.28)           & 0.97($\pm$0.09)          & 0.45($\pm$0.04)          & 0.85($\pm$0.04)          & 1.01($\pm$0.02)          & 1.04($\pm$0.05)          & 2.60($\pm$0.17)           & 3.39($\pm$0.22)           \\ 
    LFM                                           & 1.31($\pm$0.07)              & 0.90($\pm$0.10)              & 1.29($\pm$0.19)              & 1.63($\pm$0.21)              & 2.14($\pm$0.18)               & 3.68($\pm$0.25)               & 5.26($\pm$0.26)               & 1.38($\pm$0.20)              & 0.94($\pm$0.24)              & 1.33($\pm$0.15)              & 1.73($\pm$0.15)              & 2.15($\pm$0.15)              & 3.82($\pm$0.27)               & 5.66($\pm$0.28) 
    \\ \hline
    EMCDR-MF                                      & 1.12($\pm$0.22)          & 0.53($\pm$0.21)          & 0.73($\pm$0.13)          & 1.02($\pm$0.08)          & 1.04($\pm$0.05)           & 1.82($\pm$0.06)           & 3.13($\pm$0.08)           & 0.99($\pm$0.07)          & 0.66($\pm$0.02)          & 0.97($\pm$0.13)          & 1.25($\pm$0.23)          & 1.82($\pm$0.32)          & 3.13($\pm$0.32)           & 5.47($\pm$0.46)           \\
    EMCDR-NGCF                                    & 1.16($\pm$0.13)          & 0.77($\pm$0.14)          & 0.96($\pm$0.14)          & 1.41($\pm$0.21)          & 1.82($\pm$0.32)           & 2.60($\pm$0.32)           & 4.69($\pm$0.55)           & 1.07($\pm$0.22)          & 0.62($\pm$0.31)          & 1.14($\pm$0.37)          & 1.47($\pm$0.41)          & 1.56($\pm$0.55)          & 3.65($\pm$0.59)           & 5.21($\pm$0.61)           \\
    PTUPCDR                                       & 1.43($\pm$0.37)          & 1.11($\pm$0.32)          & 1.55($\pm$0.35)          & 1.82($\pm$0.36)          & 2.60($\pm$0.32)           & 4.43($\pm$0.64)           & 5.73($\pm$0.55)           & 1.42($\pm$0.20)          & 0.82($\pm$0.25)          & 2.37($\pm$0.22)          & 2.59($\pm$0.11)          & 2.16($\pm$0.32)          & 3.46($\pm$0.55)           & 6.77($\pm$0.69)           \\ \hline
    DisenCDR                                      & 2.22($\pm$0.09)          & 1.34($\pm$0.28)          & 2.31($\pm$0.29)          & 3.27($\pm$0.19)          & 2.98($\pm$0.62)           & 6.88($\pm$0.63)           & 11.42($\pm$0.27)          & 2.35($\pm$0.34)          & 1.99($\pm$0.49)          & 3.15($\pm$0.59)          & 3.86($\pm$0.44)          & 4.66($\pm$1.02)          & 9.25($\pm$1.42)           & 12.61($\pm$0.76)          \\
    UniCDR                                        & 2.97($\pm$0.11)          & 3.06($\pm$0.10)          & 3.81($\pm$0.21)          & 4.42($\pm$0.30)          & 5.94($\pm$0.26)           & 8.89($\pm$0.73)           & 11.74($\pm$1.17)          & 2.95($\pm$0.20)          & 2.97($\pm$0.23)          & 3.89($\pm$0.28)          & 4.52($\pm$0.28)          & 5.80($\pm$0.40)          & 9.46($\pm$0.62)           & 12.45($\pm$0.68)          \\
    CDRIB                                        & 4.74($\pm$0.19)    & 5.04($\pm$0.24)    & 6.44($\pm$0.32)    & 7.40($\pm$0.37)    & 9.73($\pm$0.53)     & 15.26($\pm$0.89)    & 19.79($\pm$1.10)          & 4.56($\pm$0.12)    & 4.88($\pm$0.13)    & 6.08($\pm$0.15)    & 6.92($\pm$0.21)    & 9.17($\pm$0.23)    & 13.91($\pm$0.34)    & 17.89($\pm$0.64)    \\
    DPMCDR                                       & 4.57($\pm$0.25)          & 4.81($\pm$0.29)          & 6.28($\pm$0.28)          & 7.28($\pm$0.27)          & 9.42($\pm$0.38)           & 15.25($\pm$0.35)          & 19.94($\pm$0.33)    & 4.39($\pm$0.08)          & 4.62($\pm$0.15)          & 5.87($\pm$0.15)          & 6.74($\pm$0.12)          & 8.64($\pm$0.32)          & 13.67($\pm$0.37)          & 17.78($\pm$0.23)          \\ 
    HJID                                         & {\ul 5.05($\pm$0.27)} & {\ul 5.53($\pm$0.33)} & {\ul 6.95($\pm$0.29)} & {\ul 7.96($\pm$0.29)} & {\ul 10.86($\pm$0.49)} & {\ul 16.54($\pm$0.32)} & {\ul 21.33($\pm$0.34)} & {\ul 4.91($\pm$0.10)} & {\ul 5.28($\pm$0.09)} & {\ul 6.59($\pm$0.11)} & {\ul 7.46($\pm$0.08)} & {\ul 9.90($\pm$0.05)} & {\ul 15.13($\pm$0.19)} & {\ul 19.22($\pm$0.06)} \\ \hline
    \textbf{CIDER*}                                         & \textbf{5.35($\pm$0.32)} & \textbf{5.71($\pm$0.16)} & \textbf{7.28($\pm$0.09)} & \textbf{8.26($\pm$0.34)} & \textbf{11.07($\pm$0.30)} & \textbf{17.04($\pm$0.35)} & \textbf{21.73($\pm$0.46)} & \textbf{5.00($\pm$0.08)} & \textbf{5.43($\pm$0.22)} & \textbf{6.76($\pm$0.15)} & \textbf{7.85($\pm$0.24)} & \textbf{10.81($\pm$0.21)} & \textbf{15.62($\pm$0.15)} & \textbf{19.73($\pm$0.19)} \\
    Improved(\%)                                    & 5.66\%                & 3.26\%                & 4.65\%                & 3.67\%                & 1.92\%                 & 2.91\%                 & 1.82\%                 & 1.72\%                & 2.80\%                & 2.50\%                & 4.91\%                & 8.44\%                 & 3.17\%                 & 2.59\%                               \\ \hline
    \end{tabular}
    }
\end{table*}

\subsection{Overall performance (RQ1)}
As summarized in Table \ref{tab: overall}, we present the average performance (with standard deviation) across six independent runs for each task. 
\textbf{Bold} highlights the best results, while {\ul underline} indicates second-best.
Overall, CIDER consistently surpasses state-of-the-art benchmarks across all scenarios, demonstrating substantial improvements. 
In the following section, we delve into how our model compares to existing approaches regarding Alignment and Bridge methods.

\subsubsection{Comparison with Bridge methods}
Generally, these models incorporate bespoke architectures to endow them with expressiveness for capturing user behavior.
For instance, EMCDR-NGCF and PTUPCDR employ a graph structure and a meta-network, respectively.
In addition, their knowledge transfer between domains is enabled by constructing a transformation bridge.
In particular, PTUPCDR designs personalized bridges for each user, thus performing the best among the bridge methods.
For example, in Game-Cloth tasks, PTUPCDR shows 2.6\% and 3.39\% improvements, respectively, in HR@30 of two domains compared to EMCDR-MF.
Still, these methods indiscriminately transfer the complete user representations, yet fail to account for how domain-shared and specific information should be separated.
According to the empirical results, we observe that the performance of PTUPCDR fluctuates across different runs, hampering consistent model efficacy.
For instance, in the weak correlated Video-Cloth task, PTUPCDR exhibits the highest standard deviation in HR@30, with 1.09\% for the Video domain and 0.91\% for the Cloth domain. 
Remarkably, CIDER emphasizes the significance of separating domain-relevant factors from domain-irrelevant ones when compared to Bridge methods, showing significant improvements.
For example, in the Cloth-Sport tasks, CIDER outperforms the leading bridge-based model, PTUPCDR, with NDCG improvements of 1.53\%, 2.43\%, and 3.1\%.
By imposing bijective constraints on the relationships of variant facets across domains, CIDER ensures optimal prediction results while minimizing modifications, thus maximizing the effective transfer of features.

\subsubsection{Comparison with Alignment methods}
Traditional alignment methods focus on directly applying user representations.
In CMF, users who interact with both domains reuse their representations across the domains without any adjustments, which results in unreliable performance in CDR settings. 
This stems from the assumption that all information can be directly reused between domains, which often fails to hold in practice.
Differently, LFM surpasses CMF in most evaluation metrics by assuming that the representations in both domains are generated from a shared latent space.
While providing improvements, both methods struggle to effectively differentiate between general and complex features and, thus, encounter difficulties in producing satisfactory prediction outcomes for both domains.
Empirical evidence decisively favors alignment methods with disentanglement over bridge counterparts, highlighting the utility of optimization constraints in aligning domain-shared attributes.
As an example, DPMCDR, the best performer among the Constraint family, significantly leads PTUPCDR by 4.53\% and 4.24\% in terms of NDCG@30 for the Game-Video CDR task.
However, these methods have yet to demonstrate consistent robustness across all CDR tasks, especially when domains are weakly correlated, such as Game-Sport.
In particular, DisenCDR exhibits standard deviations of 1.41\% and 1.45\% for the HR@30 in the strong-correlation Game-Video task. 
In the weak-correlation task, i.e., the Game-Sport scenario, DisenCDR still shows fluctuations of 1.43\% and 1.72\% in HR@30 metric.
To address this, CDRIB and DPMCDR employ VIB to constrain the user/item representations that effectively concentrate on domain-shared information.
For example, DPMCDR achieves 3.89\% and 3.61\% in MRR on the Game-Cloth task, outperforming both DisenCDR and UniCDR, while maintaining more stable covariance compared to CDRIB.
Nevertheless, none of these methods have sufficiently considered the feasibility or transferability of decoupling user preferences while supporting cross-domain associations.
Instead, they primarily focus on extracting information applicable to both domains, overlooking dependencies among user preferences across domains.
CIDER seeks to address both general alignable and variant, transferrable factors with a hierarchical subspace disentanglement of representations.
It builds the model upon the feature hierarchy principle and disentanglement of latent factors, enabling improved robustness when encountering domain-relevant components within each domain.
Consequently, our approach achieves superior performance across all CDR tasks, regardless of their correlations.

\begin{figure}[t]
    \centering
    \begin{subfigure}[b]{.49\linewidth}
        \centering
        \includegraphics[width=\textwidth]{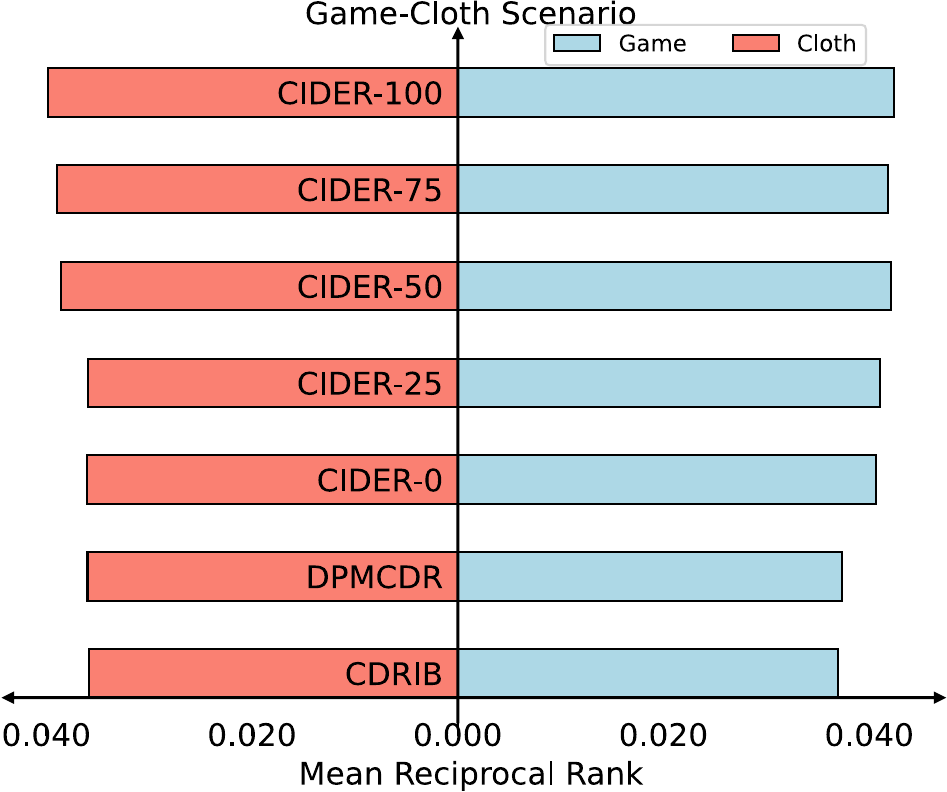}
        \label{fig:non_cs_c}
    \end{subfigure}
    \begin{subfigure}[b]{.49\linewidth}
        \centering
        \includegraphics[width=\textwidth]{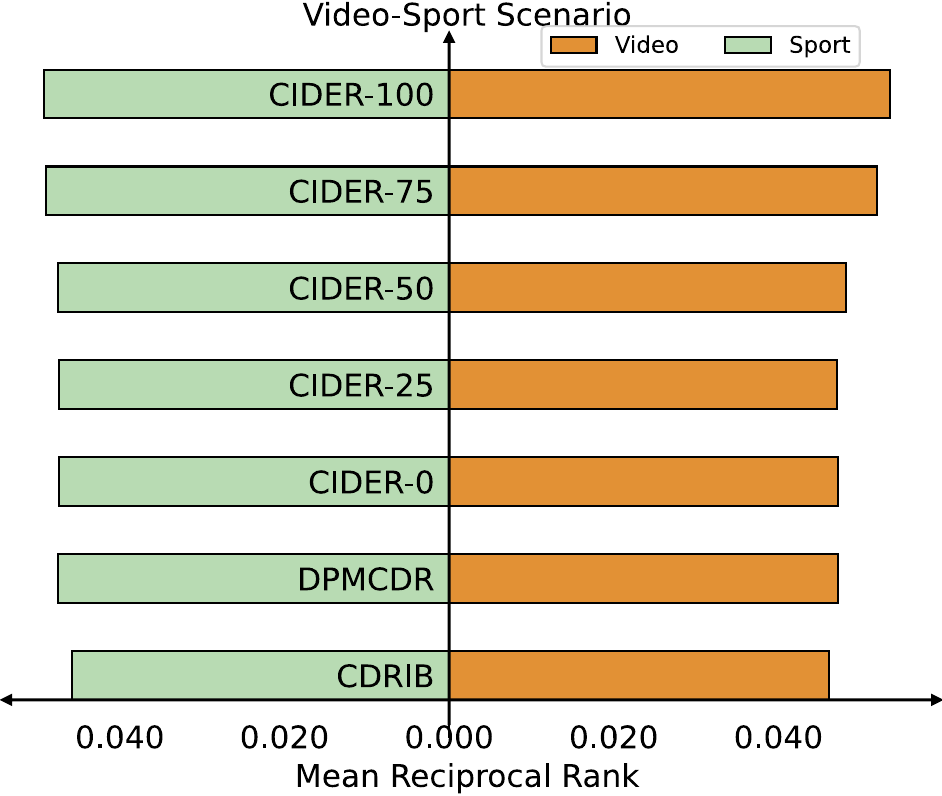}
        \label{fig:non_cs_s}
    \end{subfigure}
    \caption{Impact of different overlapping ratios.}
    \label{fig: non_overlap}
\end{figure}

\subsection{Impact of Overlapped Ratio (RQ2)}
CDR with few overlapping users demands that models precisely capture shared and adaptable user patterns from interactions across source and target domains. 
This allows for robust performance under extreme conditions, demonstrating the value of captured shared information.
To address RQ2, we apply varying ratios of overlapped users (0\%, 25\%, 50\%, 75\%, and 100\%) in the training set, keeping the testing and validation sets consistent. 
Due to space limitations, we only present the MRR results for two CDR tasks, comparing CIDER with non-overlapped CDRIB and DPMCDR.

Figure~\ref{fig: non_overlap} demonstrates DPMCDR's superiority over CDRIB across most metrics in Video-Sport tasks under non-overlapped scenarios, with only minor declines in the Game-Cloth scenario. 
Unlike CDRIB, which relies on overlapped users as anchors for cross-domain correspondence, DPMCDR eliminates the need for exact user overlap to align representations across domains. 
Thus, CDRIB surpasses DPMCDR in most overlapped CDR tasks while falling behind it in non-overlapped CDR tasks.
CIDER further advances this by hierarchically connecting user representations, avoiding strong constraints on correspondence, and enforcing joint identifiability to capture cross-domain correlations from two subspaces. 
This strategy allows CIDER to model both domain-variant and domain-invariant factors under domain shifts even without explicit cross-domain relationships.

Considering various ratios of overlapping users, CIDER series models (CIDER-0\% to CIDER-100\%) generally perform better than DPMCDR and CDRIB, particularly at higher overlapping user ratios. 
With an increasing ratio of overlapping users, CIDER series models show a clear trend of improvement in MRR across two CDR tasks. 
This suggests that a higher proportion of overlapping users enhances CIDER’s generalization and performance by providing critical information about cross-domain relationships, 
which supports the learning process for capturing bijective transformations of domain-relevant components.

\subsection{Ablation Studies (RQ3)}
In this section, 5 variants are imposed to thoroughly evaluate each proposed module.
Variant A only performs representation encoding;
Variant B implements Maximum Mean discrepancy to align representations in both shallow and deep layers;
Variant C aligns representations in two subspaces using centroid-based alignment;
Variant D applies CPA in shallow subspace and deep subspace identification in deep subspace, but removes the invertible Flow-based transformations $\boldsymbol{G}(\cdot)$;
Variant E abandons the feature hierarchy principle and applies deep subspace identification to whole representations.

\begin{table}
    \caption{Ablation study}
    \label{tab:ablation}
    \resizebox{\linewidth}{!}{
    \begin{tabular}{l|ccc|ccc}
    \hline
    \multirow{2}{*}{\textbf{Methods}} & \multicolumn{3}{c|}{\textbf{Game}}                           & \multicolumn{3}{c}{\textbf{Video}}                           \\ \cline{2-7} 
                             & MRR             & NDCG@30         & HR@30           & MRR             & NDCG@30         & HR@30           \\ \hline
    A                        & 0.0478          & 0.0720          & 0.1840          & 0.0426          & 0.0648          & 0.1704          \\
    B                        & 0.0357          & 0.0620          & 0.1783          & 0.0387          & 0.0604          & 0.1640          \\
    C                        & 0.0480          & 0.0769          & 0.2008          & 0.0463          & 0.0701          & 0.1834          \\
    D                        & 0.0449          & 0.0738          & 0.2040          & 0.0415          & 0.0689          & 0.1907          \\
    E                        & 0.0450          & 0.0717          & 0.1926          & 0.0389          & 0.0618          & 0.1681          \\ \hline
    Ours                     & \textbf{0.0492} & \textbf{0.0777} & \textbf{0.2054} & \textbf{0.0443} & \textbf{0.0701} & \textbf{0.1925} \\ \hline
    \end{tabular}
    }

    \resizebox{\linewidth}{!}{
    \begin{tabular}{l|ccc|ccc}
    \hline
    \multirow{2}{*}{\textbf{Methods}} & \multicolumn{3}{c|}{\textbf{Cloth}}                          & \multicolumn{3}{c}{\textbf{Sport}}                           \\ \cline{2-7} 
                             & MRR             & NDCG@30         & HR@30           & MRR             & NDCG@30         & HR@30           \\ \hline
    A                        & 0.0351          & 0.0553          & 0.1504          & 0.0408          & 0.0644          & 0.1755          \\
    B                        & 0.0289          & 0.0485          & 0.1442          & 0.0275          & 0.0404          & 0.1045          \\
    C                        & 0.0358          & 0.0561          & 0.1525          & 0.0432          & 0.0679          & 0.1829          \\
    D                        & 0.0307          & 0.0506          & 0.1475          & 0.0421          & 0.0687          & 0.1916          \\
    E                        & 0.0360          & 0.0558          & 0.1504          & 0.0420          & 0.0653          & 0.1741          \\ \hline
    Ours                     & \textbf{0.0361} & \textbf{0.0563} & \textbf{0.1521} & \textbf{0.0488} & \textbf{0.0748} & \textbf{0.1956} \\ \hline
    \end{tabular}
    }
    \resizebox{\linewidth}{!}{
    \begin{tabular}{l|ccc|ccc}
    \hline
    \multirow{2}{*}{\textbf{Methods}} & \multicolumn{3}{c|}{\textbf{Game}}                           & \multicolumn{3}{c}{\textbf{Cloth}}                           \\ \cline{2-7} 
                             & MRR             & NDCG@30         & HR@30           & MRR             & NDCG@30         & HR@30           \\ \hline
    A                        & 0.0376          & 0.0586          & 0.1545          & 0.0369          & 0.0556          & 0.1460          \\
    B                        & 0.0362          & 0.0566          & 0.1526          & 0.0350          & 0.0533          & 0.1407          \\
    C                        & 0.0403          & 0.0620          & 0.1645          & 0.0371          & 0.0571          & 0.1534          \\
    D                        & 0.0349          & 0.0556          & 0.1549          & 0.0363          & 0.0571          & 0.1565          \\
    E                        & 0.0397          & 0.0616          & 0.1627          & 0.0376          & 0.0570          & 0.1509          \\ \hline
    Ours                     & \textbf{0.0424} & \textbf{0.0643} & \textbf{0.1674} & \textbf{0.0387} & \textbf{0.0598} & \textbf{0.1688} \\ \hline
    \end{tabular}
    }

\end{table}

As shown in Table \ref{tab:ablation}, variant A outperforms variant B across all metrics. 
This may suggest that traditional alignment methods can be unstable in certain cases, potentially misaligning cross-domain attributes. 
However, variant C, based on our proposed CPA method, surpasses both variant A and variant B, demonstrating the advantages of aligning user interests based on fine-grained interest classes.
In combating the domain-relevant transformation, Variant D surpasses Variant B in all metrics. 
Specifically, it shows an MRR increase of 0.92\% and 0.28\% in the strongly correlated Game-Video task, with only slight drops in the weakly correlated Game-Cloth task. 
These findings suggest that domain-relevant factors dominate deep user representations, while identifiable joint distributions are essential for successful cross-domain transformation.
However, compared to Variant C, Variant D shows minimal improvement, or even a decrease in performance in some tasks, possibly due to the inability to capture variations of domain-relevant components. 
Although Variant E ensures {\it joint identifiability}, may fail to fully leverage domain-irrelevant information, leading to variable performance.
To sum up, the full CIDER consistently outperforms all other variants by establishing hierarchical relationships in user representations and ensuring {\it identifiable} joint distribution of çross-user representation, regardless of domain correlation strength.

\subsection{Parameter sensitivity (RQ 4)}
\begin{figure*}
    \centering
    \begin{subfigure}[b]{.16\linewidth}
        \centering
        \includegraphics[width=\textwidth]{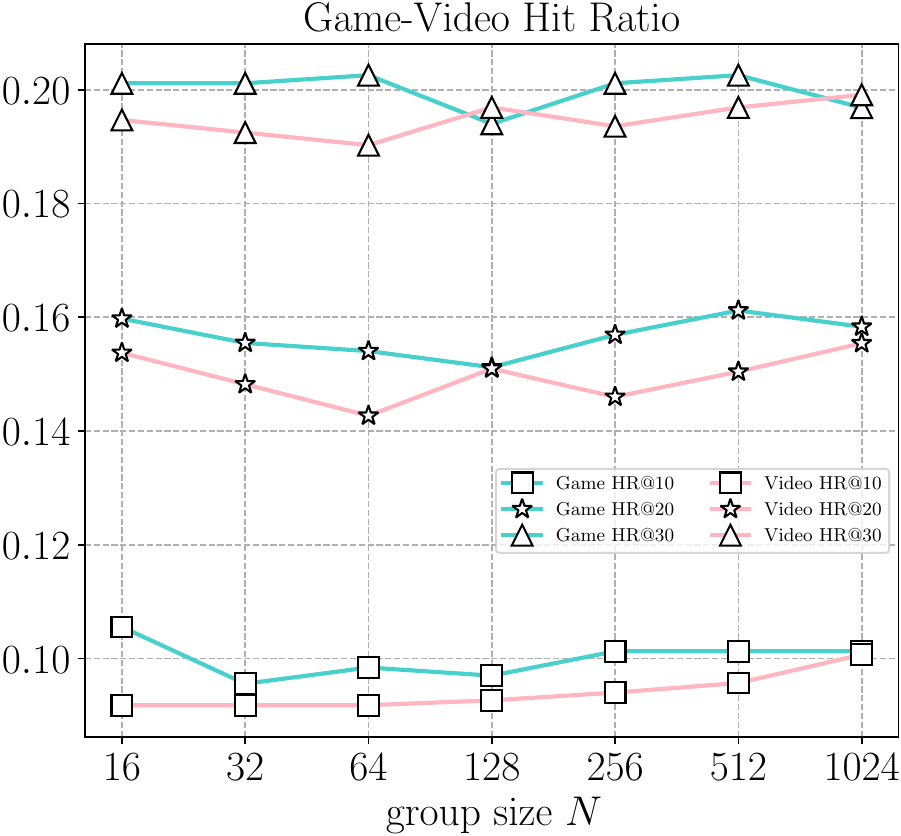}
        \label{fig:gs_groupsize}
    \end{subfigure}
    \begin{subfigure}[b]{.16\linewidth}
        \centering
        \includegraphics[width=\textwidth]{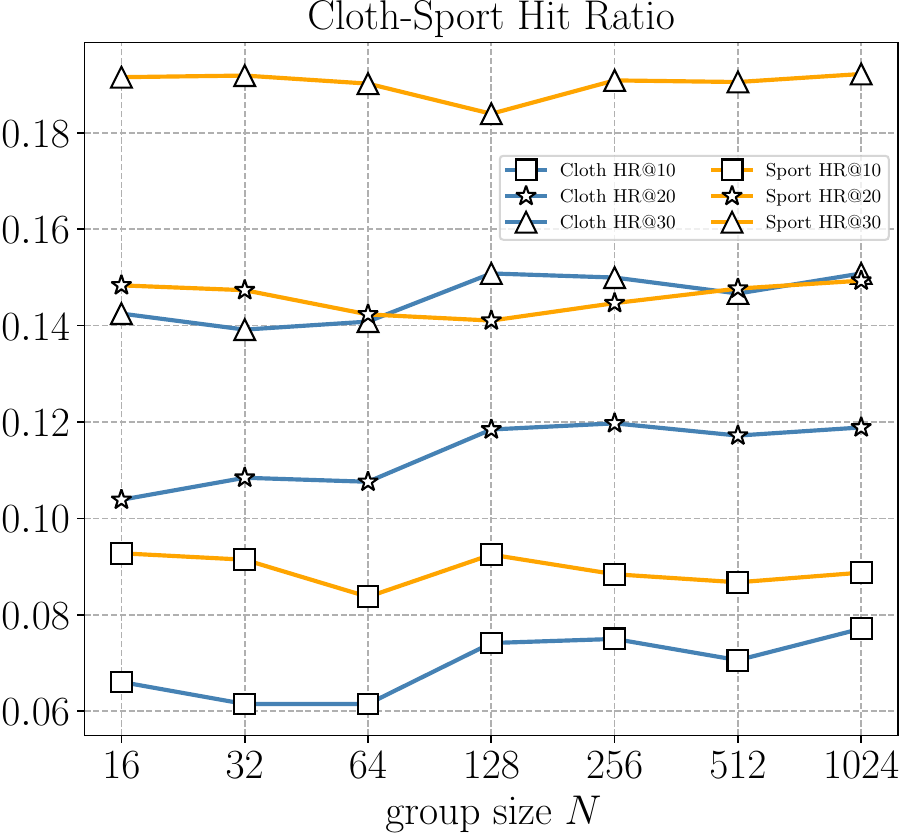}
        \label{fig:gs_flowtype}
    \end{subfigure}
    \begin{subfigure}[b]{.16\linewidth}
        \centering
        \includegraphics[width=\textwidth]{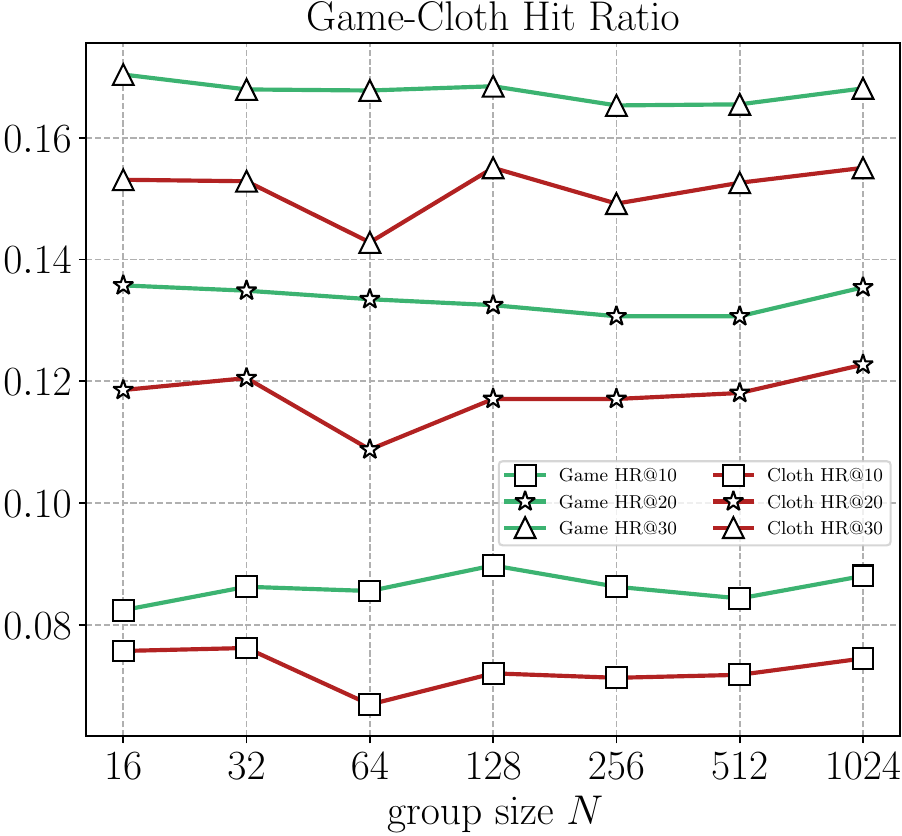}
        \label{fig:gs_flowlayer}
    \end{subfigure}
    \begin{subfigure}[b]{.16\linewidth}
        \centering
        \includegraphics[width=\textwidth]{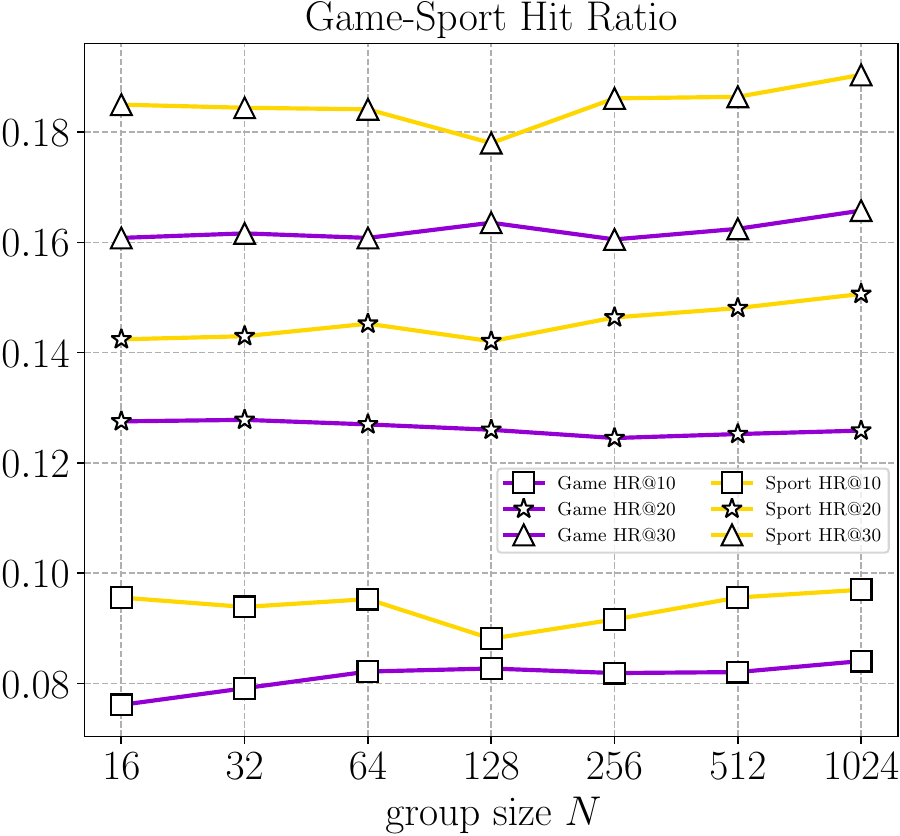}
        \label{fig:gs_align}
    \end{subfigure}
    \begin{subfigure}[b]{.16\linewidth}
        \centering
        \includegraphics[width=\textwidth]{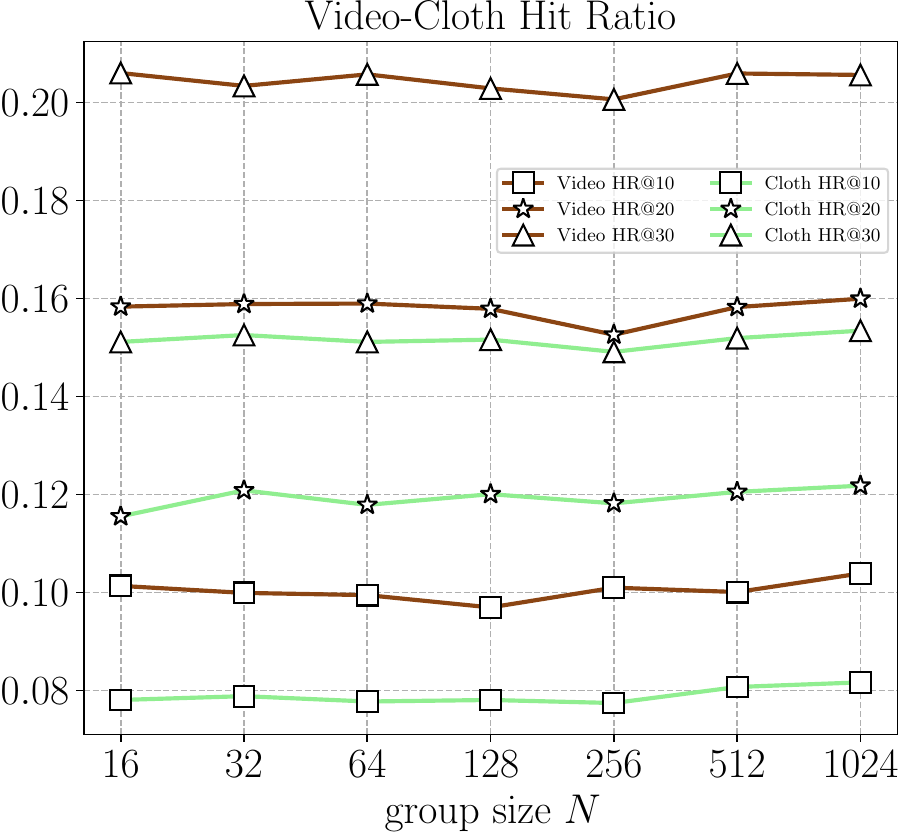}
        \label{fig:gs_gnnlayer}
    \end{subfigure}
    \begin{subfigure}[b]{.16\linewidth}
        \centering
        \includegraphics[width=\textwidth]{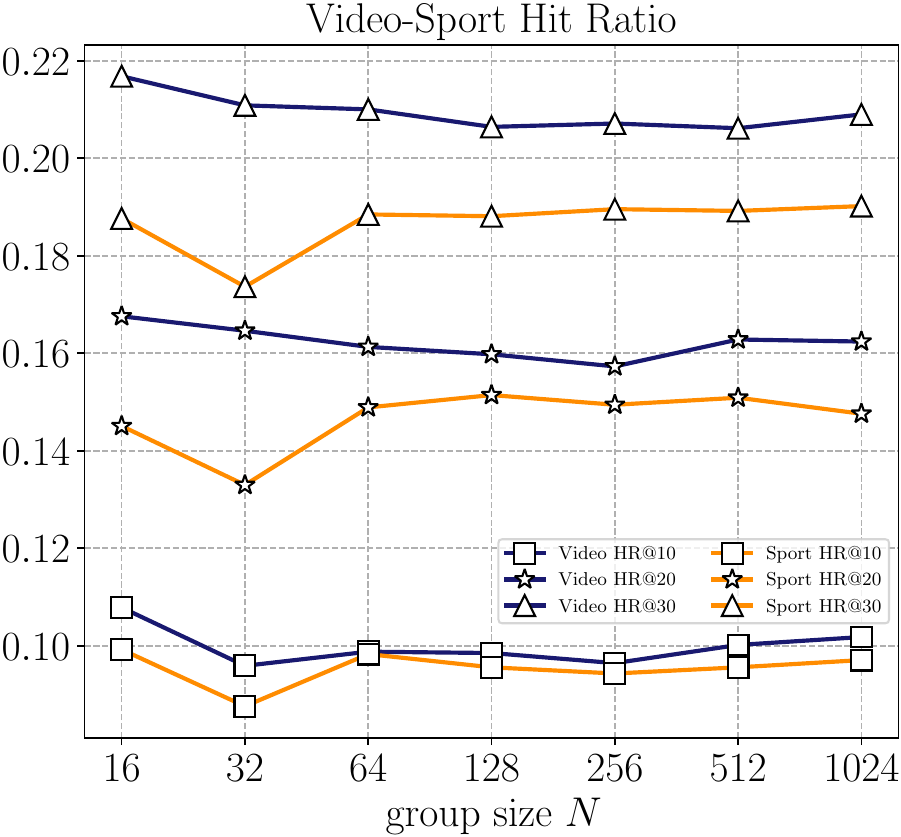}
        \label{fig:gs_gnnlayer}
    \end{subfigure}

    \begin{subfigure}[b]{.16\linewidth}
        \centering
        \includegraphics[width=\textwidth]{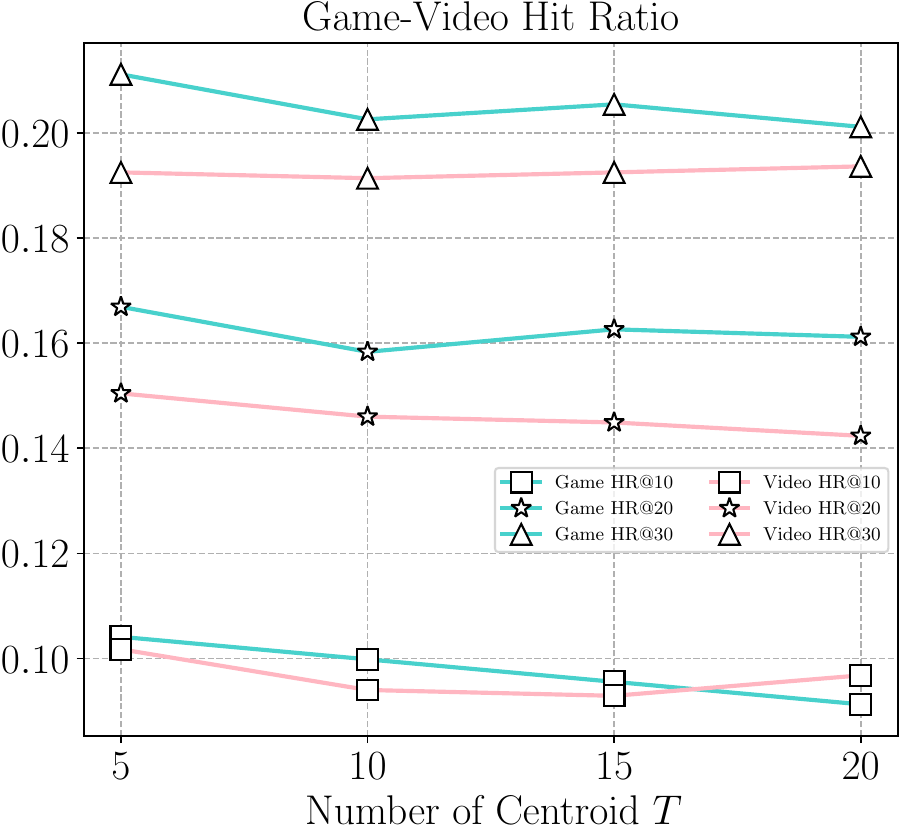}
        \label{fig:gs_groupsize}
    \end{subfigure}
    \begin{subfigure}[b]{.16\linewidth}
        \centering
        \includegraphics[width=\textwidth]{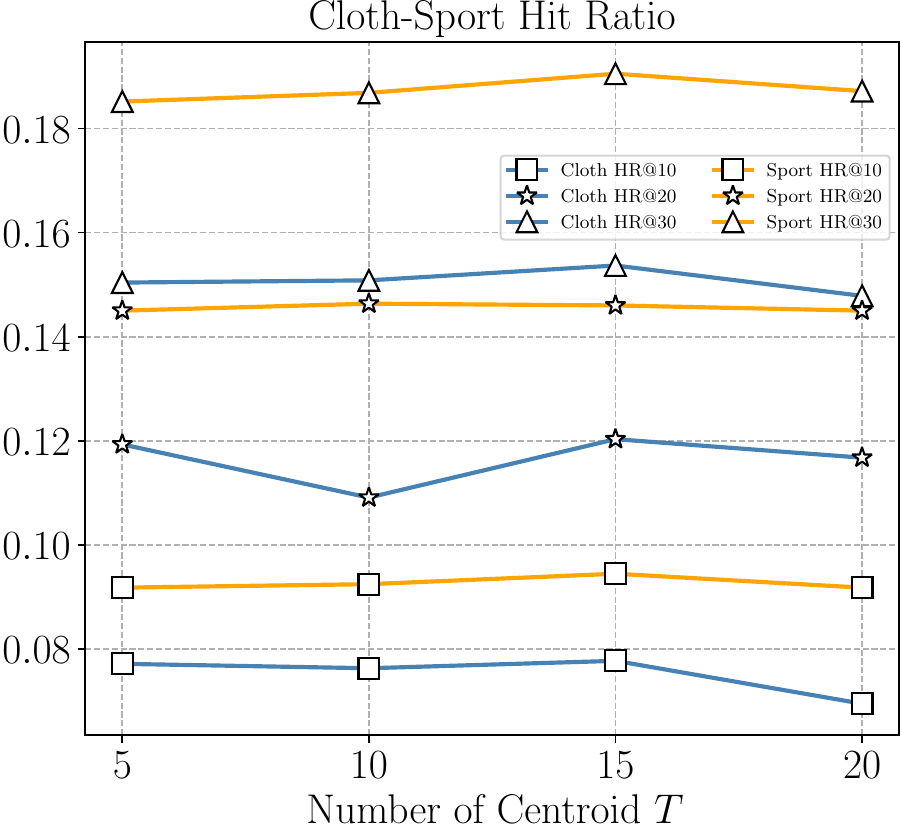}
        \label{fig:gs_flowtype}
    \end{subfigure}
    \begin{subfigure}[b]{.16\linewidth}
        \centering
        \includegraphics[width=\textwidth]{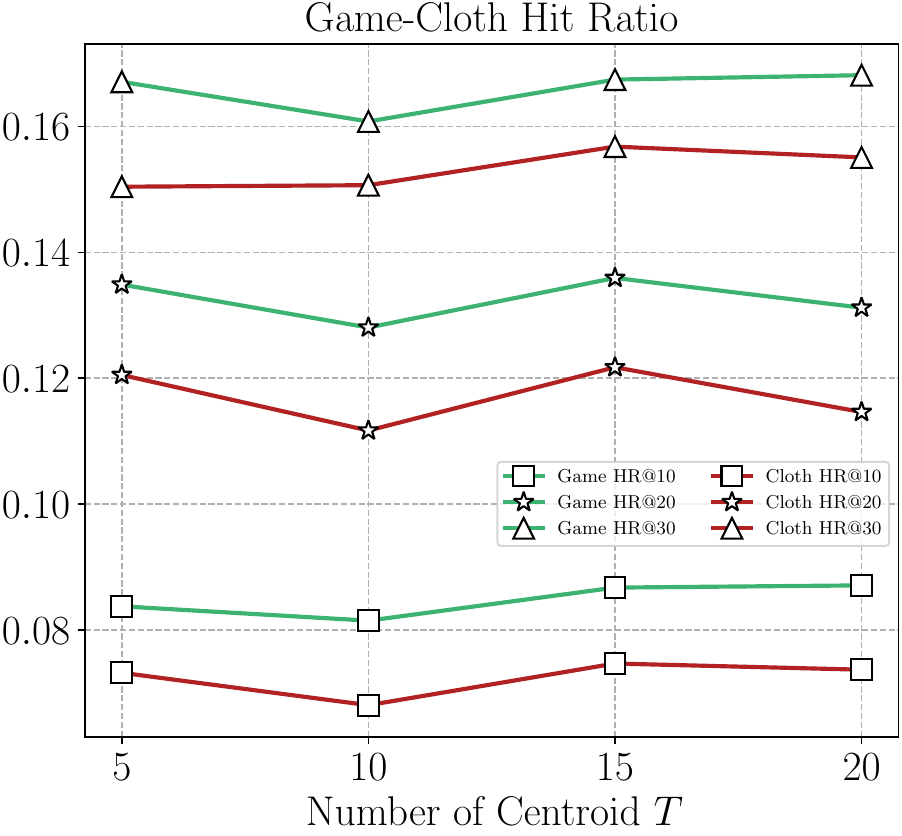}
        \label{fig:gs_flowlayer}
    \end{subfigure}
    \begin{subfigure}[b]{.16\linewidth}
        \centering
        \includegraphics[width=\textwidth]{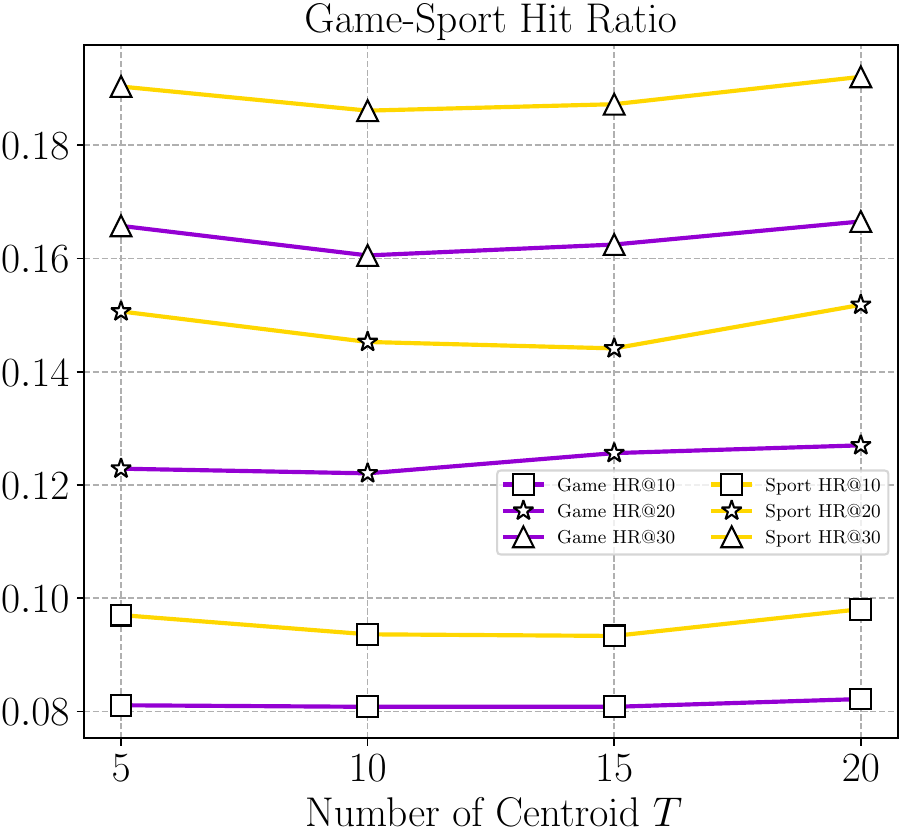}
        \label{fig:gs_align}
    \end{subfigure}
    \begin{subfigure}[b]{.16\linewidth}
        \centering
        \includegraphics[width=\textwidth]{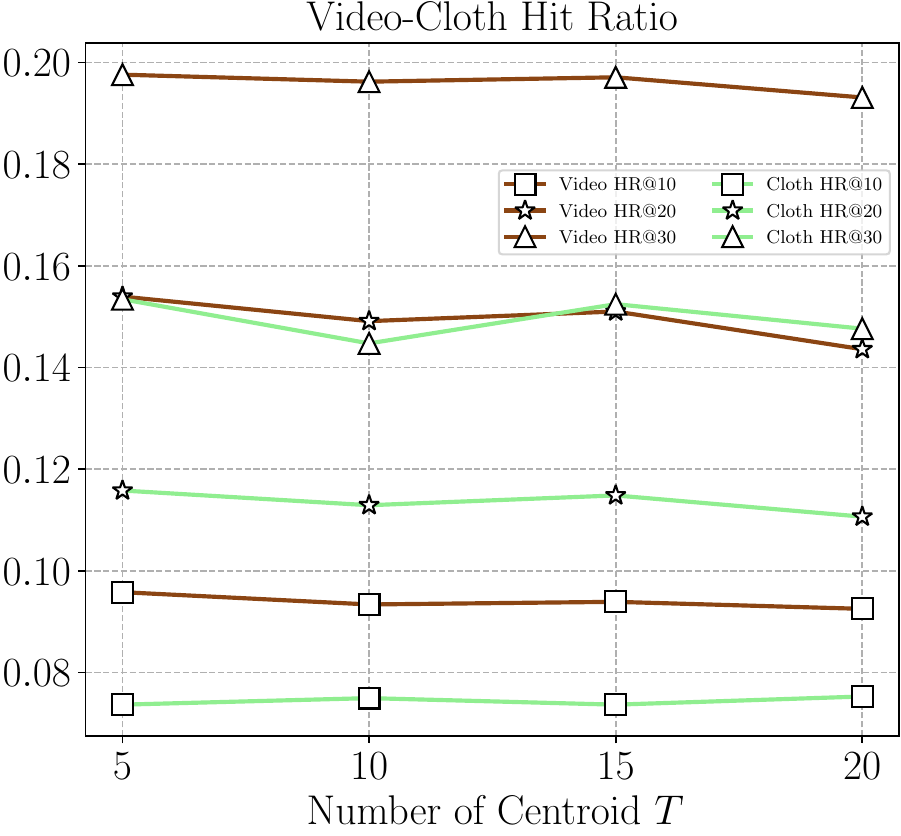}
        \label{fig:gs_gnnlayer}
    \end{subfigure}
    \begin{subfigure}[b]{.16\linewidth}
        \centering
        \includegraphics[width=\textwidth]{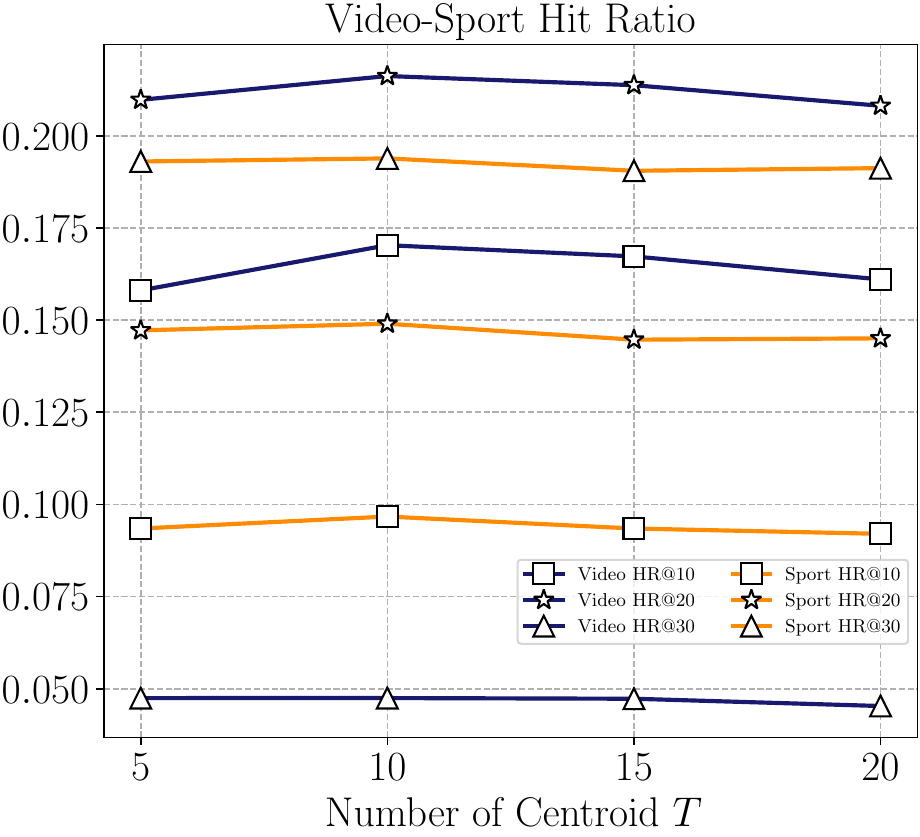}
        \label{fig:gs_gnnlayer}
    \end{subfigure}

    \begin{subfigure}[b]{.16\linewidth}
        \centering
        \includegraphics[width=\textwidth]{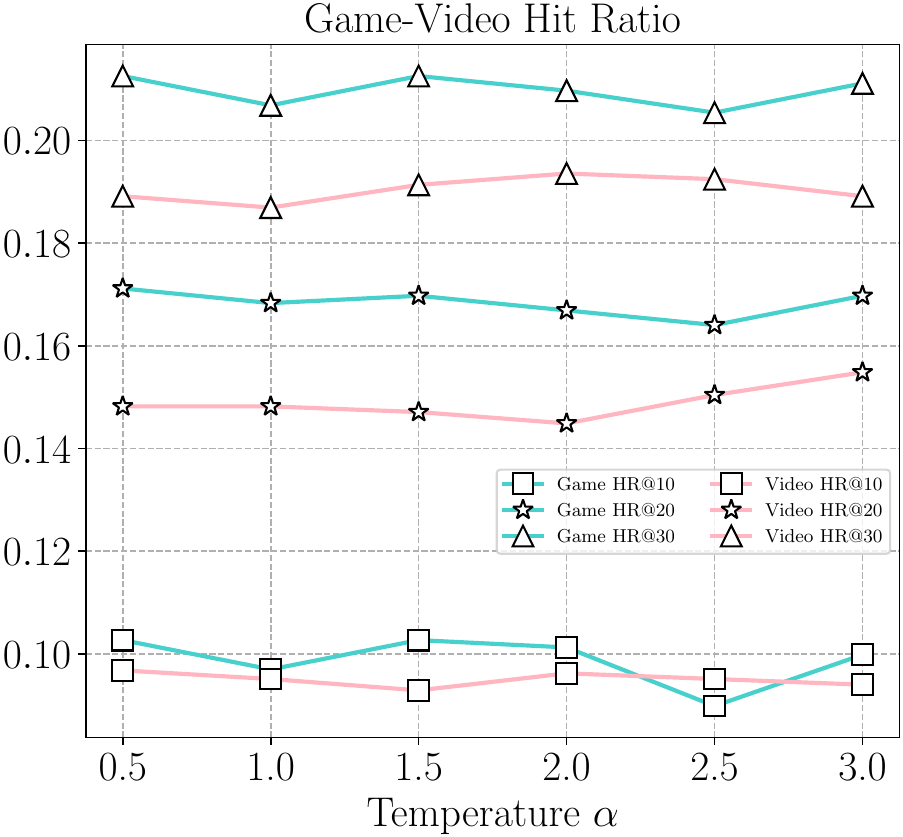}
        \label{fig:gs_groupsize}
    \end{subfigure}
    \begin{subfigure}[b]{.16\linewidth}
        \centering
        \includegraphics[width=\textwidth]{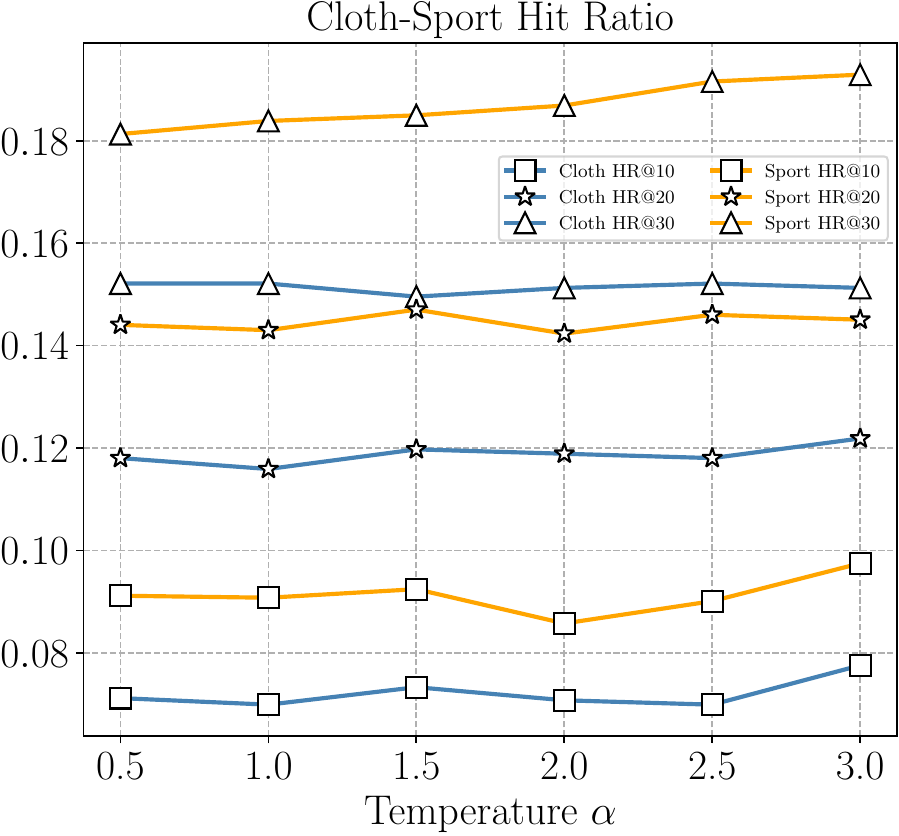}
        \label{fig:gs_flowtype}
    \end{subfigure}
    \begin{subfigure}[b]{.16\linewidth}
        \centering
        \includegraphics[width=\textwidth]{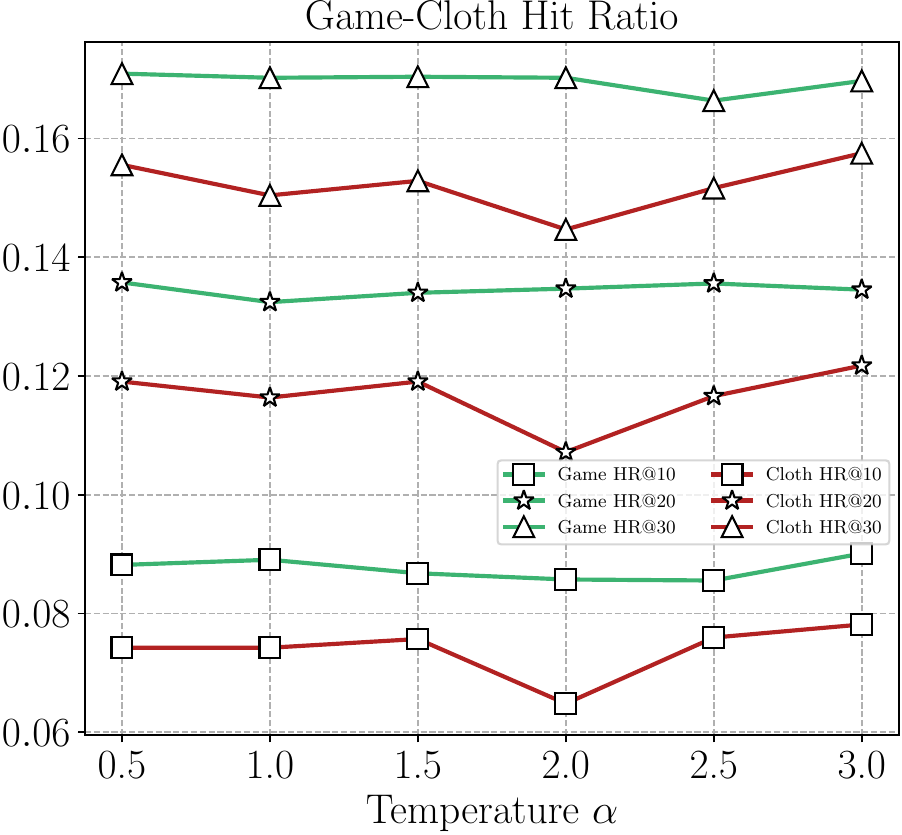}
        \label{fig:gs_flowlayer}
    \end{subfigure}
    \begin{subfigure}[b]{.16\linewidth}
        \centering
        \includegraphics[width=\textwidth]{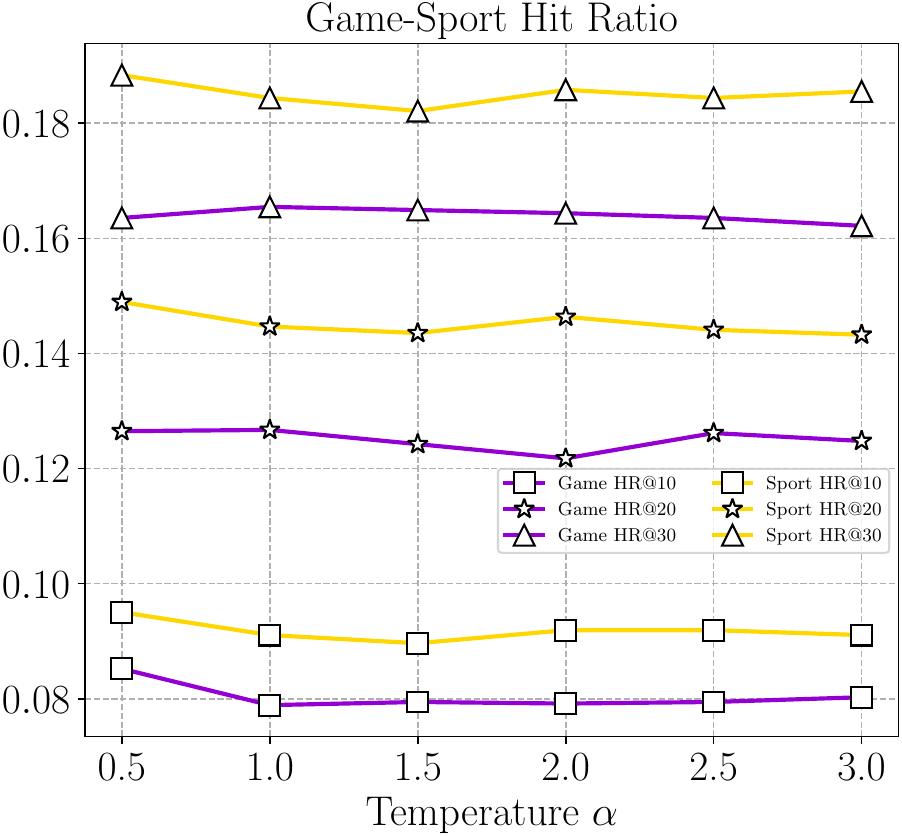}
        \label{fig:gs_align}
    \end{subfigure}
    \begin{subfigure}[b]{.16\linewidth}
        \centering
        \includegraphics[width=\textwidth]{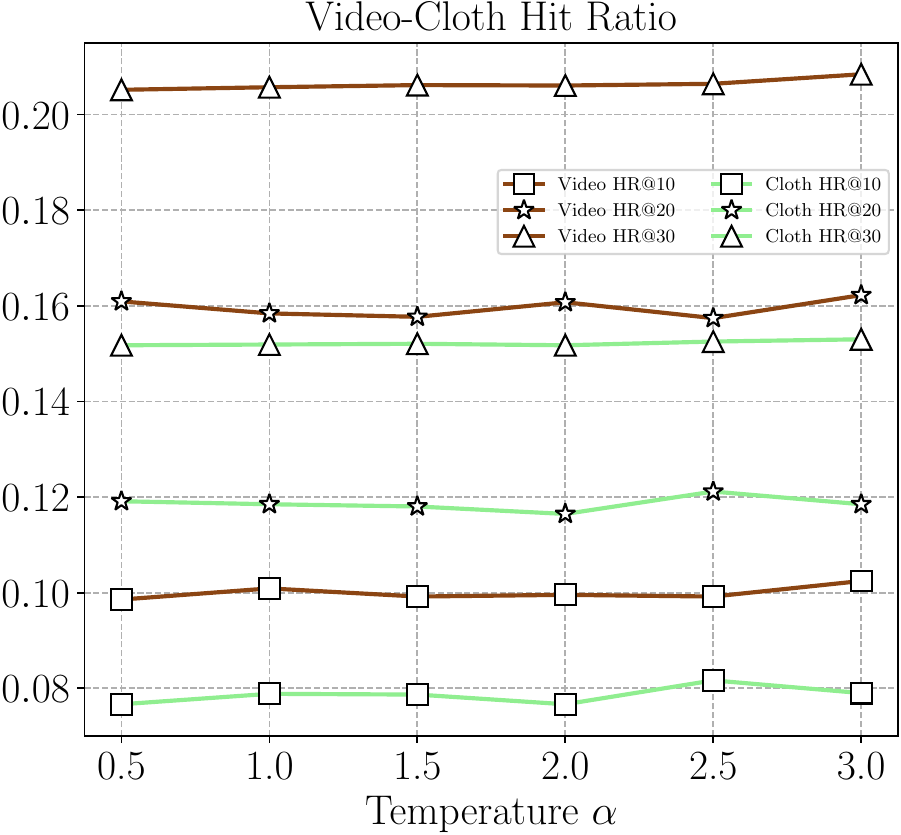}
        \label{fig:gs_gnnlayer}
    \end{subfigure}
    \begin{subfigure}[b]{.16\linewidth}
        \centering
        \includegraphics[width=\textwidth]{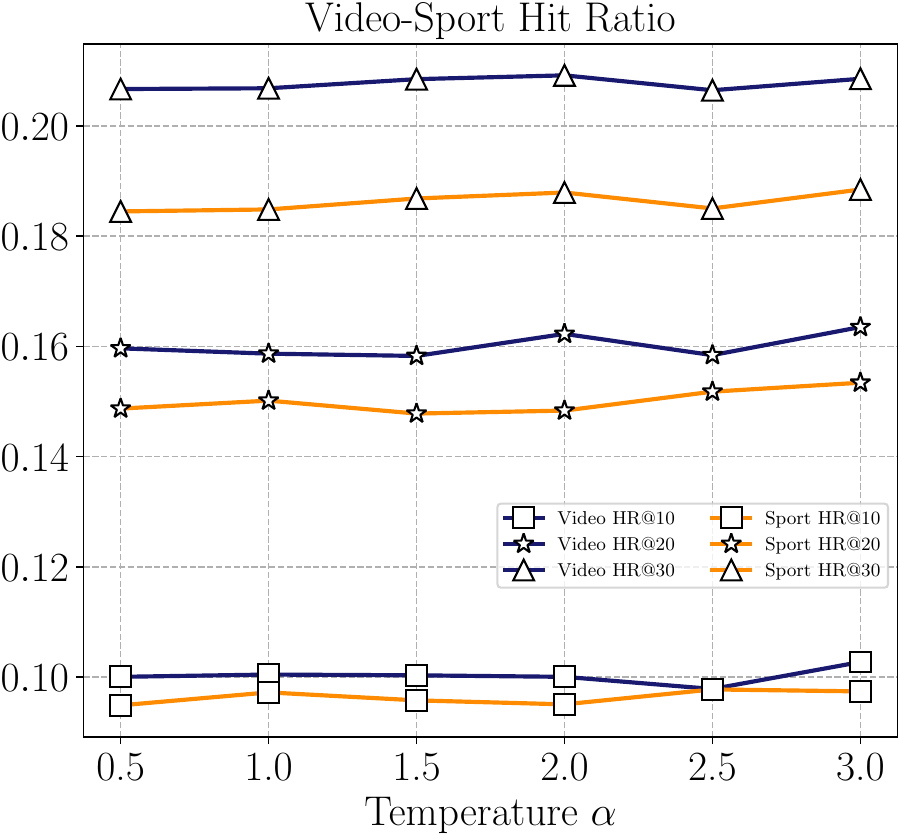}
        \label{fig:gs_gnnlayer}
    \end{subfigure}

    \begin{subfigure}[b]{.16\linewidth}
        \centering
        \includegraphics[width=\textwidth]{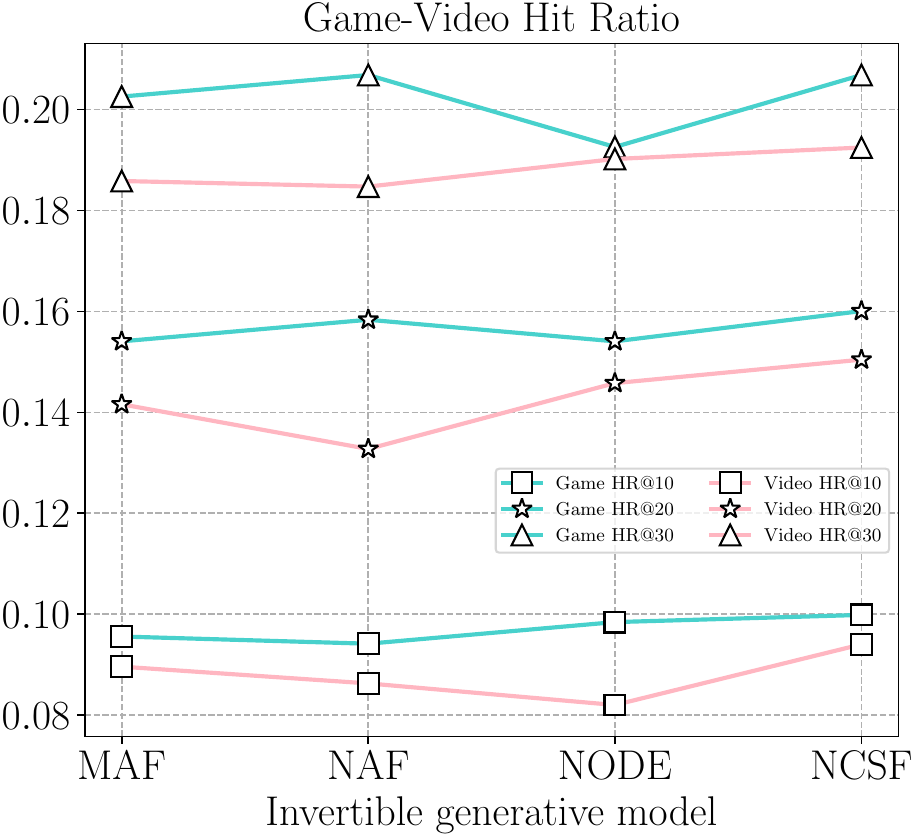}
        \label{fig:gs_groupsize}
    \end{subfigure}
    \begin{subfigure}[b]{.16\linewidth}
        \centering
        \includegraphics[width=\textwidth]{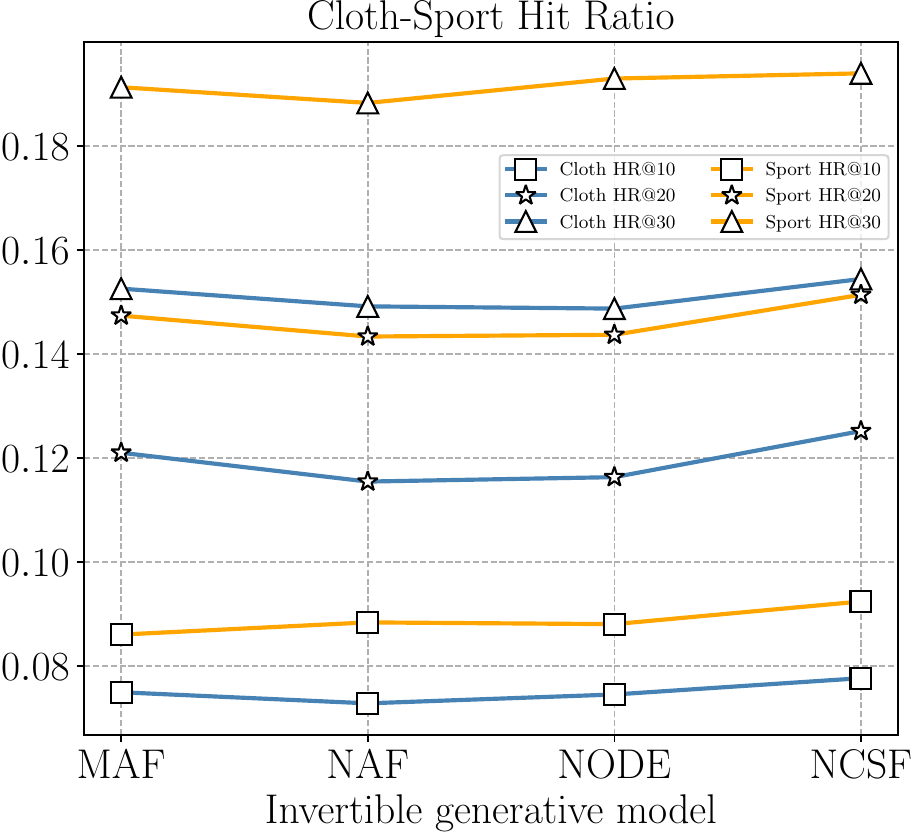}
        \label{fig:gs_flowtype}
    \end{subfigure}
    \begin{subfigure}[b]{.16\linewidth}
        \centering
        \includegraphics[width=\textwidth]{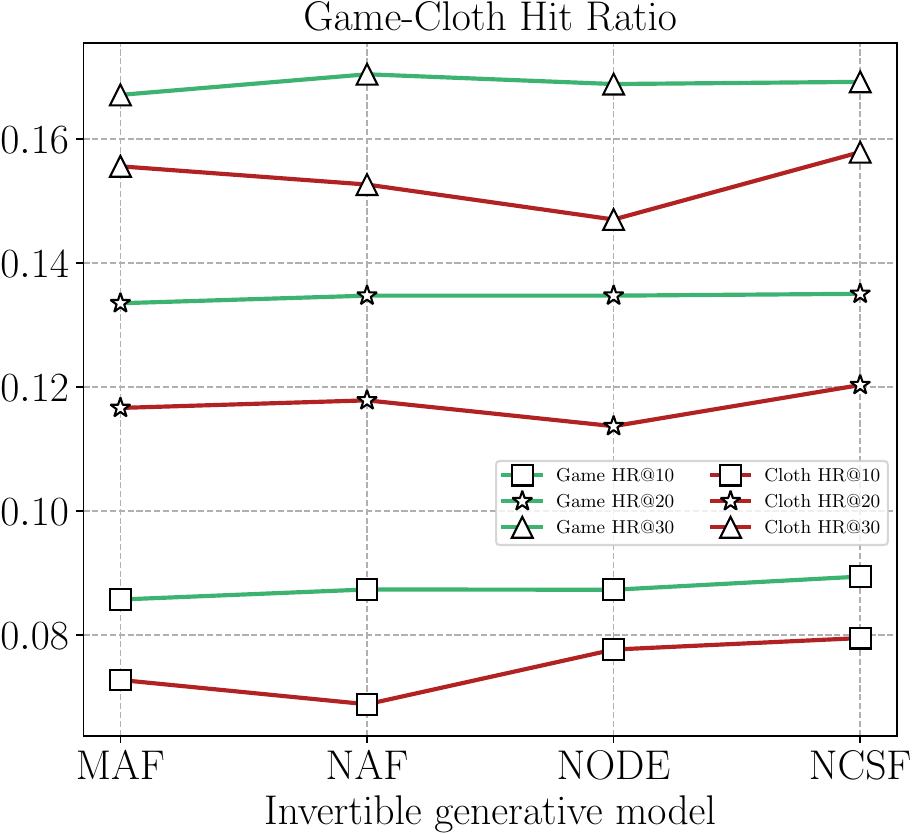}
        \label{fig:gs_flowlayer}
    \end{subfigure}
    \begin{subfigure}[b]{.16\linewidth}
        \centering
        \includegraphics[width=\textwidth]{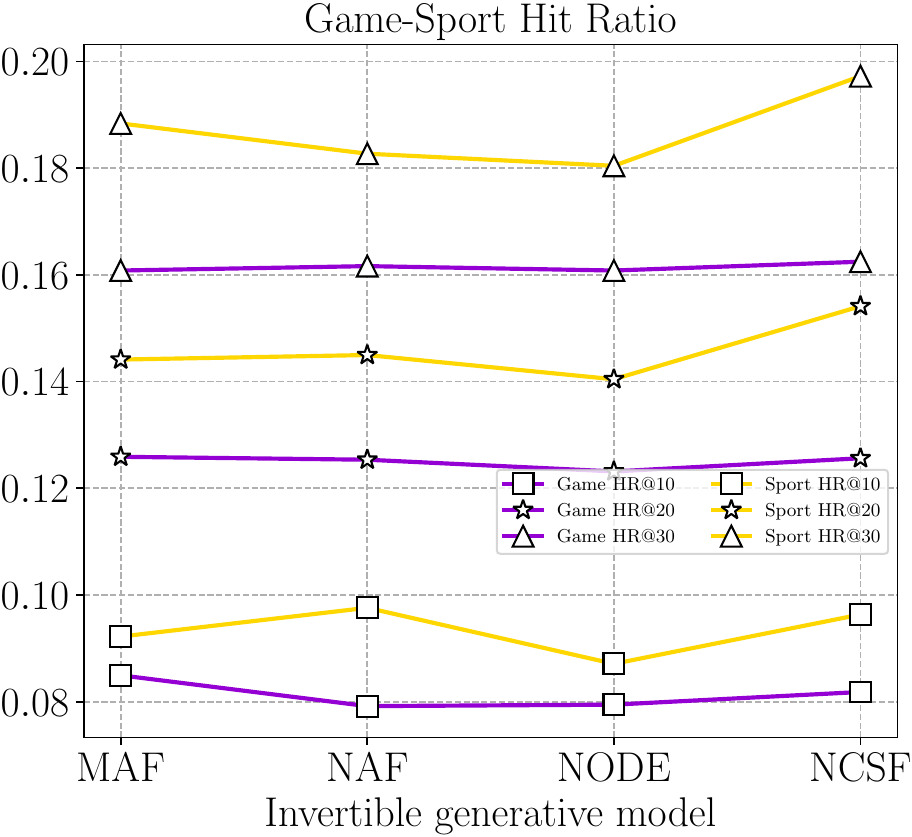}
        \label{fig:gs_align}
    \end{subfigure}
    \begin{subfigure}[b]{.16\linewidth}
        \centering
        \includegraphics[width=\textwidth]{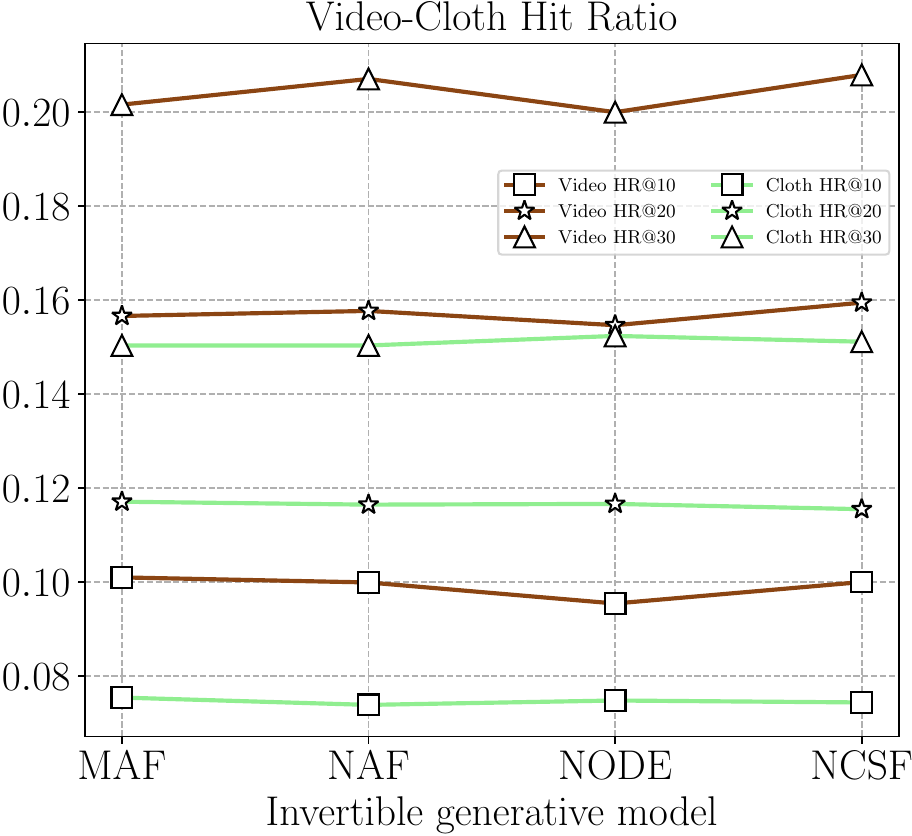}
        \label{fig:gs_gnnlayer}
    \end{subfigure}
    \begin{subfigure}[b]{.16\linewidth}
        \centering
        \includegraphics[width=\textwidth]{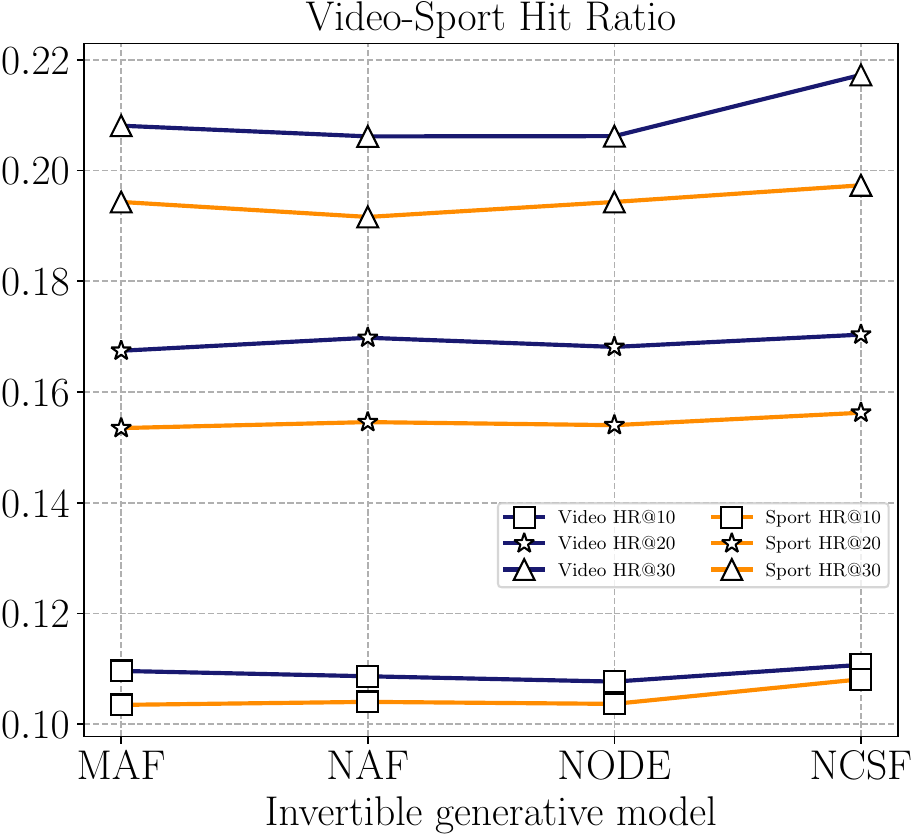}
        \label{fig:gs_gnnlayer}
    \end{subfigure}

    \begin{subfigure}[b]{.16\linewidth}
        \centering
        \includegraphics[width=\textwidth]{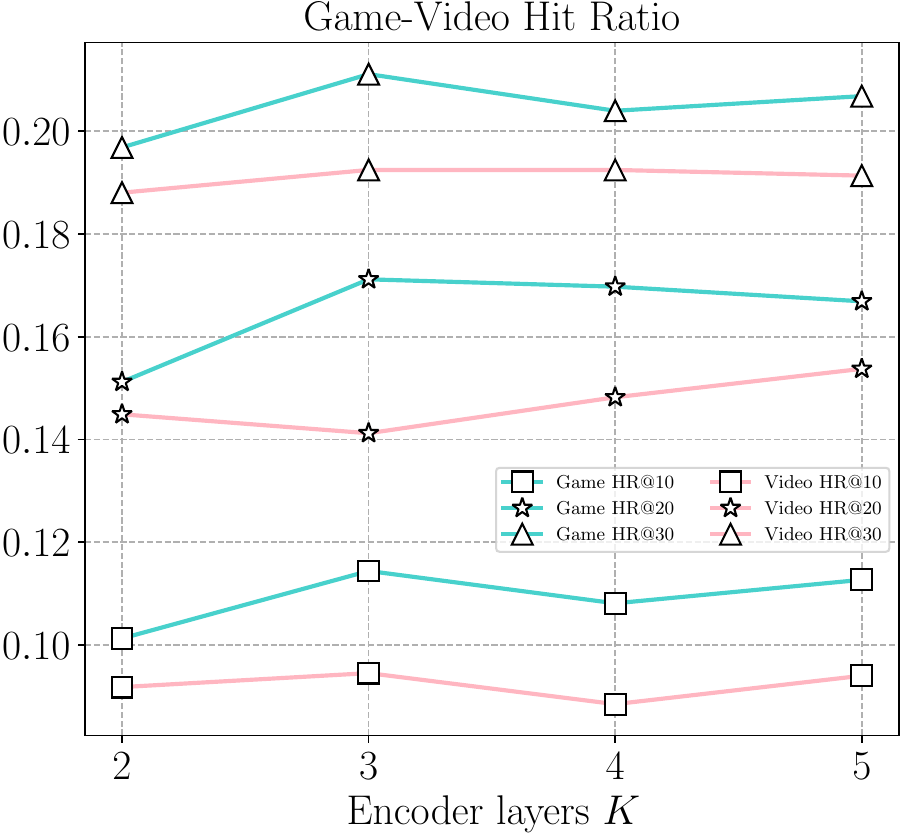}
        \label{fig:gs_groupsize}
    \end{subfigure}
    \begin{subfigure}[b]{.16\linewidth}
        \centering
        \includegraphics[width=\textwidth]{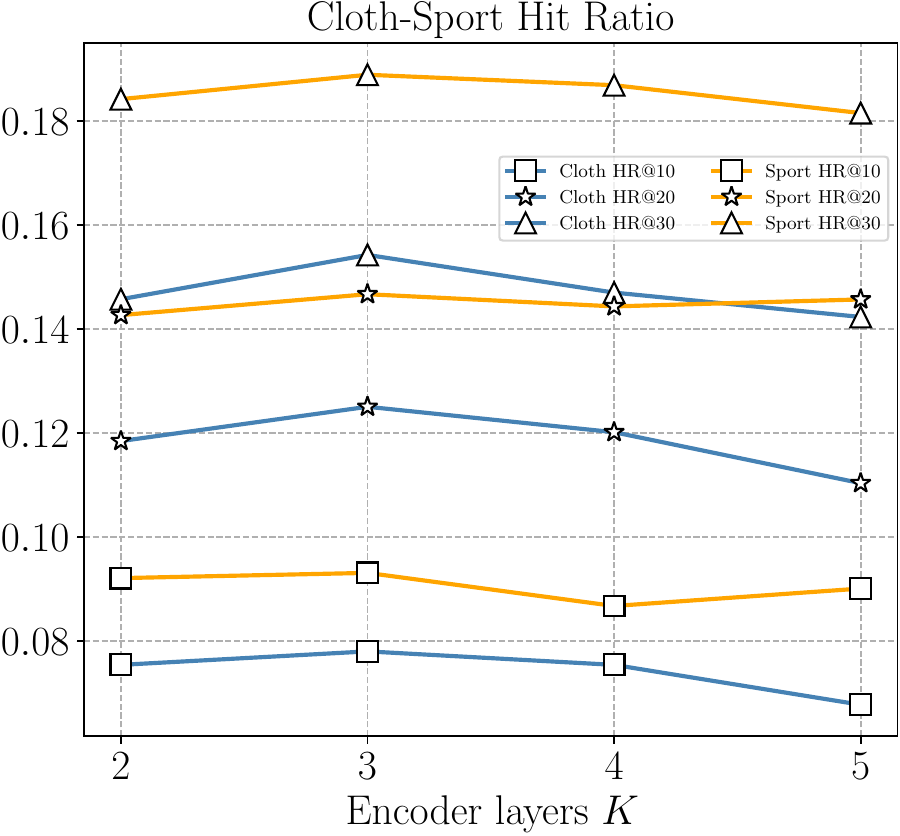}
        \label{fig:gs_flowtype}
    \end{subfigure}
    \begin{subfigure}[b]{.16\linewidth}
        \centering
        \includegraphics[width=\textwidth]{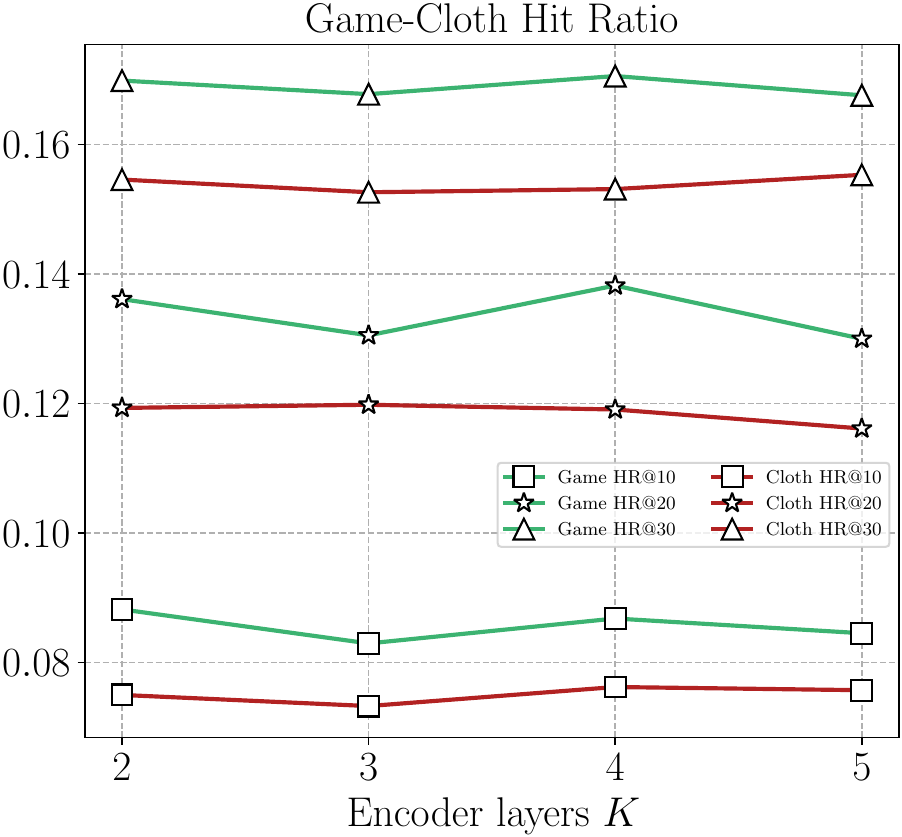}
        \label{fig:gs_flowlayer}
    \end{subfigure}
    \begin{subfigure}[b]{.16\linewidth}
        \centering
        \includegraphics[width=\textwidth]{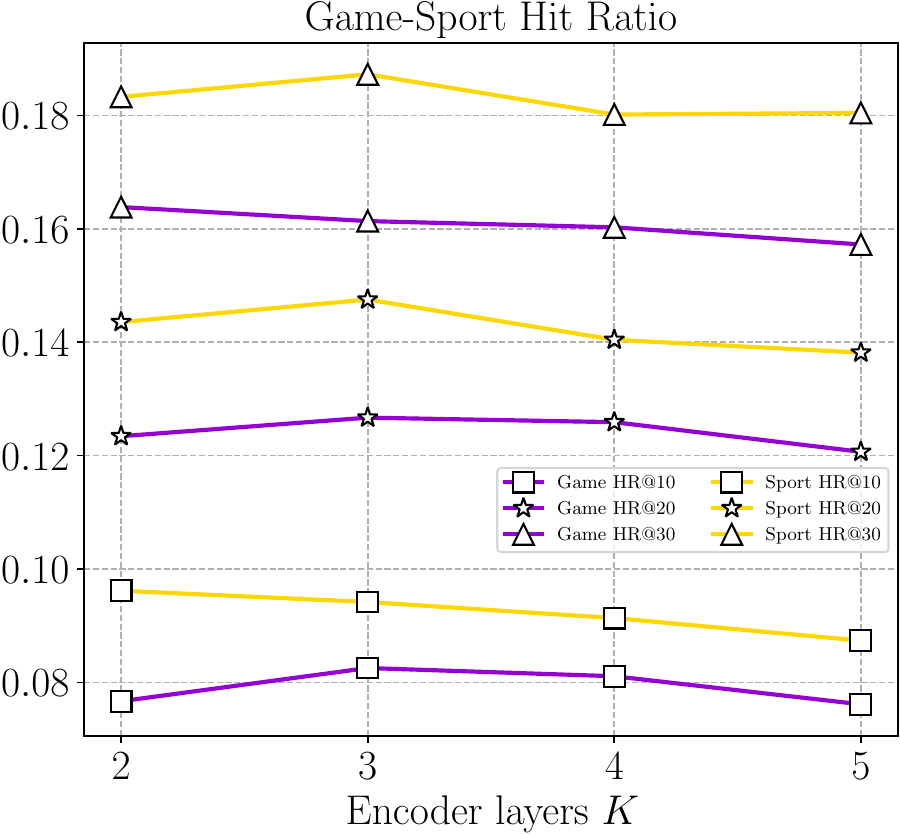}
        \label{fig:gs_align}
    \end{subfigure}
    \begin{subfigure}[b]{.16\linewidth}
        \centering
        \includegraphics[width=\textwidth]{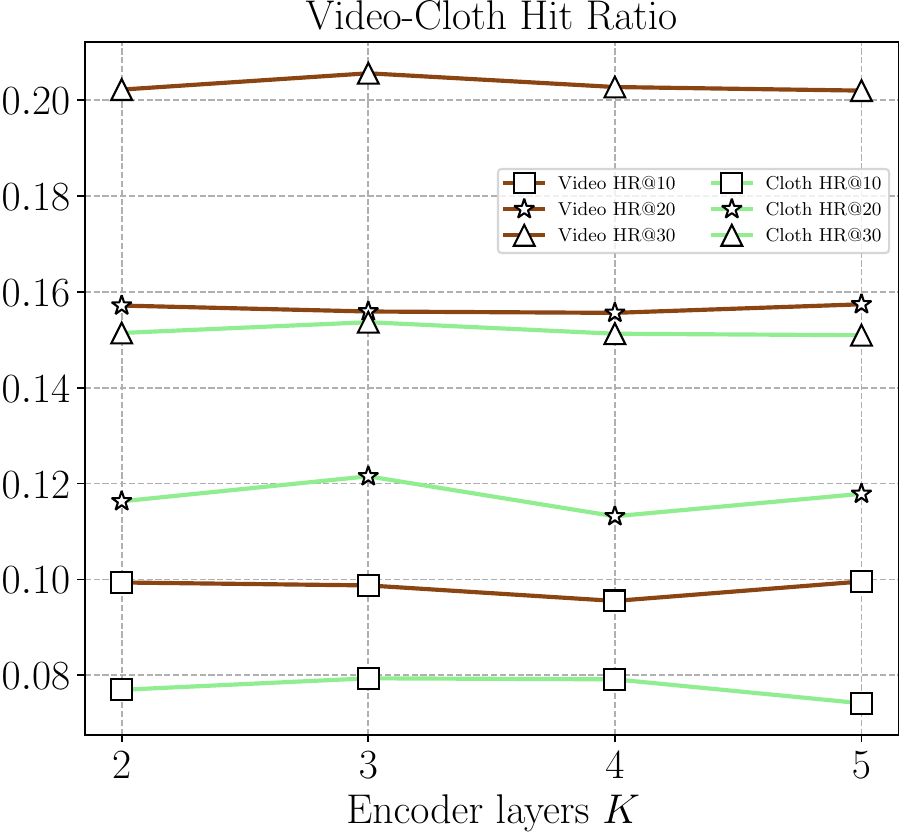}
        \label{fig:gs_gnnlayer}
    \end{subfigure}
    \begin{subfigure}[b]{.16\linewidth}
        \centering
        \includegraphics[width=\textwidth]{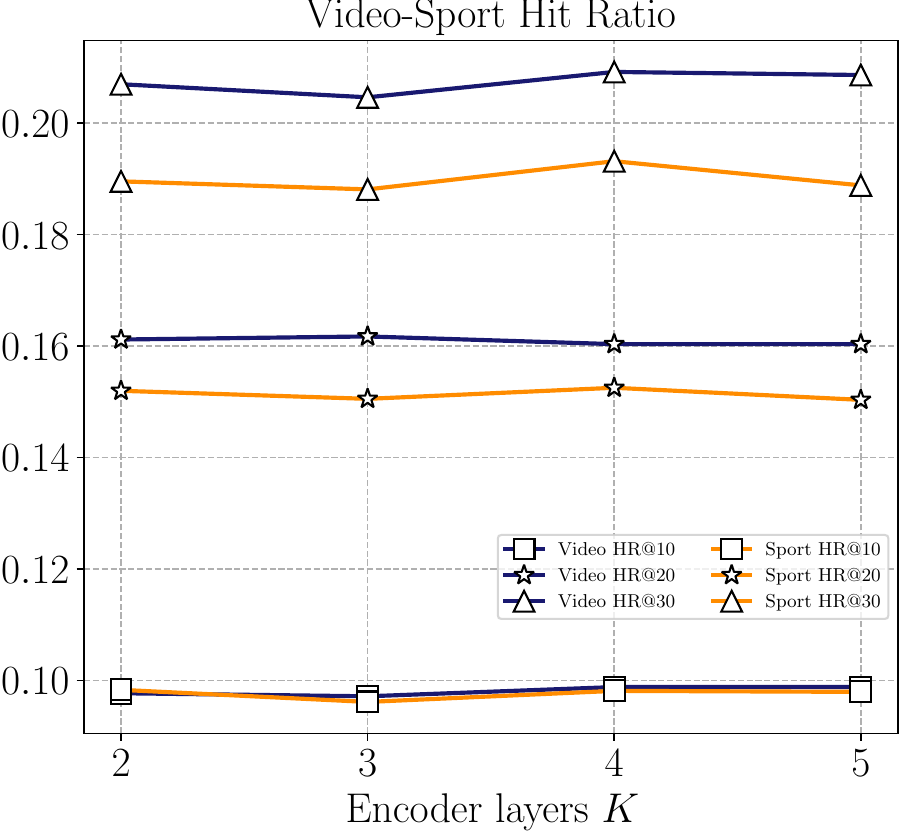}
        \label{fig:gs_gnnlayer}
    \end{subfigure}

    \begin{subfigure}[b]{.16\linewidth}
        \centering
        \includegraphics[width=\textwidth]{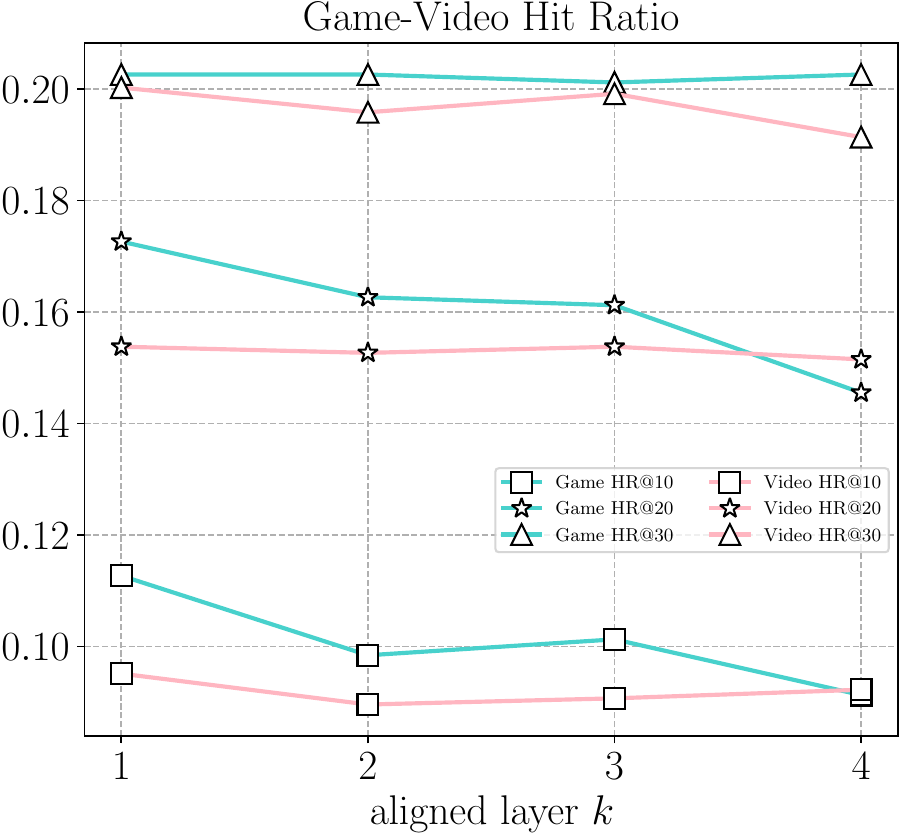}
        \label{fig:gs_groupsize}
    \end{subfigure}
    \begin{subfigure}[b]{.16\linewidth}
        \centering
        \includegraphics[width=\textwidth]{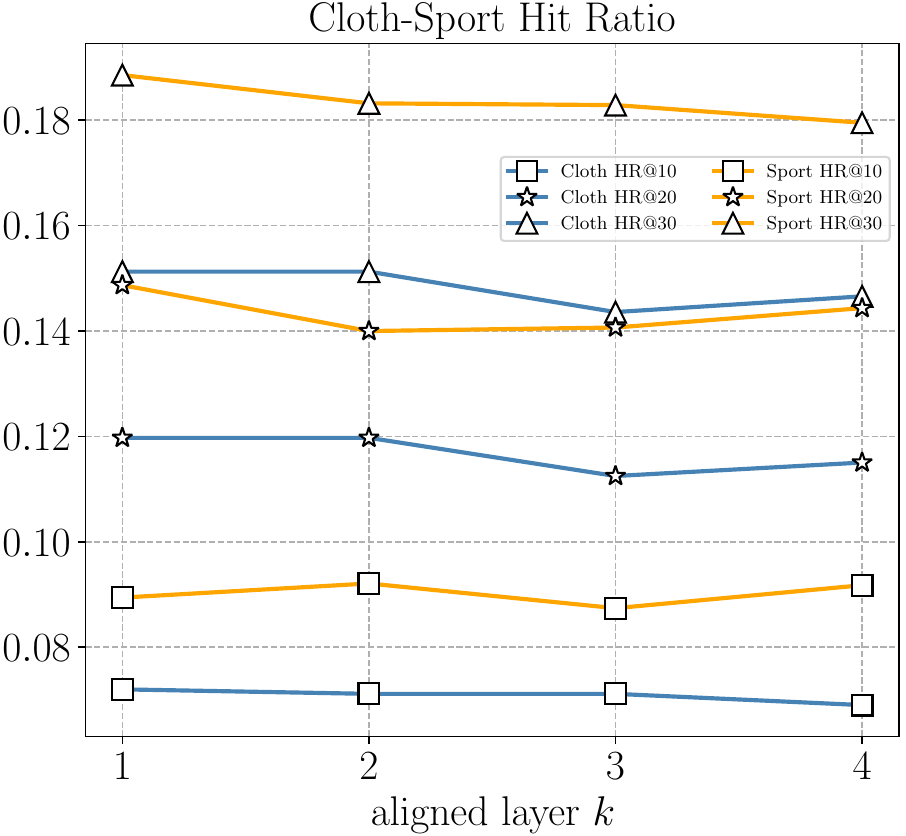}
        \label{fig:gs_flowtype}
    \end{subfigure}
    \begin{subfigure}[b]{.16\linewidth}
        \centering
        \includegraphics[width=\textwidth]{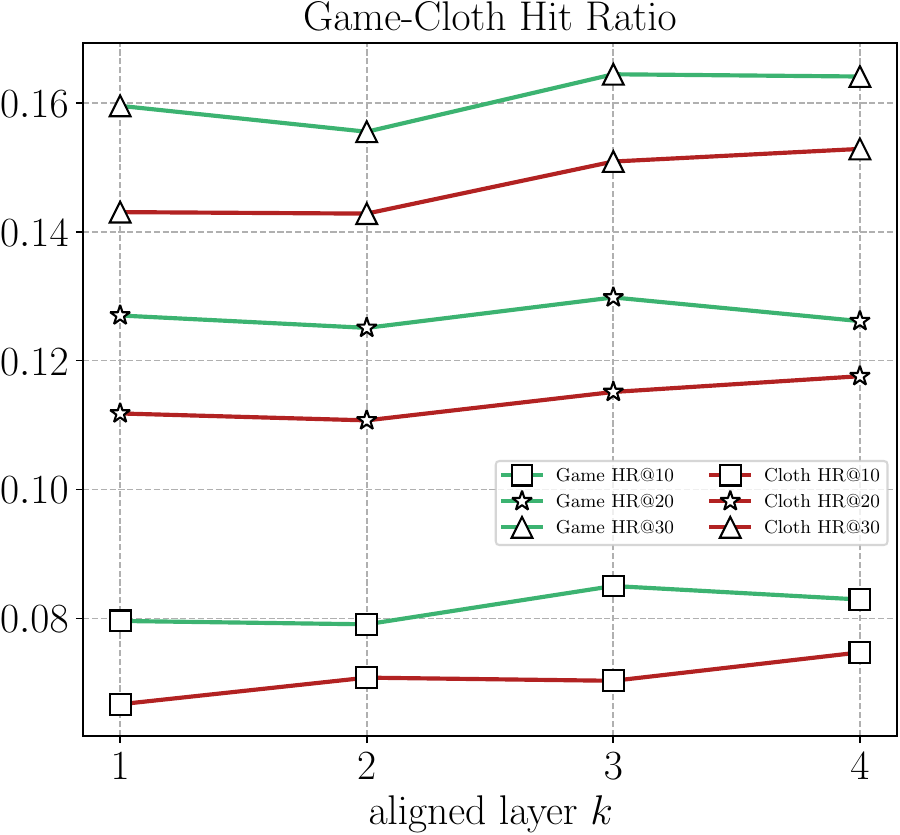}
        \label{fig:gs_flowlayer}
    \end{subfigure}
    \begin{subfigure}[b]{.16\linewidth}
        \centering
        \includegraphics[width=\textwidth]{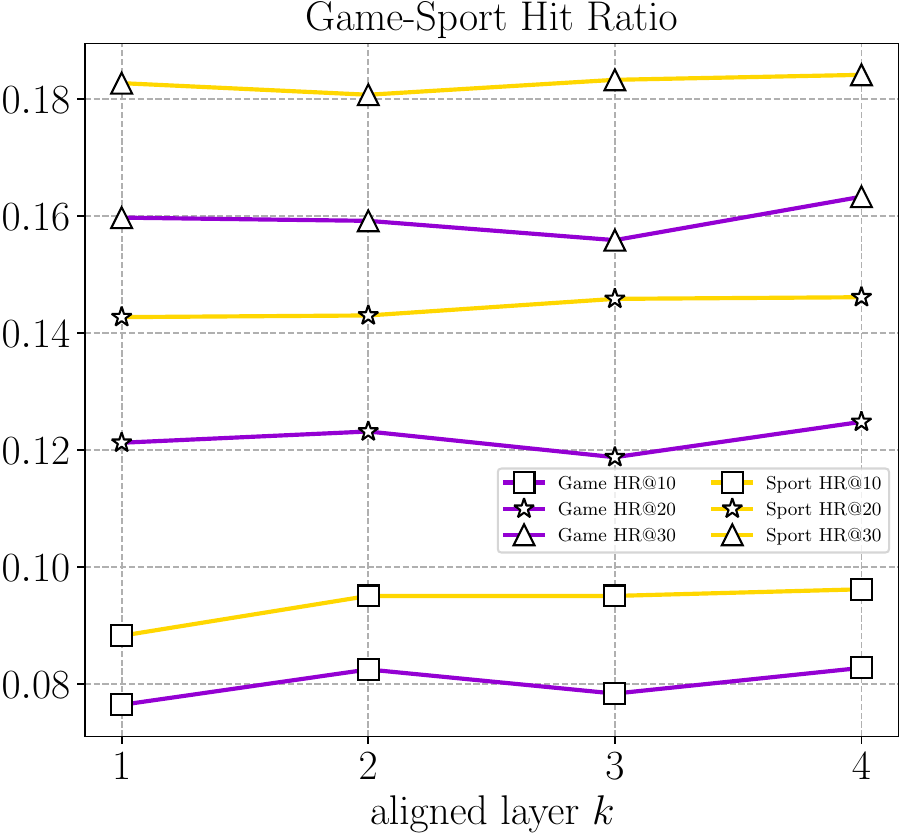}
        \label{fig:gs_align}
    \end{subfigure}
    \begin{subfigure}[b]{.16\linewidth}
        \centering
        \includegraphics[width=\textwidth]{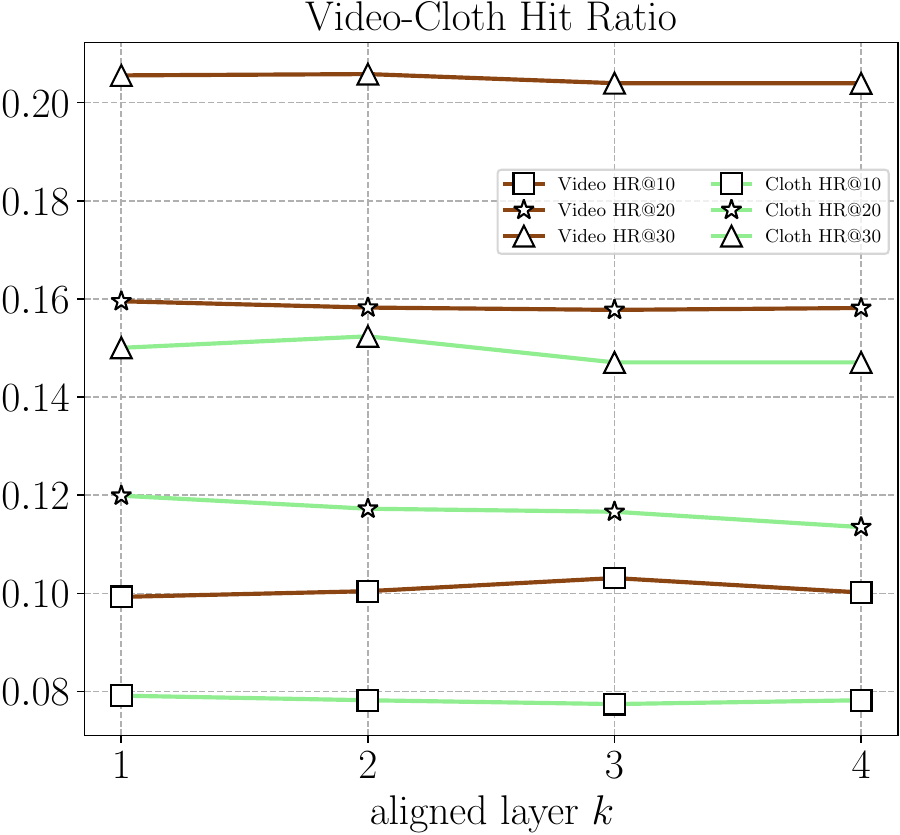}
        \label{fig:gs_gnnlayer}
    \end{subfigure}
    \begin{subfigure}[b]{.16\linewidth}
        \centering
        \includegraphics[width=\textwidth]{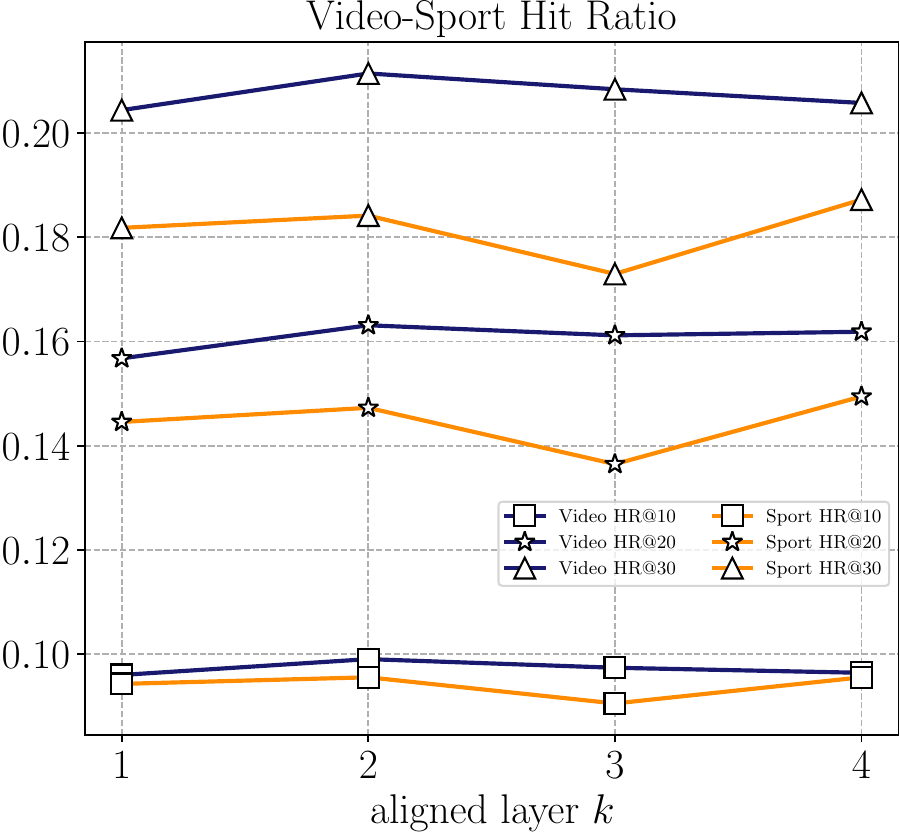}
        \label{fig:gs_gnnlayer}
    \end{subfigure}

    \begin{subfigure}[b]{.16\linewidth}
        \centering
        \includegraphics[width=\textwidth]{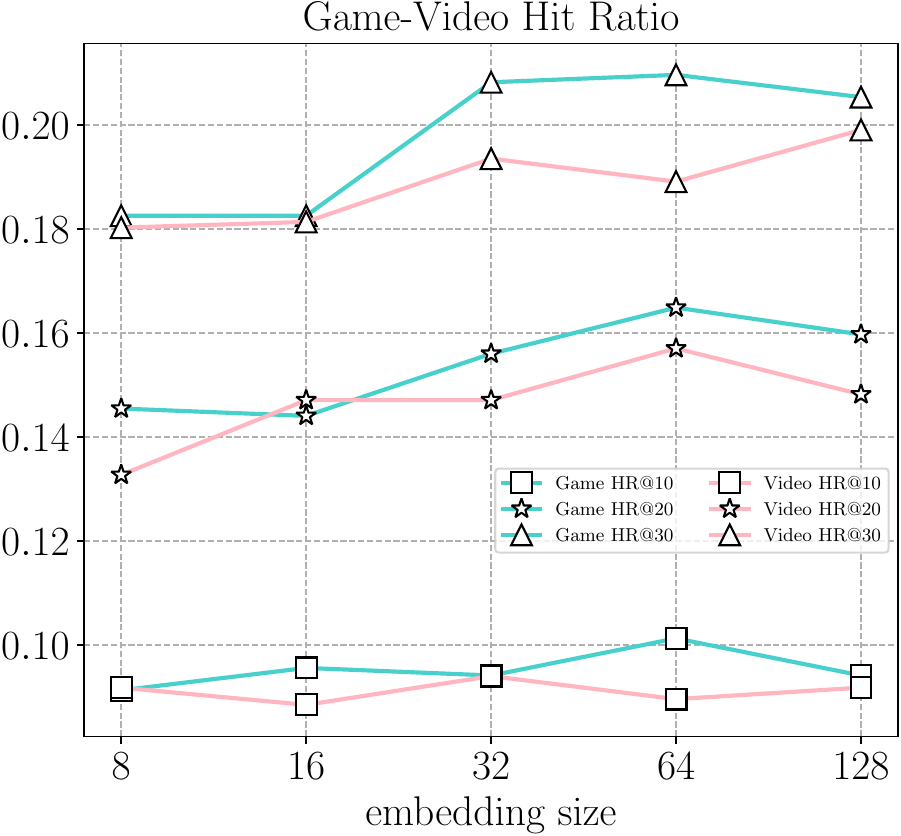}
        \label{fig:gs_groupsize}
    \end{subfigure}
    \begin{subfigure}[b]{.16\linewidth}
        \centering
        \includegraphics[width=\textwidth]{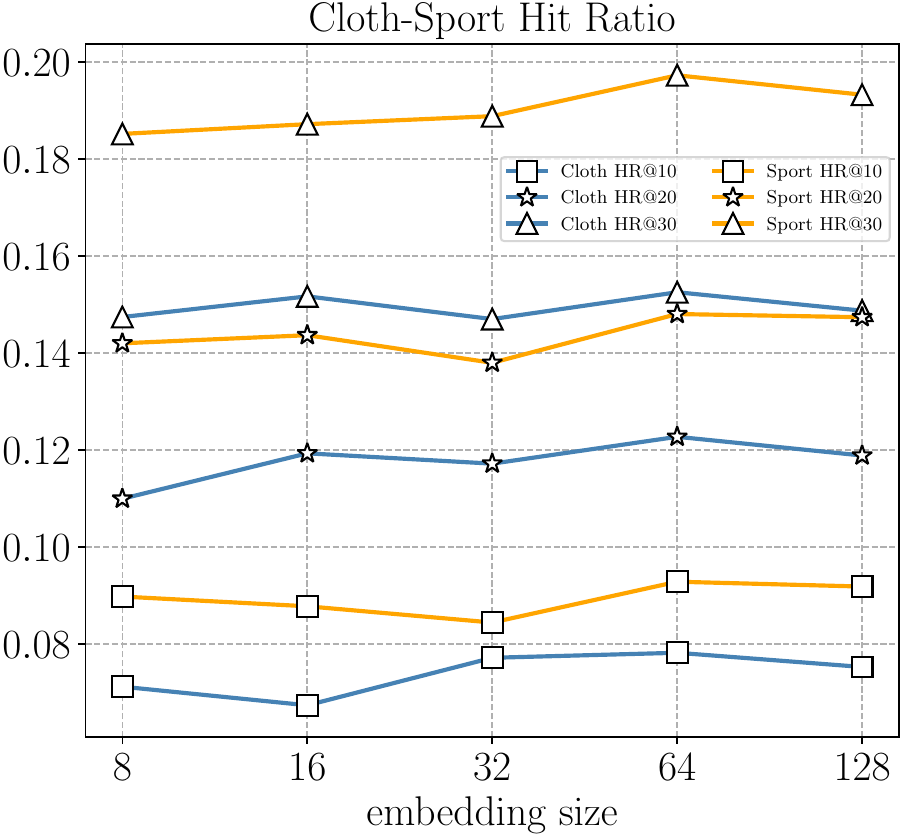}
        \label{fig:gs_flowtype}
    \end{subfigure}
    \begin{subfigure}[b]{.16\linewidth}
        \centering
        \includegraphics[width=\textwidth]{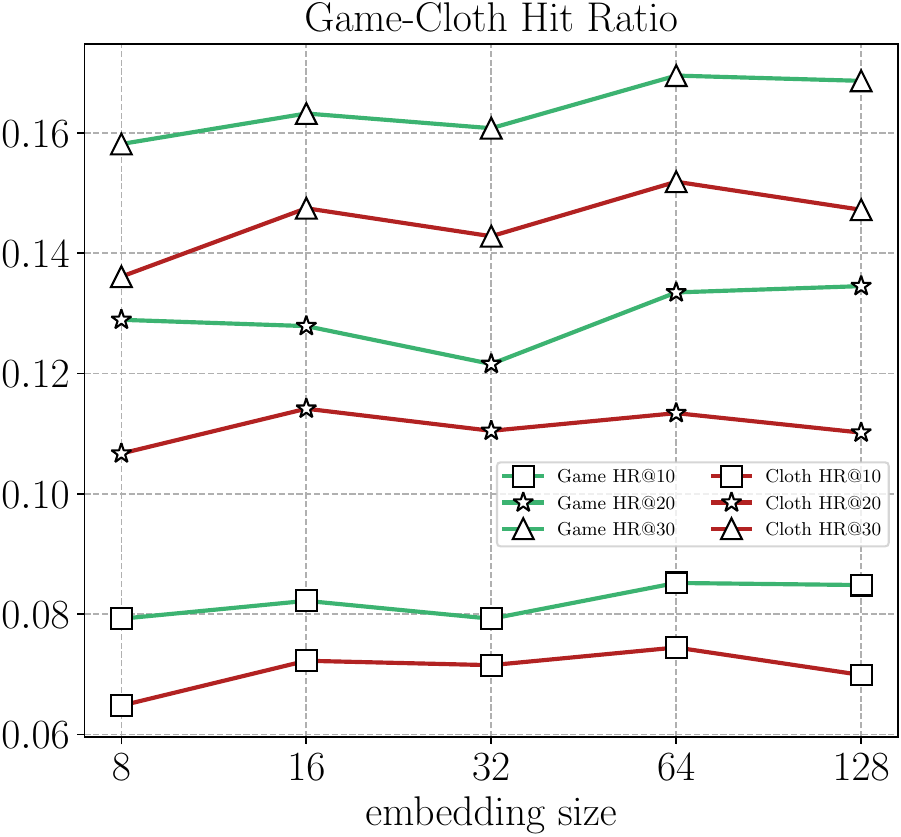}
        \label{fig:gs_flowlayer}
    \end{subfigure}
    \begin{subfigure}[b]{.16\linewidth}
        \centering
        \includegraphics[width=\textwidth]{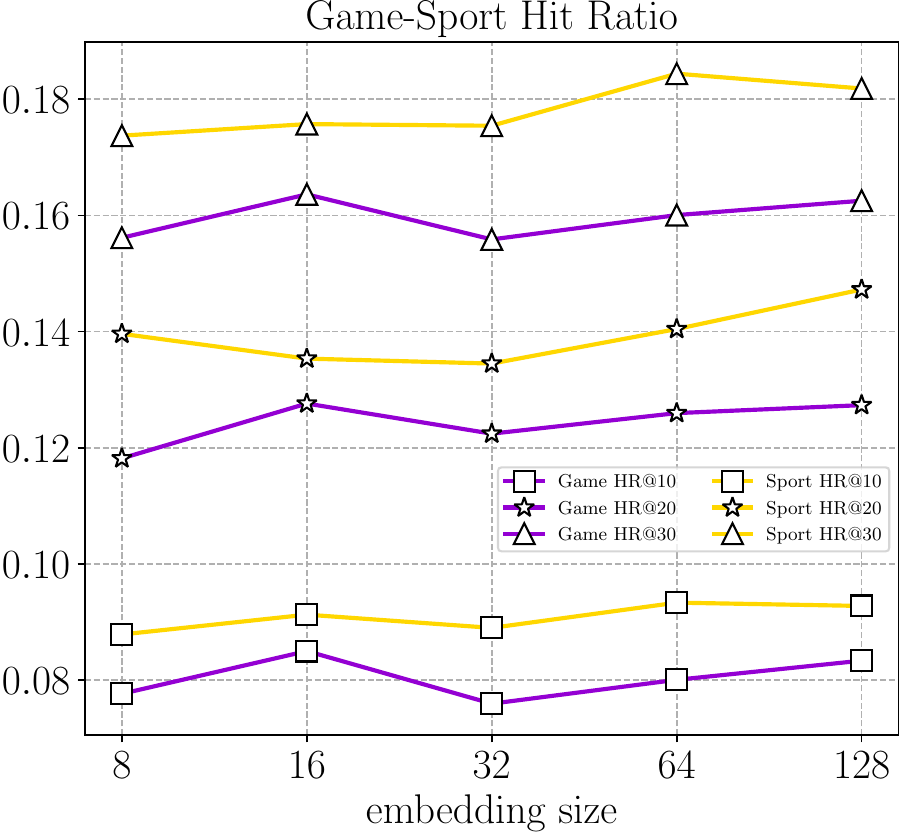}
        \label{fig:gs_align}
    \end{subfigure}
    \begin{subfigure}[b]{.16\linewidth}
        \centering
        \includegraphics[width=\textwidth]{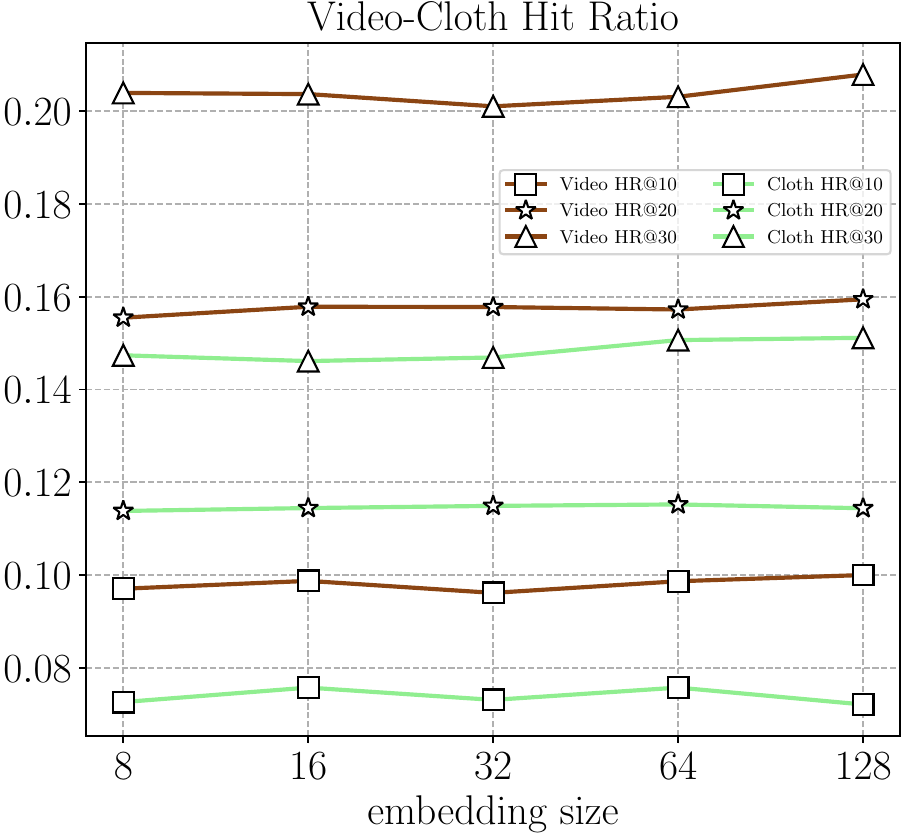}
        \label{fig:gs_gnnlayer}
    \end{subfigure}
    \begin{subfigure}[b]{.16\linewidth}
        \centering
        \includegraphics[width=\textwidth]{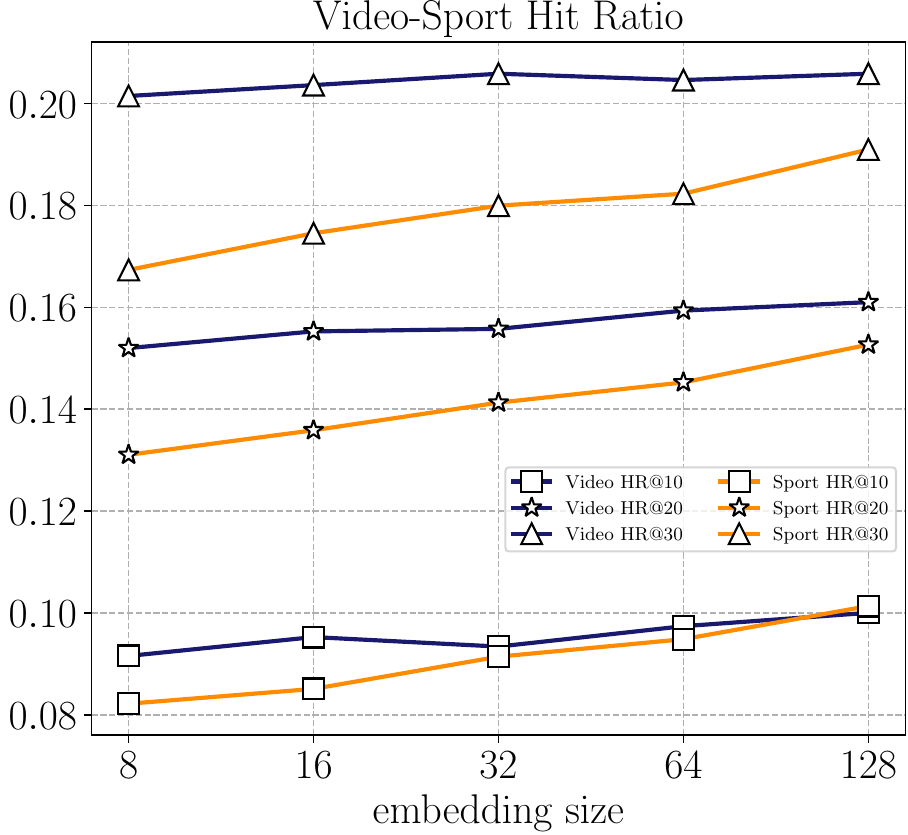}
        \label{fig:gs_gnnlayer}
    \end{subfigure}
    
    \caption{Parameter evaluation in 6 CDR tasks. Line 1: user group size $N$; line 2: the type of invertible generative model; line 3: the number of centroids $T$; line 4: the number of encoder layers $K$; line 5: the depth of shallow layers $k$; line 6: the temperature $\alpha$; line 7: embedding size.}
    \label{fig: parameters}
\end{figure*}

We search the best value of the following 7 parameters: user group size $N$, the invertible generative model, the number of centroids $T$, the number of encoder layers $K$, the depth of shallow layers $k$, the temperature $\alpha$, and embedding size throughout the encoder neural network. The results in 6 CDR tasks are shown in Fig.~\ref{fig: parameters}.

\textbf{A smaller group size yields the best performance for CIDER}. 
Notably, five CDR tasks achieve optimal performance across most metrics with a group size of N=16, while Game-Sport performs best at N=1024. 
When the group size is small, the model converges progressively to the optimal solution over multiple iterations, making it a highly efficient strategy.
Conversely, our model can also effectively infer the target joint distribution with a large enough number of random samples, which may account for the reliable performance seen with larger group sizes.

In the \textbf{Shallow Subspace Alignment} module, two important hyperparameters — namely, the number of interest classes (centroids) $T$ and the temperature $\alpha$ — control the convergence of CIDER. 
We observe that \textbf{different domains require different numbers of centroids, indicating the presence of diverse interest classes across contexts.} For instance, the Game-Video and Video-Cloth tasks achieve optimal results with $T$=5 centroids.
Game-Sport and Game-Cloth require the most centroids to align shallow user representations, suggesting a more granular division of interests in these domains.
\textbf{The temperature $\alpha$ regulates the sharpness of assigning user representations to specific centroids.} 
Lower temperatures result in the probability distribution spread more evenly across centroids, whereas higher temperatures lead to sharper assignments, with users being assigned to centroids with greater confidence. 
CIDER achieves the best performance with the highest temperature, $\alpha$=3, in most metrics across the six CDR tasks, indicating that \textbf{user representations are well-separated and strongly associated with specific centroids.}

In the deep subspace, four invertible transformations, MAF, NAF, NODE, and NCSF, are compared.
All Flow-based models employed three layers for fairness. 
\textbf{CIDER provides the most stable performance with NCSF for bijective generations in all tasks.}
As the layers of representation encoders will establish an initial basis for CIDER, we then investigate the optimal number of \textbf{encoder layers $K$}.
Since the FH principle is employed and the 1-layer encoder is not suitable for hierarchical structures, we only conduct experiments in the range of in \{2, 3, 4, 5\}.
Empirical results indicate that \textbf{a 3-layer VBGE configuration yields superior performance across multiple metrics}.
Subsequently, we fix $K$=5 and examine the \textbf{most effective shallow layers $k$} for shallow subspace alignment within \{1, 2, 3, 4\}.
The optimal value of $k$ varies among different CDR tasks. 
\textbf{Higher aligned layers possess superior results}, encouraging CIDER to decouple user representations into shallow-layer and deep-layer representations at higher levels.

Finally, we assess the \textbf{embedding size $d$} across \{8, 16, 32, 64, 128\} with 3-layer VBGEs and NCSF.
Since we concatenate the outputs of each layer, the size of representations spans across \{24, 48, 128, 192, 384\}.
\textbf{Optimal embedding sizes vary across different scenarios.}
The metrics exhibit fluctuation, with the optimal performance achieved at $d$=64 for Game-Video, Cloth-Sport, Game-Cloth, and Game-Sport tasks, and $d$=128 for Video-Cloth and Video-Sport tasks.

\section{Conclusion and Future Works}
We have introduced a hierarchical user preference modeling framework that resolves challenges in establishing cross-domain joint identifiability by maintaining cross-domain consistency. 
Specifically, we separate user representations into shallow domain-irrelevant features and deep domain-relevant features, in line with the FH principle. 
To align domain-irrelevant features, we applied centroid-based alignment to probabilistically coordinate general, domain-agnostic interest classes within the shallow subspace.
We further ensured {\it joint identifiability} stable and variant components by constructing a bijective mapping relationship across domains, guided by a causal data generation graph regarding cross-domain user preferences and behaviors. 
Empirical evaluations on real-world datasets highlight CIDER's superior performance over current methods, establishing a new state-of-the-art.
Future work will aim to uncover robust relationships and extend model generalization to multi-modal contexts.

\section{Acknowledgment}
This work was supported by the Australian Research Council Projects with Nos. LP210301259 and DP230100899, Macquarie University Data Horizons Research Centre, and Applied AI Research Centre.

\bibliographystyle{IEEEtran}
\bibliography{Reference.bib}

\vfill

\end{document}